\documentclass[longauth]{aa}  

\usepackage{graphicx}
\usepackage[varg]{txfonts}
\usepackage[colorlinks=true,linkcolor=red,citecolor=blue,urlcolor=blue]{hyperref}
\usepackage{color}

\usepackage{nicefrac}

\newcommand{\rev}[1]{#1}
\newcommand{\revrev}[1]{#1}

\bibpunct{(}{)}{;}{a}{}{,} 

\begin{document}

\title{Unveiling wide-orbit companions to K-type stars in Sco-Cen with \textit{Gaia} EDR3\thanks{Based on observations collected at the European Organisation for Astronomical Research in the Southern Hemisphere under ESO programs 099.C-0698(A), 0101.C-0341(A), 1101.C-0092(E), and 0104.C-0265(A).}}

\author{
Alexander~J.~Bohn\inst{1}
\and Christian~Ginski\inst{2}
\and Matthew~A.~Kenworthy\inst{1}
\and Eric~E.~Mamajek\inst{3,4}
\and Tiffany~Meshkat\inst{5}
\and Mark~J.~Pecaut\inst{6}
\and Maddalena~Reggiani\inst{7}
\and Christopher~R.~Seay\inst{1}
\and Anthony~G.~A.~Brown\inst{1}
\and Gabriele~Cugno\inst{8}
\and Thomas~Henning\inst{9}
\and Ralf~Launhardt\inst{9}
\and Andreas~Quirrenbach\inst{10}
\and Emily~L.~Rickman\inst{11,12}
\and Damien~Ségransan\inst{11}
}

\institute{Leiden Observatory, Leiden University, PO Box 9513, 2300 RA Leiden, The Netherlands\\
              \email{ajbohn.astro@gmail.com}
              \and Sterrenkundig Instituut Anton Pannekoek, Science Park 904, 1098 XH Amsterdam, The Netherlands
              \and Jet Propulsion Laboratory, California Institute of Technology, 4800 Oak Grove Drive, M/S 321-100, Pasadena CA 91109, USA
              \and Department of Physics \& Astronomy, University of Rochester, Rochester NY 14627, USA
              \and IPAC, California Institute of Technology, M/C 100-22, 1200 East California Boulevard, Pasadena CA 91125, USA
              \and Rockhurst University, Department of Physics, 1100 Rockhurst Road, Kansas City MO 64110, USA
              \and Institute of Astronomy, KU Leuven, Celestijnenlaan 200D, B-3001 Leuven, Belgium
              \and ETH Zürich, Institute for Particle Physics and Astrophysics, Wolfgang-Pauli-Str. 27, 8093 Zürich, Switzerland
              \and Max-Planck-Institut für Astronomie, Königstuhl 17, 69117 Heidelberg, Germany
              \and Landessternwarte, Zentrum für Astronomie der Universität Heidelberg, Königstuhl 12, 69117 Heidelberg, Germany
              \and Observatoire Astronomique de l’Université de Genève, 51 Ch. des Maillettes, 1290 Versoix, Switzerland
              \and European Space Agency (ESA), ESA Office, Space Telescope Science Institute, 3700 San Martin Drive, Baltimore, MD 21218, USA
       }

\date{Received November 15, 2020 / Accepted <date>}

\abstract 
{%
Detection of low-mass companions to stellar hosts is important to test formation scenarios of these systems.
Companions at wide separations are particularly intriguing objects as they are easily accessible for variability studies of rotational dynamics and cloud coverage of these brown dwarfs or planetary-mass objects.
} 
{%
We aim to identify new low-mass companions to young stars using the astrometric measurements provided by the \textit{Gaia} space mission.
\rev{When possible}, high-contrast imaging data collected with VLT/SPHERE \rev{is used for confirmation}.
} 
{%
\rev{We identify companion candidates from a sample of K-type, pre-main sequence stars in the Scorpius Centaurus association using the early version of the third data release of the \textit{Gaia} space mission.}
Based on the provided positions, proper motions, and magnitudes, we identify all objects within a predefined radius whose differential proper motions are consistent with a gravitationally bound system.
\rev{As the ages of our systems are known}, we derive companion masses through comparison with evolutionary tracks.
For \rev{seven identified} companion \rev{candidates} we use additional data collected with VLT/SPHERE and VLT/NACO to assess the accuracy of the properties of the companions based on \textit{Gaia} photometry alone.
}
{%
We identify \rev{110} comoving companions that have a companionship likelihood of more than 95\%. 
Further color-magnitude analysis confirms their Sco-Cen membership.
We identify \rev{ten} especially intriguing companions that have masses in the brown dwarf regime down to $20\,M_\mathrm{Jup}$.
Our high-contrast imaging data confirm \rev{both astrometry and photometric masses derived from \textit{Gaia} alone}.
\rev{We discover a new} brown dwarf companion, TYC~8252-533-1~B, \rev{with a projected separation of approximately 570\,au from its Sun-like primary}.
It is likely to be located outside the debris disk around its primary star and SED modeling of \textit{Gaia}, SPHERE, and NACO photometry provides a companion mass of $52^{+17}_{-11}\,M_\mathrm{Jup}$. 
} 
{%
We show that the \textit{Gaia} database can identify low-mass companions at wide separations from their host stars.
For K-type Sco-Cen members \textit{Gaia} can detect sub-stellar objects at projected separations larger than 300\,au and is sensitivity limited beyond 1,000\,au with a lower mass limit down to $20\,M_\mathrm{Jup}$.
\rev{A similar analysis of} other star-forming regions could significantly enlarge the sample size of such objects and test formation and evolution theories of planetary systems.
}


\maketitle

\section{Introduction}
\label{sec:introduction}

Whereas direct, adaptive optics (AO)-assisted imaging is the best tool to detect and characterize faint companions close to the diffraction limits of current 10\,m-class telescopes, these instruments usually cannot identify comoving objects at separations that are larger than several arcseconds.
The Gemini Planet Imager \citep[GPI;][]{Macintosh2014} and the Spectro-Polarimetric High-contrast Exoplanet REsearch instrument \citep[SPHERE;][]{beuzit2019} has a field of view with a radial extent of approximately 1\farcs4 and 5\farcs5 with respect to the primary star.
Even though several remarkable discoveries have been made with these instruments \citep[e.g.,][]{macintosh2015,chauvin2017b,keppler2018}, it is reasonable to assume that a non-negligible fraction of wide-orbit companions remain undiscovered as they are located outside the field of view of the respective detectors.

Studying this unexplored population of wide-orbit objects is crucial for obtaining a complete census of the occurrence rates of sub-stellar companions.
Dynamical and spectroscopic monitoring of these companions will help significantly in understanding the underlying formation mechanisms and test the efficiency of several proposed formation channels.
Due to the low amount of flux contamination from the primary star, spectroscopic analysis of these widely separated companions is relatively easy and can be conducted with non-AO instruments.
Constraining elemental abundances of the atmospheres and comparison to the stellar properties provides important clues as to whether the companion has formed in situ \citep[e.g.,][]{kroupa2001,chabrier2003}, inside a protoplanetary disk closer to the star \citep[e.g.,][]{pollack1996,boss1997}, or was even captured \citep[e.g.,][]{kouwenhoven2010,malmberg2011}.
Even transiting exomoons could be detected around these low-mass companions by monitoring their light curves.

Due to limits of the field of view of high-contrast imagers, comoving objects at these larger separations need to be identified using other techniques.
The \textit{Gaia} mission \citep{gaia2016} of the European Space Agency and especially its \rev{early version of the third} data release \citep[\textit{Gaia} \rev{EDR3};][]{gaia2020} are best suited to identify this population of low-mass companions at wide orbital separations.
Future data releases of the \textit{Gaia} mission might even provide partial orbital solutions for some of the identified companions.

In Sect.~\ref{sec:methods} of this article we describe how we use \textit{Gaia} EDR3 to search for comoving objects at wide separations and in to K-type stars in the Scorpius-Centaurus association \citep[Sco-Cen;][]{dezeeuw1999}.
Sco-Cen is composed of the three subgroups Upper Scorpius (US), Upper Centaurus-Lupus (UCL), and Lower Centaurus Crux (LCC) with mean distances of 145 pc, 140 pc, and 118 pc, respectively \citep{dezeeuw1999}.
Its youth -- with average ages of 10\,Myr, 16\,Myr, and 15\,Myr for the US, UCL, and LCC subgroups respectively \citep{pecaut2016} -- and its proximity made this association subject to several studies for young, directly imaged exoplanets \citep{rameau2013,chauvin2017b,keppler2018,haffert2019}.
The solar-type stars that are observed within the Young Suns Exoplanet Survey \citep[YSES;][]{bohn2020a} constitute a subsample of the larger selection of K-type Sco-Cen members, which is studied in this article.
We further analyze \rev{this sample of preselected companion candidates}, derive object masses and companionship probabilities in Sect.~\ref{sec:results_analysis}.
In Sect.~\ref{sec:hci_results} we present the \rev{results from complementary imaging data that were collected for seven of our preselected candidate companions.
}
\rev{These data} can be used to assess the quality of our strictly \textit{Gaia}-based results.
Lastly, we discuss our findings in Sect.~\ref{sec:discussion} and present our conclusions and further implications of this work in Sect.~\ref{sec:conclusions}.

\section{Data and methods}
\label{sec:methods}

\rev{We introduce the sample of K-type pre-main sequence stars that is objective to our analysis in Sect.~\ref{subsec:sample_selection} and we reassess their membership status in light of \textit{Gaia} EDR3 astrometric measurements.
The selection algorithm that we apply to identify comoving companion candidates to the stars of our sample in the \textit{Gaia} catalogue is presented in Sect.~\ref{subsec:methods_gaia_edr3}.
As several of the companion candidates were also imaged within the scope of YSES, we have additional near infrared imaging data to assess the properties of these companions.
Our high-contrast observations and data reduction methods are described in Sect.~\ref{subsec:methods_data_reduction}.
}

\subsection{\rev{Sample selection}}
\label{subsec:sample_selection}

\rev{Our analyses are based on} a sample of \rev{493} K-type pre-main sequence stars located in Sco-Cen.\footnote{Compiled from Table 7 of \citet{pecaut2016}.}
The subsample of solar-type stars ($0.8\,M_{\odot}<M<1.2\,M_{\odot}$) in the LCC sub-group of Sco-Cen is used within YSES to search for planetary-mass companions.
These targets and their selection are extensively described in \citet{pecaut2016}.
As their criteria for Sco-Cen membership are amongst other parameters based on kinematically constrained parallaxes, we assessed if the trigonometric \textit{Gaia} parallaxes confirm the membership status of these stars.
We applied the same thresholds for Sco-Cen membership as presented in \citet{dezeeuw1999} and \citet{pecaut2016}, which are given by $4\,\mathrm{mas}<\varpi<20\,\mathrm{mas}$, $\mu<55\,\mathrm{mas}\,\mathrm{yr}^{-1}$, $\mu_{\alpha *}<10\,\mathrm{mas}\,\mathrm{yr}^{-1}$, $\mu_\delta<30\,\mathrm{mas}\,\mathrm{yr}^{-1}$.
This analysis indicated that only ten targets (2.0\%) from the initial catalog \rev{do not comply with these astrometric requirements for Sco-Cen membership} (see Table~\ref{tbl:dismissed_targets}).
\begin{table}
\caption{
Targets that \rev{were excluded from our analysis}.
}
\label{tbl:dismissed_targets}
\def\arraystretch{1.2}
\setlength{\tabcolsep}{12pt}
\small
\centering
\begin{tabular}{@{}ll@{}}
\hline \hline
2MASS ID & Reason\\
\hline
11143442-4418240 & $\varpi=(1.94\pm0.01)\,\mathrm{mas}<4\,\mathrm{mas}$\\
\rev{11472064-4953042} & \rev{No astrometric measurements}\\
12063292-4247508 & $\mu=(56.5\pm0.1)\,\mathrm{mas\,yr}^{-1}>55\,\mathrm{mas\,yr}^{-1}$\\
\rev{12124890-6230317} & \rev{No astrometric measurements}\\
\rev{12253370-7227480} & \rev{No astrometric measurements}\\
\rev{13010856-5901533} & \rev{No astrometric measurements}\\
\rev{13032904-4723160} & \rev{No astrometric measurements}\\
\rev{13405585-4244505} & \rev{No astrometric measurements}\\
\rev{15151856-4146354} & $\sigma_\varpi=0.55\,\mathrm{mas}>0.50\,\mathrm{mas}$\\
\rev{15250358-3604455} & $\sigma_\varpi=0.57\,\mathrm{mas}>0.50\,\mathrm{mas}$\\
\rev{15272286-3604087} & \rev{No astrometric measurements}\\
15280322-2600034 & $\varpi=(1.76\pm0.06)\,\mathrm{mas}<4\,\mathrm{mas}$\\ 
15364094-2923574\tablefootmark{a} & \rev{Optical triple} \\
15374917-1840449 & $\varpi=(0.76\pm0.03)\,\mathrm{mas}<4\,\mathrm{mas}$\\
\rev{15455225-4222163} & \rev{No astrometric measurements}\\
\rev{15463111-5216580} & \rev{No astrometric measurements}\\
\rev{15494499-3925089} & \rev{No astrometric measurements}\\
\rev{15550624-2521102} & \rev{No astrometric measurements}\\
15564402-4242301 & \rev{No astrometric measurements}\\
\rev{16003134-2027050} & \rev{No astrometric measurements}\\
\rev{16015149-2445249} & \rev{No astrometric measurements}\\
\rev{16021045-2241280} & \rev{No astrometric measurements}\\
16025243-2402226 & $\varpi=(0.82\pm0.02)\,\mathrm{mas}<4\,\mathrm{mas}$\\ 
\rev{16025396-2022480} & \rev{No astrometric measurements}\\
\rev{16034536-4355492} & \rev{No astrometric measurements}\\
\rev{16062196-1928445} & \rev{No astrometric measurements}\\
\rev{16070356-2036264} & \rev{No astrometric measurements}\\
\rev{16120140-3840276} & \rev{No astrometric measurements}\\
\rev{16131738-2922198} & \rev{No astrometric measurements}\\
16141107-2305362 & $\varpi=(0.50\pm0.11)\,\mathrm{mas}<4\,\mathrm{mas}$\\ 
16195068-2154355 & \rev{No astrometric measurements}\\ 
\rev{16240632-2456468} & \rev{No astrometric measurements}\\
\rev{16271951-2441403} & \rev{No astrometric measurements}\\
\rev{16384946-2735294} & $\varpi=(3.73\pm0.72)\,\mathrm{mas}<4\,\mathrm{mas}$\\
\hline
\end{tabular}
\tablefoot{
The targets in the table were excluded from the analysis, either because these
are not consistent with Sco-Cen membership according to the criteria listed in \citet{dezeeuw1999} and \citet{pecaut2016} \rev{or the \textit{Gaia} database contained insufficient or uncertain astrometric information.}
\tablefoottext{a}{%
\rev{2MASS~J15364094-2923574 is resolved by \textit{Gaia} into an optical triple, comprised of a foreground ($\varpi$ $\simeq$ 16 mas), high proper motion, physical binary (Gaia EDR3 6209195913018942976 \& Gaia EDR3
6209195913018942720) and a third background interloper (Gaia EDR3
6209195913015396608).}
}
}
\end{table}
\begin{figure*}
\resizebox{\hsize}{!}{\includegraphics{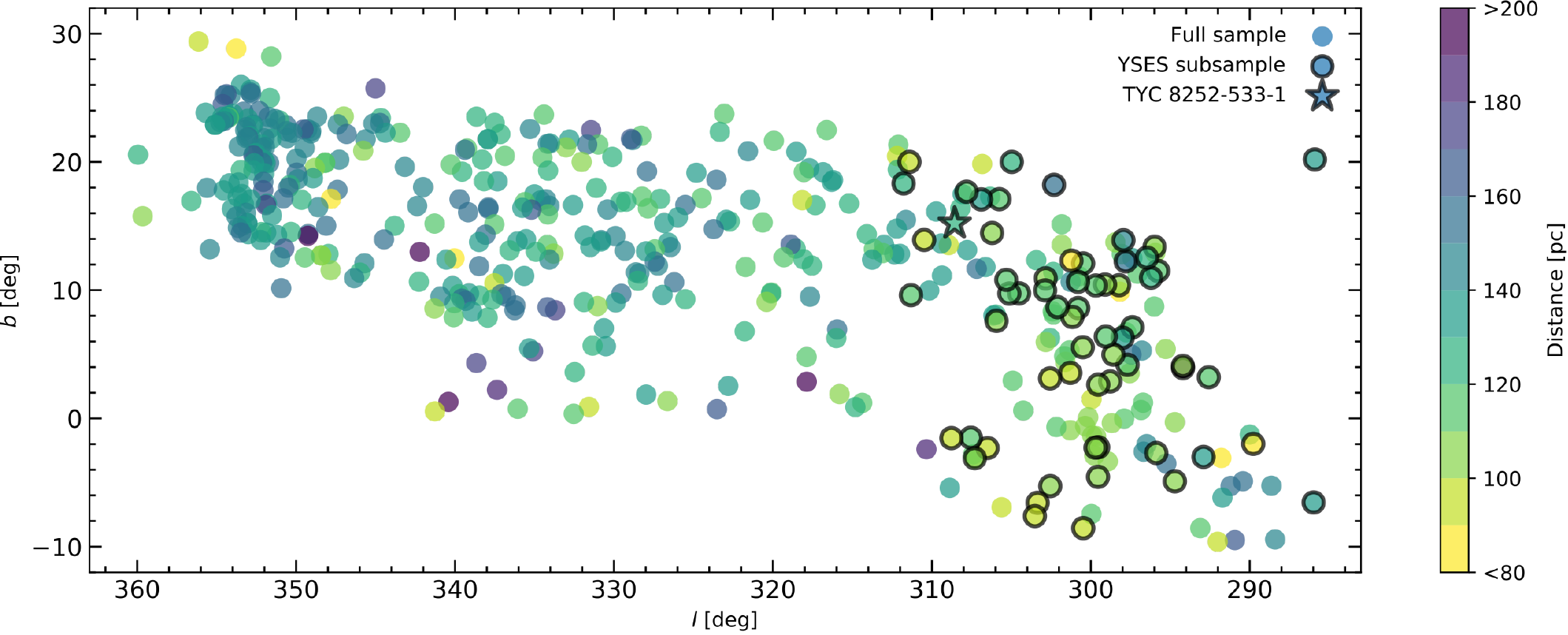}}
\caption{
\rev{Input sample of K-type pre-main sequence stars in Sco-Cen.
We show the sky positions of the targets in galactic coordinates and the color of the markers indicates the distance to the objects.
Members of the YSES subsample are highlighted by the black outlines around the markers. 
The star highlights TYC~8252-533-1, whose brown dwarf companion is analyzed in Sect.~\ref{subsec:bd_companion_yses}.
}
}
\label{fig:scocen_sample}
\end{figure*}
Whilst this hypothesis needs to be confirmed by further measurements, we removed the corresponding targets from the full sample for the scope of this work.
Furthermore, \rev{no astrometric measurements were available in the \textit{Gaia} database for 24 additional stars from the initial catalogue (see Table~\ref{tbl:dismissed_targets}).
As our identification of comoving companion candidates relies on precise astrometric measurements for all stars from the input sample, we discarded these insufficiently characterized objects as well.
}
\rev{The final input catalogue of stars used in this study is comprised of 459} K-type Sco-Cen members.
\rev{
The positions and distances of these targets are visualized in Fig.~\ref{fig:scocen_sample}.
}

\subsection{\rev{preselection of companion candidates in \textit{Gaia} EDR3}}
\label{subsec:methods_gaia_edr3}

\rev{Our preliminary companionship assessment is based on the data products provided by \textit{Gaia} EDR3.
We corrected the parallax measurements for the zero point bias as discussed in \citet{lindegren2020a}}\footnote{\rev{The applied correction formula for the parallactic zero point bias is available at \url{https://gitlab.com/icc-ub/public/gaiadr3_zeropoint}.}} \rev{and performed the correction of the $G$ band magnitudes of sources with 6-parameter astrometric solutions as described by \citet{riello2020}}.\footnote{\rev{The applied correction formula for the $G$ band magnitudes is available at \url{https://github.com/agabrown/gaiaedr3-6p-gband-correction}.}}
We compiled a first list of potential companions in four steps.
\rev{For each target from our input catalogue}
\begin{enumerate}
    \item we found all targets within \rev{a projected physical separation of $\rho_\mathrm{cutoff}=10\,000$\,au},
    \item we dismissed data without any parallax measurements or a measurement with an uncertainty larger than 0.5\,mas,
    \item we checked for the remaining objects, if the \rev{parallaxes deviated by less than 20\% from that of the host},
    \item and we \rev{required the proper motions in right ascension and declination to be consistent within a \rev{10\,km\,s$^{-1}$} interval}.
\end{enumerate}
\rev{Similar selection methods are used by \citet{fontanive2019}, \citet{fontanive2021}, and within the scope of the COol Companions ON Ultrawide orbiTS \citep[COCONUTS;][]{zhang_zj2020,zhang_zj2021} program.}
A detailed discussion of these preselection criteria can be found in Sect.~\ref{subsec:discussion_preselection_criteria}.
As we are predominantly interested in the identification of sub-stellar companions, we do not require a radial velocity (RV) measurement, since brown dwarf members of \rev{Sco-Cen} are too distant and thus too faint to allow an RV measurement in the \textit{Gaia} database.
\rev{The radial velocity measurements reported in \textit{Gaia} EDR3 are not newly derived for this catalogue, but adopted from the second data release of the \textit{Gaia} mission \citep[Gaia DR2;][]{gaia2018}, for which an average limiting magnitude to facilitate RV measurements of $\sim$13\,mag is reported \citep{cropper2018}.}

After applying the four selection criteria as described before, we identified \rev{172} companion candidates to \rev{148} stars of our sample (for a detailed list see Table~\ref{tbl:companions_full_1} in Appendix~\ref{sec:identified_companions}) within a cutoff \rev{separation of 10,000\,au}.
When studying the list of identified companion \rev{candidates} we realized that twelve of the identified companion \rev{candidates} were part of the input catalog.
\rev{These six pairs of potential stellar multiples that were} directly obtained from the list of K-type Sco-Cen members in \citet{pecaut2016} \rev{are listed in Table~\ref{tbl:duplicates}}.
\begin{table}
\caption{
\rev{
Duplications amongst our list of preselected candidate companions.
}
}
\label{tbl:duplicates}
\def\arraystretch{1.2}
\setlength{\tabcolsep}{12pt}
\small
\centering
\begin{tabular}{@{}ll@{}}
\hline \hline
2MASS ID & \textit{Gaia} EDR3 ID\\
\hline
13335329-6536473 & 5863747467220735744\\
13335481-6536414 & 5863747467220738432\\
\hline
15241147-3030582 & 6208136391829015936\\
15241303-3030572 & 6208136396126956928\\
\hline
16023814-2541389 & 6235806323497438592\\
16023910-2542078 & 6235806259081172736\\
\hline
16123916-1859284 & 6245821092014031616\\
16124051-1859282 & 6245821126373768832\\
\hline
16265700-3032232 & 6037784004457590528\\
16265763-3032279 & 6037784008760031872\\
\hline
16320058-2530287 & 6045791575844270208\\
16320160-2530253 & 6045791953801392128\\
\hline
\end{tabular}
\tablefoot{
We show targets that were both listed in the input catalog and the full selection of candidate companions presented in Appendix~\ref{sec:identified_companions}.
The horizontal lines separate individual pairs of potential candidate companions.
}
\end{table}
\begin{table*}
\caption{
\rev{
Candidate triple systems to K-type stars in Sco-Cen.
The list originates from our preselected candidates companions presented in Appendix~\ref{sec:identified_companions}.
}
}
\label{tbl:triples}
\def\arraystretch{1.2}
\small
\centering
\begin{tabular}{@{}llllll@{}}
\hline \hline
\multicolumn{2}{c}{Primary} & \multicolumn{2}{c}{Secondary} & \multicolumn{2}{c}{Tertiary}\\
2MASS ID & \textit{Gaia} EDR3 ID & 2MASS ID & \textit{Gaia} EDR3 ID & 2MASS ID &  \textit{Gaia} EDR3 ID\\
\hline
11554295-5637314 & 5343603288120259072 & - & 5343603288130858112 & - & 5343603180734334592\\
12094184-5854450 & 6071087597518919040 & - & 6071087597518919808 & - & 6071087597497876480\\
12123577-5520273 & 6075815841096386816 & - & 6075816592695096576 & - & 6075816596995760640\\
12474824-5431308 & 6073980172067600640 & - & 6073980240787079680 & - & 6073980172067600000\\
13071310-5952108 & 6056115131031531264 & - & 6056115135337482496 & - & 6056115169731484288\\
13335481-6536414 & 5863747467220738432 & 13335329-6536473 & 5863747467220735744 & - & 5863747462890662144\\
13540743-6733449 & 5850443307764629376 & - & 5850443303440130816 & - & 5850443303446183040\\
15113968-3248560 & 6207460471351260160 & - & 6207460436991521280 & - & 6207460471351260928\\
- & 6200310514738629504 & 15171083-3434194 & 6200310519037175040 & - & 6200310484677437184\\
15241147-3030582 & 6208136391829015936 & 15241303-3030572 & 6208136396126956928 & - & 6208136396122878976\\
15370214-3136398 & 6208381582919629568 & - & 6208381587215253504 & - & 6208381587220236160\\
15451286-3417305 & 6014696841553696768 & - & 6014696841553696896 & - & 6014696875913435520\\
16065795-2743094 & 6042124910722287744 & - & 6042124915024285440 & - & 6042124807643059968\\
16085427-3906057 & 5997035351934438784 & - & 5997035317574700544 & - & 5997035416337166976\\
16114387-2526350 & 6049537165279381632 & - & 6049532737170518016 & - & 6049537169580902272\\
16135801-3618133 & 6022499010422921088 & - & 6022499006128860928 & - & 6022499010440068864\\
16161423-2643148 & 6042418858284146688 & - & 6042418862581174016 & - & 6042418828224074752\\
16204468-2431384 & 6049266036882633856 & - & 6049266002522895232 & - & 6049266414839756416\\
- & 6024816059387932416 & - & 6024816020718366464  & 16235484-3312370 & 6024816025017339392\\
16265763-3032279 & 6037784008760031872 & 16265700-3032232 & 6037784004457590528 & - & 037784004457597568\\
16345314-2518167 & 6047289699098449920 & - & 6047289699098450944 & - & 6047289733458188032\\
\hline
\end{tabular}
\tablefoot{
\rev{
The targets are classified as potential primary, secondary or tertiary of the system by increasing $G$ band magnitudes.
Only for targets from our input catalog we list the corresponding 2MASS identifiers.
}
}
\end{table*}
\rev{Three of these candidate multiple systems from the input catalog even host a third potential companion that was identified by our preselection for both of the respective Sco-Cen K-type stars:
these are Gaia EDR3 5863747462890662144, Gaia EDR3 6208136396122878976, and Gaia EDR3 6037784004457597568 that are preselected as candidate companions to the potential stellar multiples composed of Gaia EDR3 5863747467220735744 (2MASS J13335329-6536473) and Gaia EDR3 5863747467220738432 (2MASS J13335481-6536414), Gaia EDR3 6208136391829015936 (2MASS J15241147-3030582) and Gaia EDR3 6208136396126956928 (2MASS J15241303-3030572), and Gaia EDR3 6037784004457590528 (2MASS J16265700-3032232) and Gaia EDR3 6037784008760031872 (2MASS J16265763-3032279), respectively.
}
After removing these duplicates from our list of preselected companions, \rev{163} potential companions to \rev{142} Sco-Cen members remained.\footnote{To report the full output of our algorithm, we did not remove these duplications from the table presented in Appendix~\ref{sec:identified_companions}. All further analysis, however, is performed on the cleaned sample of \rev{163} individual companion candidates.}

Amongst these newly identified candidate binary systems we found \rev{21} potential systems of even higher multiplicity.
\rev{In addition to the three previously mentioned candidate triples comprising two stars from our input catalog each, we found 18 further targets that were associated with two individual candidate companions according to our preselection.
All of these potential triple systems are listed in Table~\ref{tbl:triples}.
}

In addition, we identify HD~98363 (2MASS J11175813-6402333, Gaia~EDR3 5240643988513309952) among this sample of preselected candidate companions.
\citet{bohn2019} showed that HD~98363 is the $1.9\,M_\sun$ primary to the solar-mass star Wray~15-788 (2MASSJ11175186-6402056, Gaia~EDR3 5240643988513310592) from our target stars.
This demonstrates that the input stars of the algorithm do not necessarily have to be the primary of the system, as the detected companions can be equal or even higher mass.
\rev{%
Furthermore, we identify seven candidate companions that were also observed with SPHERE as part of YSES.
The identifiers of all primary stars with companion candidates from our preselection that have been imaged within the scope of YSES are listed in Table~\ref{tbl:yses_targets}.
}
\begin{table}
\caption{
\rev{
Candidate companion systems with complementary SPHERE data from YSES.
}
}
\label{tbl:yses_targets}
\def\arraystretch{1.2}
\setlength{\tabcolsep}{12pt}
\small
\centering
\begin{tabular}{@{}ll@{}}
\hline \hline
2MASS ID & \textit{Gaia} EDR3 ID\\
\hline
12195938-5018404 & 6126648698878768384\\
12391404-5454469 & 6074374346993398144\\
12505143-5156353 & 6075310478057303936\\
12560830-6926539 & 5844909156504879360\\
13130714-4537438 & 6088027047281877120\\
13233587-4718467 & 6083750638577673088\\
13335481-6536414 & 5863747467220738432\\
\hline
\end{tabular}
\end{table}
\begin{figure*}
\resizebox{\hsize}{!}{\includegraphics{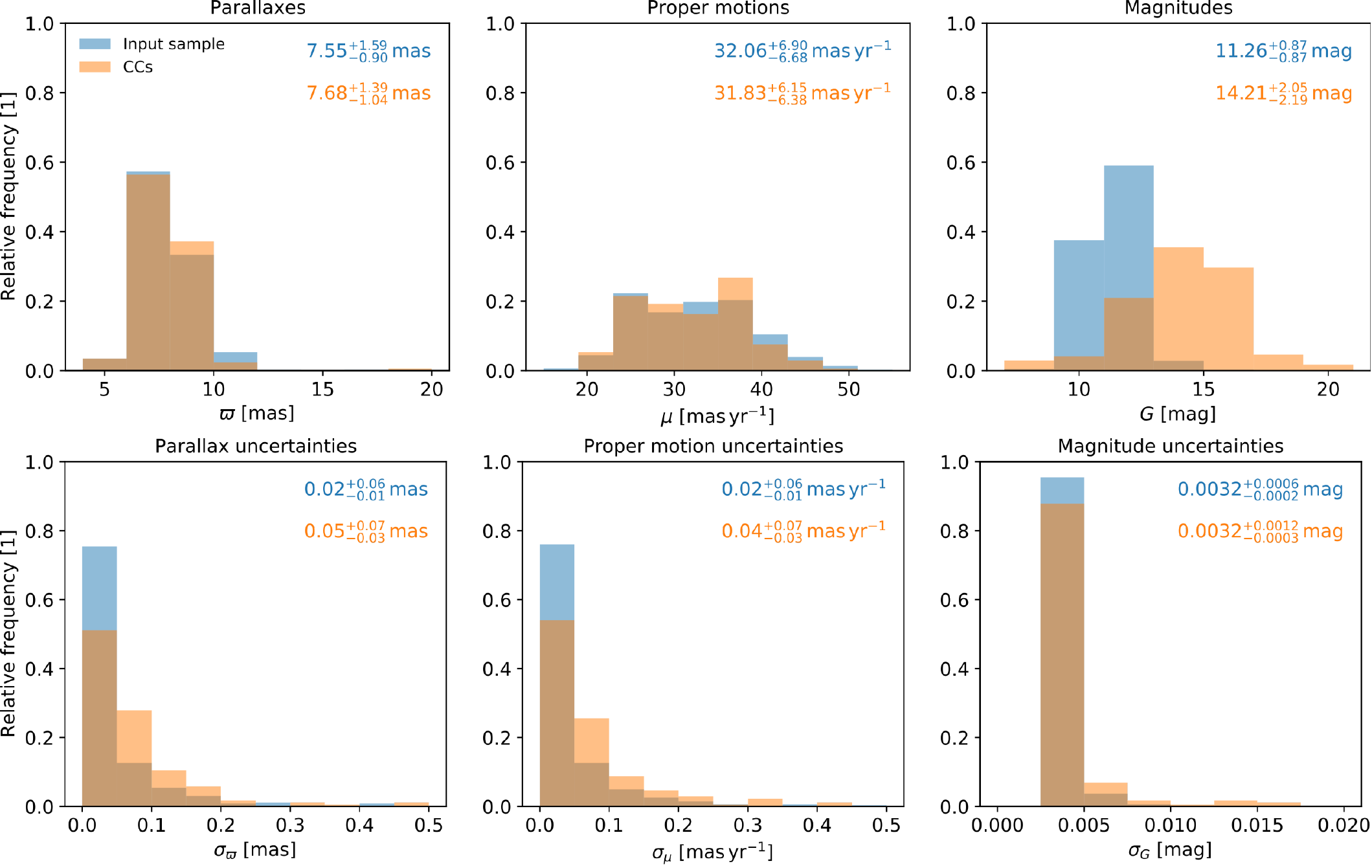}}
\caption{
\rev{%
Astrometric and photometric properties of the input sample and the preselected list of candidate companions (CCs).
In the top panel we present the distributions of the object parallaxes, proper motions, and mangitudes from both samples.
In the lower panel the uncertainties are shown.
The colored values in the upper right of each panel indicate the median and the 68\% confidence intervals of the associated distributions.
}
}
\label{fig:sample_properties}
\end{figure*}
A detailed analysis of \rev{these} objects \rev{that combines \textit{Gaia} and high-contrast imaging data} is performed in Sect.~\ref{sec:hci_results}.
The planetary-mass companions that \citet{bohn2020a,bohn2020c,bohn2021} detected for YSES~1 (TYC~8998-760-1, 2MASS J13251211-6456207, Gaia~EDR3 5864061893213196032) and \rev{YSES~2 (TYC 8984-2245-1, 2MASS J11275535-6626046, Gaia EDR3 5236792880333011968)}, however, are not listed in the \textit{Gaia} archive.
This is explained by the large contrast of \rev{all three} companions that are found at angular separations of less than 3\farcs5, which places them directly inside the red dashed exclusion zone highlighted in Fig.~\ref{fig:separation_magnitude}.

\rev{
We further compared the astrometric and photometric properties of the targets from our input sample to the preselection of candidate companions.
The parallaxes, proper motions, and magnitudes among both groups and their corresponding uncertainties are presented in Fig~\ref{fig:sample_properties}.
}
\rev{%
The distributions of parallaxes and proper motions are very similar among our input sample and the detected candidate companions.
We derive median parallaxes of $7.55^{+1.59}_{-0.90}$\,mas and $7.68^{+1.39}_{-1.04}$\,mas and proper motions of $32.06^{+6.90}_{-6.68}$\,mas\,yr$^{-1}$ and $31.83^{+6.15}_{-6.38}$\,mas\,yr$^{-1}$ for the former and latter sample, respectively.\footnote{\rev{The uncertainties represent the 68\% confidence intervals of the distributions.}}
This is not surprising as our pre-selection criteria are supposed to identify objects with astrometric properties that are similar to those of the K-type Sco-Cen members from \citet{pecaut2016}.
The $G$ band magnitude distributions of our input catalog and identified candidate companions are distinct, with median values of $11.26\pm0.87$\,mag and $14.21^{+2.05}_{-2.19}$\,mag, respectively.
This is also expected, as the input catalog contains only stars of the same spectral type that are all members of the same association;
hence no large range of the stellar magnitudes is covered by this sample.
The companion candidates, however, exhibit fainter magnitudes that extend down to the \textit{Gaia} limiting magnitude of $G\approx21$\,mag.
Due to the fainter nature of our candidate companions, their astrometric uncertainties are on average larger by a factor of 2 than those of the stars from our input catalog (see bottom left and middle panel of Fig.~\ref{fig:sample_properties}).
The median magnitude uncertainties among both samples are comparable though, yet there are a few objects with magnitude uncertainties as large as 0.02\,mag in our candidate companion sample.
These outliers are associated to the faintest targets from our preselection algorithm.
}

This preselected sample can be refined by comparing the differential projected velocity of a candidate companion to the maximum allowed speed for a gravitationally bound orbit based on the masses of both bodies and their separation.
A detailed application of this refinement procedure is described in Sect.~\ref{subsec:companionship_assessment}.

\subsection{High-contrast imaging observations}
\label{subsec:methods_data_reduction}

\rev{%
Seven of our preselected companion candidates were also imaged with SPHERE as part of YSES.
The identifiers of these systems are listed in Table~\ref{tbl:yses_targets}.
For a potential brown dwarf companion around} TYC~8252-533-1 \rev{we collected additional data with} NACO \citep{lenzen2003,rousset2003} at ESO's Very Large Telescope.
A detailed overview of \rev{all} observations and the weather conditions is provided in Table~\ref{tbl:yses_observations} in Appendix~\ref{sec:hci_observing_conditions}.
The reduction is detailed in \citet{bohn2020a} with a custom processing pipeline that is based on version 0.8.1 of the \texttt{PynPoint} package \citep{stolker2019}.
As the companions are reasonably bright and widely separated from the star, we did not perform any PSF subtraction when evaluating their \rev{astrometry and} photometry.

\section{Gaia results and analysis}
\label{sec:results_analysis}

\rev{%
In this Section we analyze the results that were derived from our preselection described in Sect.~\ref{subsec:sample_selection}.
In Sect.~\ref{subsec:raw_astro_photo} we show the object magnitudes of identified candidate companions as a function of angular separation with respect to the primary star.
As we know the ages of all our targets, we can convert these flux measurements to mass estimates for the companion candidates.
This analysis is detailed in Sect.~\ref{subsec:mass_estimation}.
We further assess the colors of our candidate companions and evaluate whether these are consistent with the proposed Sco-Cen membership in Sect.~\ref{subsec:color_magnitude_analysis}.
We analyze the relative motions of all identified objects and probe if these are in agreement with a gravitationally bound orbit around the primary star.
This assessment is presented in Sect.~\ref{subsec:companionship_assessment}.
In Sect.~\ref{subsec:high_confidence_companions} we present the properties of the candidates that exhibit the highest likelihoods to be bound companions.
}

\subsection{Raw \textit{Gaia} astrometry and photometry}
\label{subsec:raw_astro_photo}

\begin{figure}
\resizebox{\hsize}{!}{\includegraphics{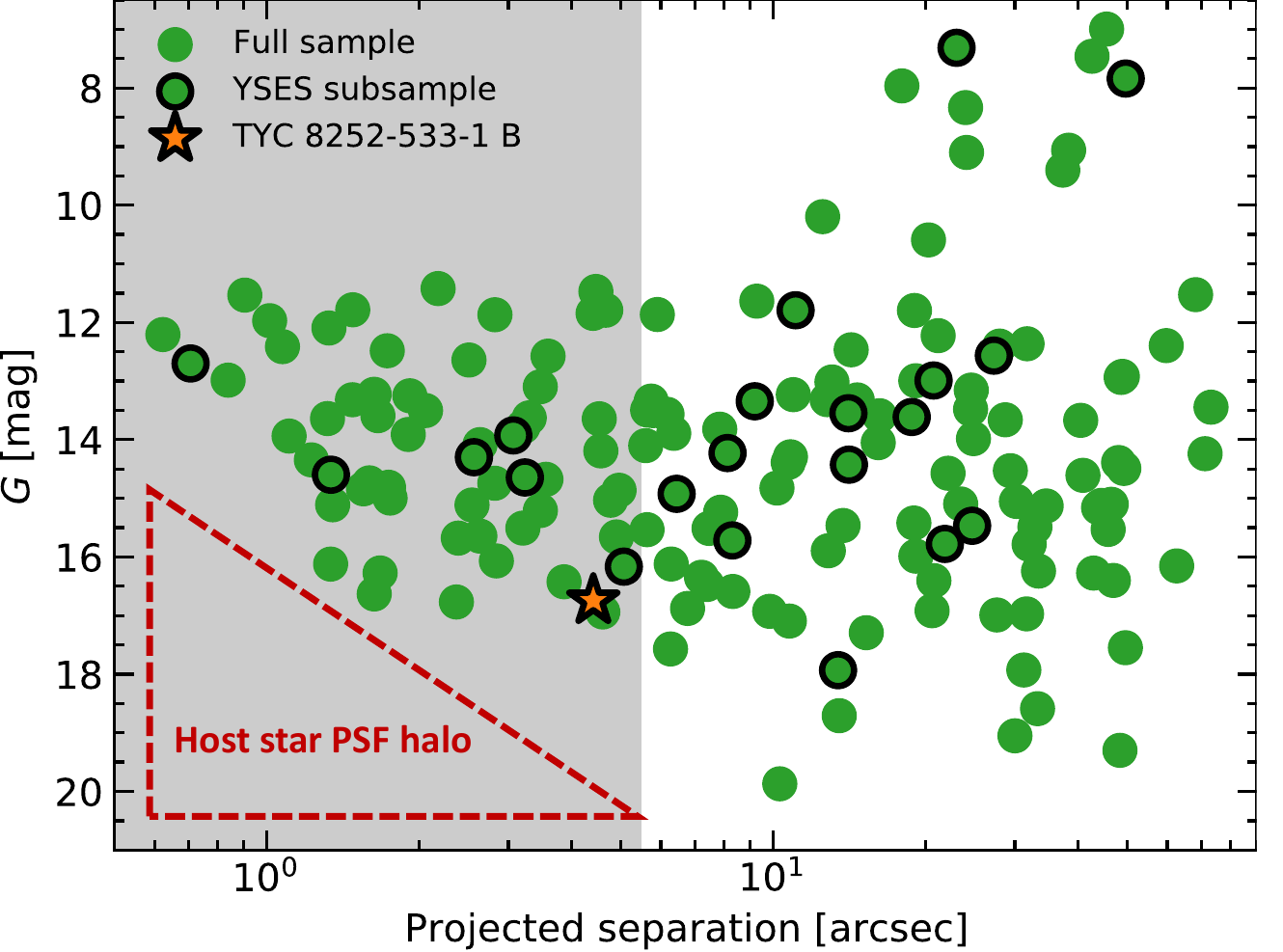}}
\caption{
Identified companion \rev{candidates} to K-type Sco-Cen members.
Members of the YSES subsample are highlighted by the black outline around the marker.
The orange star highlights the brown dwarf companion TYC~8252-533-1~B that is analyzed in Sect.~\ref{subsec:bd_companion_yses}.
The apparent \textit{Gaia} $G$ band magnitude is presented as a function of projected separation in arcseconds.
Due to the PSF halo of the primary, the detectable magnitude threshold increases with larger separations from the host.
This area of reduced sensitivity is indicated by the red, dashed triangle. 
\rev{The gray-shaded region represents the field of view of the SPHERE/IRDIS detector.}
}
\label{fig:separation_magnitude}
\end{figure}

In Fig.~\ref{fig:separation_magnitude} we present the apparent $G$ band magnitudes of the identified companion candidates as a function of projected separation.
The detection sensitivity improves with increasing separation from the target, which can be attributed to the PSF halo of the host star.
There is a clear exclusion zone (as indicated by the red, dashed triangle in Fig.~\ref{fig:separation_magnitude}) in which the presented technique is not able to detect any low mass comoving objects.
This region is however highly complementary to the parameter space that is covered by the field of view of current generation of high-contrast imaging instruments.
\rev{%
The gray-colored background of Fig.~\ref{fig:separation_magnitude} shows the field of view of the SPHERE/IRDIS camera with a radial extent of $\leq$\,5\farcs5.
The seven companions that were observed within the scope of YSES are located in this part of the parameter space.
}
For separations that are larger than approximately 10 arcseconds the \textit{Gaia} contrast to the primary becomes background limited and we are sensitive to objects with an apparent $G$ band magnitude down to 20\,mag.
This is in good agreement with the limiting magnitude of $G=21$\,mag that is required for objects to appear with a five-parameter astrometric solution in the EDR3 catalog \citep{gaia2020}.

\subsection{Mass estimation}
\label{subsec:mass_estimation}

To characterize the identified companion \rev{candidates} in further detail, we determined the absolute magnitudes of the objects using the parallax measurement provided by \textit{Gaia} \rev{EDR3}.
In many cases, the parallaxes of the identified companion candidates have larger uncertainties than the parallax measurements of the stars from our input catalog, which have a median error of only 0.02\,mas \rev{(see Fig.~\ref{fig:sample_properties})}.
We therefore used the latter parallax values to estimate the distances to the detected objects. \rev{Even though this requires true companionship between both objects, this estimate is a reasonable assumption as all candidate companions are Sco-Cen members according to the astrometric membership criteria from \citet{dezeeuw1999} and \citet{pecaut2016} (see Sect.~\ref{subsec:sample_selection}).}
As \citet{pecaut2016} provide age estimates\footnote{We note that these age estimates rely on pre-\textit{Gaia} parallaxes that were derived kinematically and which might differ from the value reported in \textit{Gaia} EDR3. A reassessment of these parameters -- especially for the identified systems with low-mass companions -- is advisable, but beyond the scope of this work.} for all systems but one, we converted the absolute magnitudes of the companions to object masses by evaluation of BT-Settl models \citep{allard2012,baraffe2015} at the corresponding system age.
The remaining system without an age measurement, SZ~65 (2MASS J15392776-3446171, Gaia EDR3 6013399894569703040), is a member of the Lupus star forming region, so we assigned it an age of 2\,Myr in accordance with the average age of this cloud complex \citep{comeron2008,alcala2014}.
For all ages we assumed an uncertainty of 2\,Myr with a youngest possible age of 1\,Myr.
The BT-Settl models we used are valid for objects with masses smaller than $1.2\,M_\sun$, which is equivalent to an apparent $G$ band magnitude of approximately 11\,mag at the average distance of Sco-Cen and at an age of 15\,Myr.
For the brightest objects from Fig.~\ref{fig:separation_magnitude}\rev{, which were outside the magnitude range supported by the BT-Settl models,} we used MIST isochrones \citep{dotter2016,choi2016} to convert their $G$ band photometry to a stellar mass.
As the object masses strongly depend on the underlying system age and only a single photometric measurement is used for the derivation, we do not claim that the provided mass estimates are very precise, but instead indicate whether it is a brown dwarf or a stellar companion.
The derived mass estimates for the companion candidates are listed in Table~\ref{tbl:companions_full_1} in Appendix~\ref{sec:identified_companions}.

\begin{figure}
\resizebox{\hsize}{!}{\includegraphics{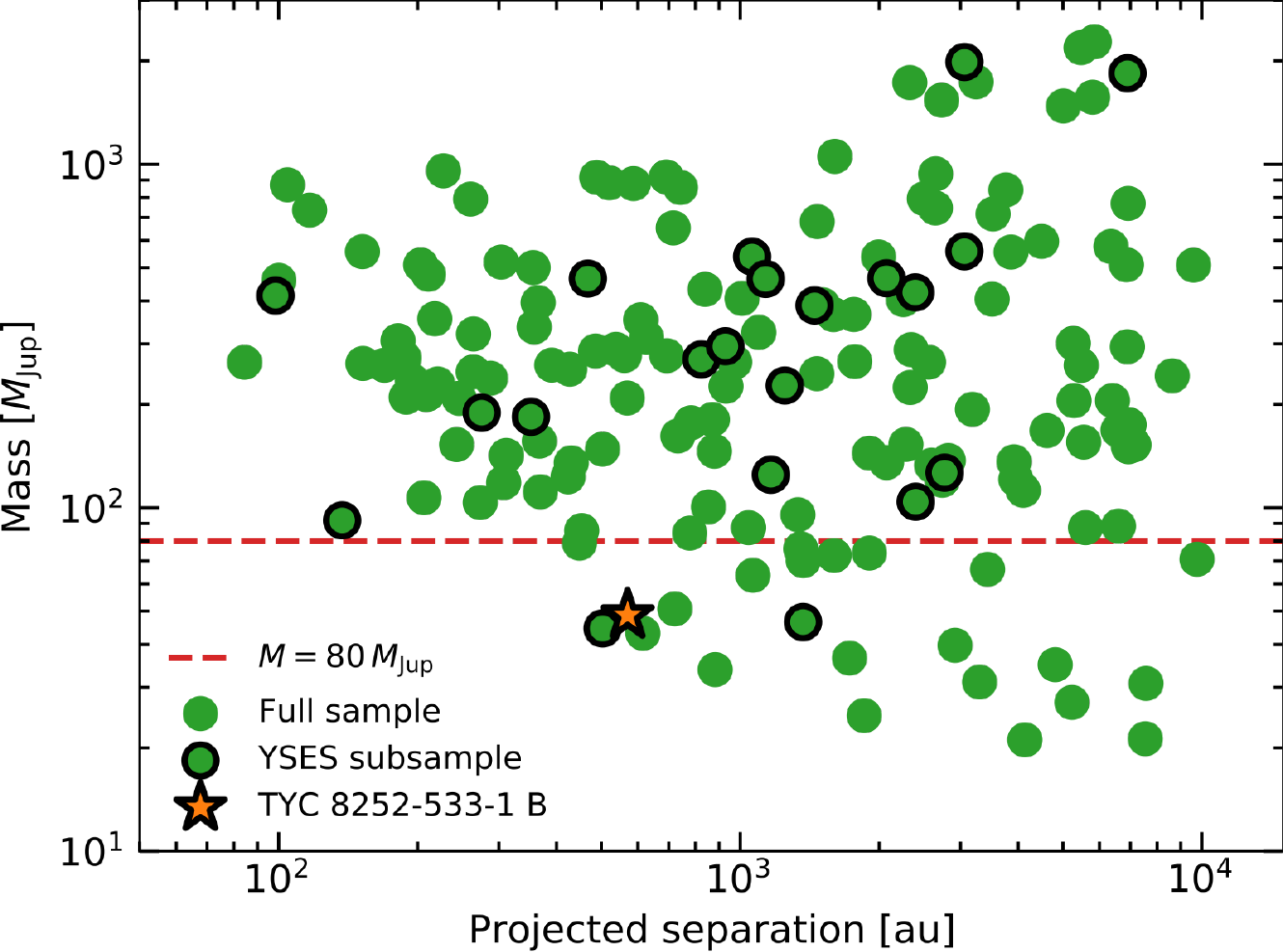}}
\caption{
Identified companion \rev{candidates} to K-type Sco-Cen members II.
Members of the YSES subsample are highlighted by the black outline around the marker.
The orange star highlights TYC~8252-533-1~B analyzed in Sect.~\ref{subsec:bd_companion_yses}.
The companion mass is presented as a function of projected separation in Astronomical Units.
Mass conversion was performed by comparison to BT-Settl and MIST models evaluated at the system age.
The dashed red line at $80\,M_\mathrm{Jup}$ indicates the threshold between brown dwarfs and stellar mass companions.
}
\label{fig:separation_mass}
\end{figure}

We present the companion masses as a function of projected physical separation in Fig.~\ref{fig:separation_mass}.
The conversion from projected angular separations to projected physical separations was performed using the distance estimate based on the parallax measurement of the primary.
We do not include the difference in \textit{Gaia} parallaxes between primary and companion to determine a three-dimensional separation, as these measurements have typical errors of \rev{0.02\,mas} and \rev{0.05\,mas} for stars from our input catalog and companion candidates, respectively \rev{(see Fig.~\ref{fig:sample_properties})}.
For two average Sco-Cen members with parallaxes of 7.5\,mas, Gaussian error propagation provides an uncertainty for the radial separation in the order of \rev{1\,pc} or approximately \rev{200,000\,au}.
This uncertainty is much larger than our measured projected separations and would dominate the uncertainty of the derived parameter.
We proceed using projected separations as a lower limit for the three-dimensional distances.

In Fig.~\ref{fig:separation_mass} the region of limited sensitivity due to the PSF halo of the primary is still clearly visible and it implies that for separations closer than approximately 300\,au we are only sensitive to stellar-mass companions.
Farther away from the host, however, we can detect comoving brown dwarf companions with masses smaller than $80\,M_\mathrm{Jup}$.
The lower mass threshold is reached for separations larger than approximately 1000\,au and we are sensitive to objects masses as low as $20\,M_\mathrm{Jup}$.
Exoplanets at Sco-Cen distance and age and with masses below the deuterium burning limit of approximately $13\,M_\mathrm{Jup}$ are thus marginally too faint to be confidently detected with \textit{Gaia} \rev{EDR3}.

\subsection{Color-magnitude analysis}
\label{subsec:color_magnitude_analysis}

\begin{figure}
\resizebox{\hsize}{!}{\includegraphics{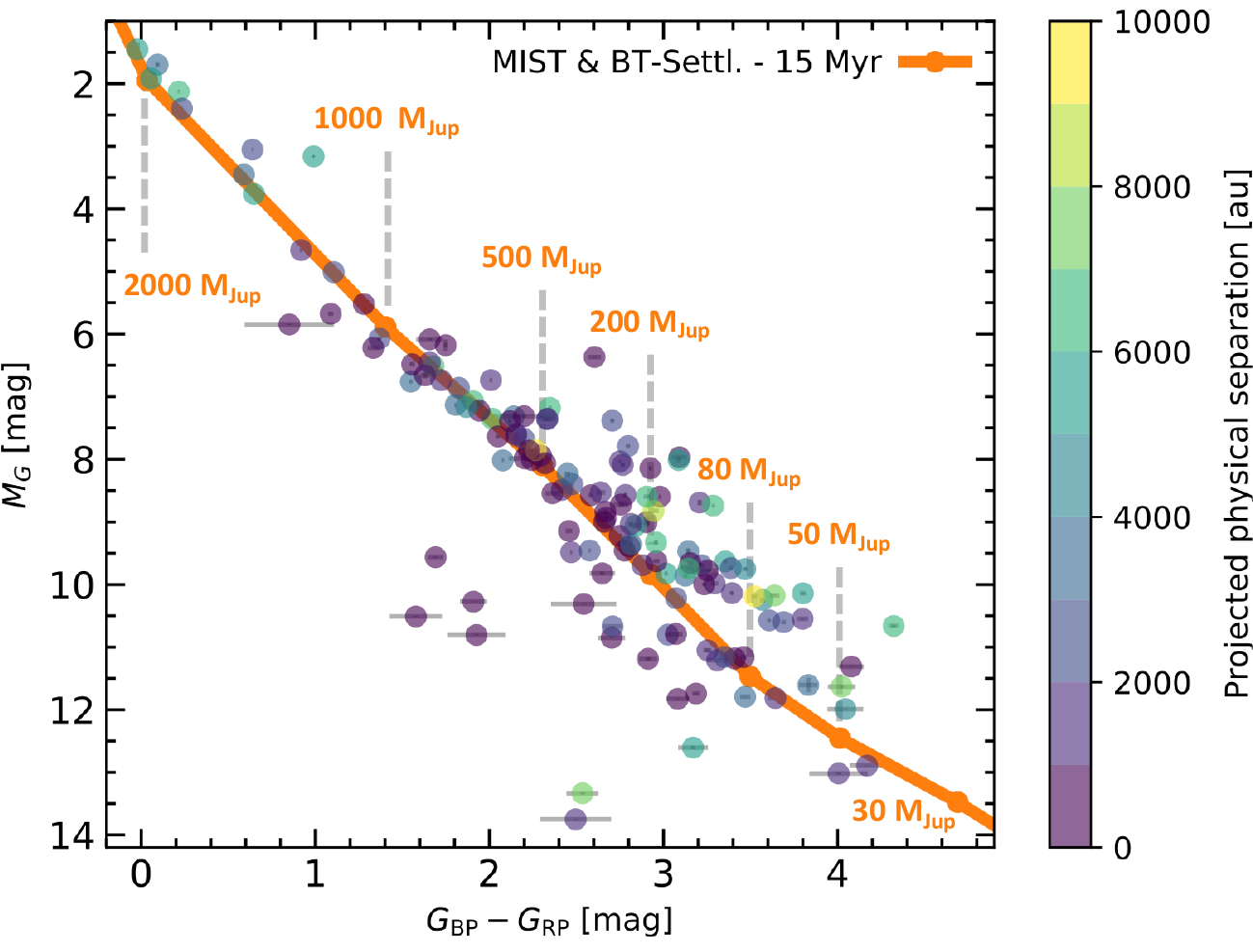}}
\caption{
Color-magnitude diagram of identified companion \rev{candidates} to K-type stars in Sco-Cen.
The colors of the markers indicate the projected separation between host and companion.
The orange line represents a combination of MIST and BT-Settl isochrones for an age of 15\,Myr.
}
\label{fig:cmd}
\end{figure}

To confirm the young and (sub-)stellar nature of the detected objects, we evaluated all candidate companions in a color-magnitude diagram as presented in Fig.~\ref{fig:cmd}.
The color is estimated using the \textit{Gaia} filters for the blue ($G_\mathrm{BP}$) and red part ($G_\mathrm{RP}$) of the $G$ band continuum.
As visualized in Fig~\ref{fig:cmd}, most of the candidate companions have colors that are in very good agreement with synthetic MIST and BT-Settl models for young objects of 15\,Myr, which is a reasonable average for the presented Sco-Cen sample \citep{pecaut2016}.
This corroborates the Sco-Cen membership of the analyzed candidates and each of these is very likely to be a young member of the association, even if it turns out not to be a companion to any of the stars from our input catalog.
\rev{%
There are two outliers at the faint end of the sequence that appear to be bluer than predicted by the evolutionary models.
This is, however, not a contradiction with Sco-Cen membership of these objects.
}
The corresponding objects, Gaia~EDR3 5232514298301348864 and Gaia~EDR3 5997035416337166976, are the faintest among our full sample of identified companion candidates with a $G$ band magnitude of 19.9\,mag \rev{and 19.3\,mag, respectively}.
Their $G_\mathrm{BP}$ and $G_\mathrm{RP}$ photometric measurements exhibit a corrected flux excess factor $C^*$ of 0.62 and 0.28, respectively, which is \rev{different from $C^*=0$ that is associated with} trustworthy photometry \citep{riello2020}.
This large uncertainty most likely originates from the faint $G_\mathrm{BP}$ flux of the objects, whose true value might be below \textit{Gaia}'s sensitivity limits in this channel.
Hence it is certainly possible that Gaia~EDR3 5232514298301348864 and Gaia~EDR3 5997035416337166976 are much redder than indicated in Fig.~\ref{fig:cmd} and consistent with the colors that are predicted for a low-mass brown dwarf companion at Sco-Cen age.
This conclusion is corroborated by the better-constrained $G-G_\mathrm{RP}$ colors of both objects.
\rev{These values of 1.5\,mag and 1.3\,mag} are in good agreement with late M spectral types.

It should be noted that not all of our identified companion candidates have $G_\mathrm{BP}$ and $G_\mathrm{RP}$ flux measurements, as can be seen in Table~\ref{tbl:companions_full_1} in Appendix~\ref{sec:identified_companions}.
These \rev{27} targets without any color information are not included in the plot shown in Fig.~\ref{fig:cmd} and their photometric Sco-Cen membership compatibility thus cannot be assessed properly.
For \rev{23} of these insufficiently characterized companions the lack of precise color measurements is explained by their proximity and contrast with respect to the target star.
\rev{All of these objects exhibit angular separations of less than 3\arcsec.}
The remaining four candidates without any color information are at larger separations from the target star, but in all cases these have an additional \textit{Gaia} source of equal or higher brightness at very close separations ($<2\arcsec$);
these are most likely affecting the flux measurements.
For Gaia~EDR3 6022499010440068864 and Gaia~EDR3 6208381587215253504 these close contaminants are members of the same candidate multiple systems that we have preselected.

\subsection{Companionship assessment}
\label{subsec:companionship_assessment}

We further analyzed whether the identified companion candidates are gravitationally bound multiple systems based on the available astrometric data.
In the general description of the gravitational two-body problem, a system is considered bound when its total kinetic energy
\begin{equation}
    T = \frac{\mu}{2}\left(\dot{r}^2+r^2\dot{\phi}^2\right)
\end{equation}
is smaller than the potential energy 
\begin{equation}
    V = \frac{1}{r}GM\mu\,,
\end{equation}
of both bodies.
In this formalism $r$ denotes the separation of the two bodies; $\dot{r}$ and $\dot{\phi}$ are the radial and azimuthal velocities with respect to the center of mass. 
The sum of both individual object masses $M_1$ and $M_2$ is $M$ and we define the reduced mass as
\begin{equation}
\mu=\frac{M_1M_2}{M_1+M_2}=\frac{M_1M_2}{M}\,.
\end{equation}
These conditions imply that a gravitationally bound system requires
\begin{equation}
    \left(\dot{r}^2+r^2\dot{\phi}^2\right) < \frac{2GM}{r}\,.
    \label{eqn:velocity_lim}
\end{equation}
Even though we do not know the objects' true three-dimensional separations and differential velocities, equation~\eqref{eqn:velocity_lim} allows us to assess which systems cannot be bound.
This is possible because our projected differential velocity $v_\mathrm{proj}$ is strictly smaller or equal to the amplitude of the true differential velocity $\sqrt{\dot{r}^2+r^2\dot{\phi}^2}$.
Likewise, the right-hand side of equation~\eqref{eqn:velocity_lim} is bounded above by substituting $r$ in the expression with our measured projected separation $\rho\leq r$ as
\begin{equation}
    \frac{2GM}{r}\leq\frac{2GM}{\rho}=:v_\mathrm{max}^2\,,
    \label{eqn:def_v_max}
\end{equation}
where we defined this new expression as the maximum allowed differential velocity for a system to be bound.

Hence we evaluated the ratio of our measured projected velocity difference $v_\mathrm{proj}$ and $v_\mathrm{max}$ to assess which systems are most likely not bound as per
\begin{align}
    \label{eqn:v_proj_v_max}
    \frac{v_\mathrm{proj}}{v_\mathrm{max}}&=v_\mathrm{proj}\sqrt{\frac{\rho}{2GM}}\\
    &=
    \begin{cases}
    <1\qquad\text{System can be gravitationally bound or}\\
    \geq 1\qquad\text{System most likely not gravitationally bound.}\\
    \end{cases}\nonumber
\end{align}
This expression is completely symmetric in $M_1$ and $M_2$, so within this formalism it does not matter which star is considered to be the primary of the system.
Whereas a ratio greater than or equal to one is a good indicator that the studied system cannot be physically bound, a value smaller than unity does not necessarily indicate that the opposite scenario is true, due to the unknown contributions to separation and differential velocity along our line of sight.
This metric is valid for regular binary systems but it is not suitable to assess systems with higher orders of multiplicity, such as the potential triple systems we have found around some of our targets.
A proper treatment for these special cases, however, is not straightforward and will thus be neglected in our further analysis.
\rev{%
The fraction of triple or higher-order multiple systems is assumed to be approximately 25\,\% among solar-type multiple systems \citep[e.g.,][]{mayor1987,eggleton2008,duchene2013}.
It is thus not surprising that we have identified 21 candidate triple systems out of 142 candidate multiples from our preselection.
This smaller fraction of $\approx$15\,\% does not violate these statistical constraints as some triple systems will certainly be missed either due to the cutoff separation or additional components below the resolution limit of \textit{Gaia}.
}

We calculated the differential proper motions between target and companion from the values listed in the \textit{Gaia} \rev{EDR3} database (see Appendix~\ref{sec:identified_companions}).
To convert this angular movement to a differential projected velocity $v_\mathrm{proj}$ in km\,s$^{-1}$ we used the inverted parallax of the star from our input catalog.
For the derivation of $v_\mathrm{max}$ we adopted the mass measurements for the stars of our sample from \citet{pecaut2016}.
For SZ~65, which was missing this quantity in \citet{pecaut2016}, we used a mass of $0.7\,M_\sun$ based on previous work by \citet{alcala2017}.
The masses of the identified companions were estimated as previously determined by \textit{Gaia} photometry and for the objects that were heavier than the maximum valid mass from our BT-Settl models we used MIST isochrones to derive a stellar mass.

\begin{figure}
\resizebox{\hsize}{!}{\includegraphics{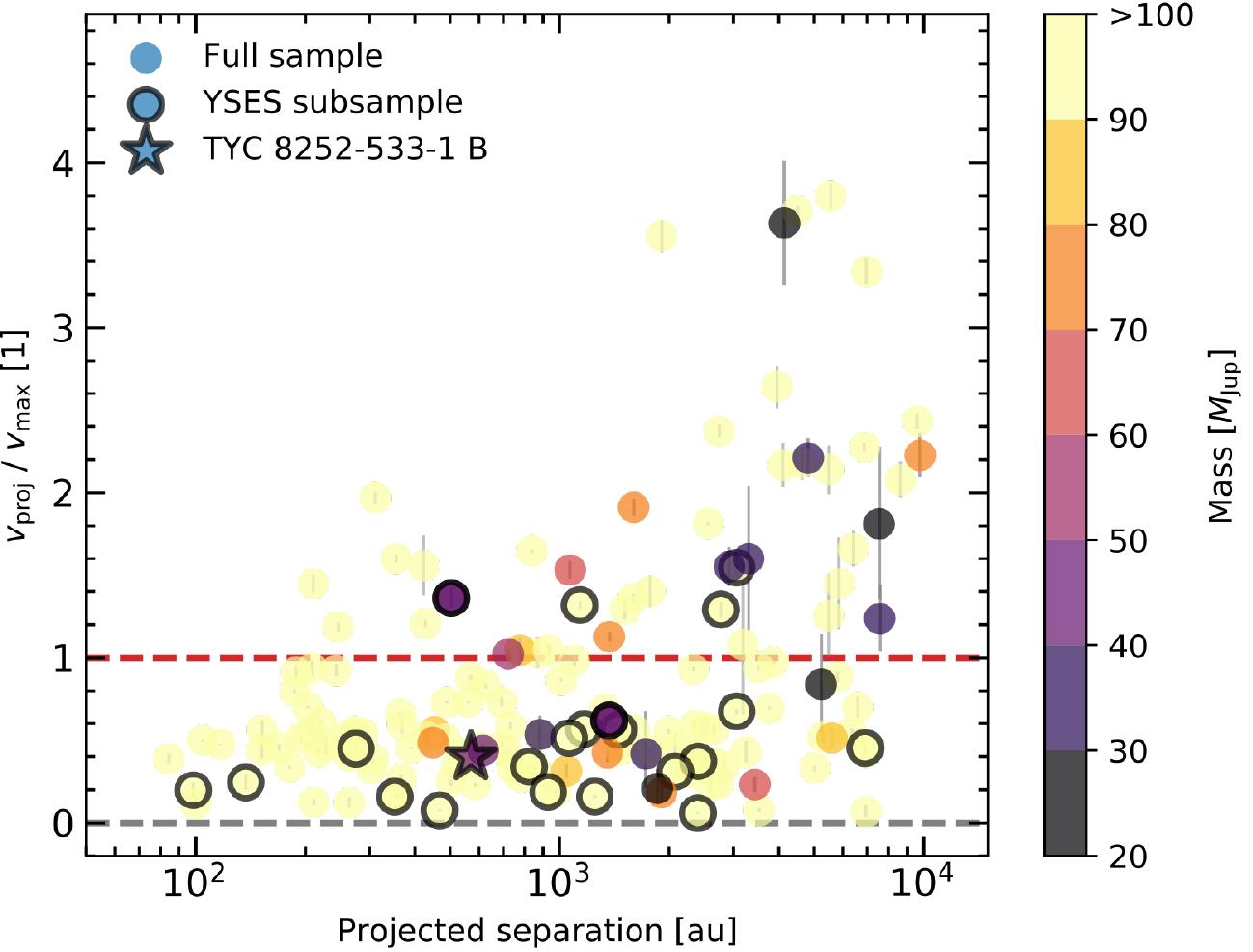}}
\caption{%
Relative velocities of identified companions of a sample of K-type Sco-Cen members.
We show the ratio of the differential projected velocities $v_\mathrm{proj}$ to the maximum velocity that allows a gravitationally bound system $v_\mathrm{max}$ as defined in equation~\eqref{eqn:v_proj_v_max}.
This parameter is a measure of whether our preselected candidate companions are part of a binary system;
a value greater than unity renders this scenario as unlikely.
These two different regimes are separated by the dashed red line.
The colors of the markers indicate the mass of the corresponding candidate companion.
To focus primarily on the sub-stellar objects in our sample the color bar is cut off at masses that are higher than $100\,M_\mathrm{Jup}$.
Markers with a black outline refer to members of the YSES subsample and TYC~8252-533-1~B is highlighted by the star.
}
\label{fig:separation_escape_velocity}
\end{figure}

The resulting ratios of $v_\mathrm{proj}$ to $v_\mathrm{max}$ that we derived for all objects are listed in Table~\ref{tbl:companions_full_1} in Appendix~\ref{sec:identified_companions}.
The error budget in the calculation of these quantities is dominated by the uncertainty of the differential velocities of both objects.
\rev{The uncertainties of} $v_\mathrm{proj}/v_\mathrm{max}$ \rev{were propagated by a bootstrapping approach}.
\rev{We repeated the velocity calculation} 10,000 times, drawing the initial proper motions of both objects from Gaussian distributions that are centered around the \textit{Gaia} \rev{EDR3} values, and their corresponding uncertainties served as standard deviation for these initial normal distributions.
The derived $v_\mathrm{proj}$ to $v_\mathrm{max}$ ratios represent the median of the corresponding posterior distribution and the surrounding 68\% confidence interval as an estimate of the uncertainties.
We visualize these derived parameters as a function of angular separation and companion mass in Fig.~\ref{fig:separation_escape_velocity}.

Most of the identified companion candidates are consistent with $v_\mathrm{proj}/v_\mathrm{max}<1$, which indicates that for these objects a gravitationally bound orbit is not ruled out by \textit{Gaia} \rev{EDR3} data.
To quantify the likelihood of these objects to be a real companion to the target of our sample, we derive the probability of a $v_\mathrm{proj}$ to $v_\mathrm{max}$ ratio that is smaller than unity, based on our 10,000 samples for each posterior distribution.
This value $p_\mathrm{dyn}^\mathrm{C}$ \rev{represents an upper limit for the dynamical companionship probability} as the complete three dimensional astrometry of all objects is not available.

\rev{%
As this value does not account for comoving Sco-Cen members that might have similar projected velocities as a target from our input catalog despite not being gravitationally associated, we further derived a statistical companionship likelihood $p_\mathrm{stat}^\mathrm{C}$.
For each companion candidate we checked how many sources were found by our preselection algorithm, when choosing a cutoff separation that is twenty times as large as the projected separation of the candidate \citep[see e.g.,][]{fontanive2021}.
This number provides an estimate of how many other Sco-Cen objects with similar parallaxes and proper motions are in the immediate surroundings of the stars from our input catalog.
\revrev{We obtained the adjusted number of objects, $N_{20}$, by counting the objects within the twenty times larger reference region while excluding the originally enclosed companion candidates.}
\revrev{Based on this number} we derived a statistical probability to have one of these sources in the 400 times smaller search region.
The statistical companionship likelihood is then calculated as
\begin{equation}
    p_\mathrm{stat}^\mathrm{C}=1-\frac{N_{20}}{400}\;.
\end{equation}
Our final companionship probability considers both these dynamical and statistical terms and is derived as
\begin{equation}
    p^\mathrm{C}= p_\mathrm{dyn}^\mathrm{C}p_\mathrm{stat}^\mathrm{C}\;.
\end{equation}
These values are provided in Table~\ref{tbl:companions_full_1} in Appendix~\ref{sec:identified_companions}.
}

We find that only \rev{27\%, 31\%, and 33\%} of our identified companion candidates have a $p^\mathrm{C}$ value that is smaller than 0.1, 0.5, and 0.95, respectively.
Even though the true membership probabilities might be lower than the values we have derived here, this test strongly suggests that our preselection based on \textit{Gaia} astrometry is a viable method to detect comoving companions to stars that are listed with position and proper motion measurements in \textit{Gaia} \rev{EDR3}.

We do not detect any trends of objects without any color measurement having significantly higher likelihoods to be unbound.
This is expected as the missing color information does not arise from the companion candidates but by the proximity to another object of equal or higher brightness.
The \rev{majority of the} aforementioned \rev{23} companions without color data that are closer than \rev{$3\arcsec$} to the target star from our input catalog exhibits $p^\mathrm{C}>0.5$ and \rev{19} out of these even $p^\mathrm{C}>0.95$.
As expected due to their small projected separations of less than 300\,au, these are very likely gravitationally bound stellar binaries.
\rev{Only three objects without color measurements at separations that are smaller than 3\arcsec are found to have $p^\mathrm{C}<0.1$}

\subsection{\rev{high confidence companions}}
\label{subsec:high_confidence_companions}

When applying a conservative cutoff of $p^\mathrm{C}>0.95$ to the full list of preselected companion candidates, \rev{110} objects around \rev{104} targets have a high probability to be gravitationally associated to the star from the input catalog.
A detailed list of these high confidence companions and their main parameters is presented in Table~\ref{tbl:high_priority_companions_1}.
\begin{sidewaystable*}
\caption{
\rev{\textit{Gaia} companion candidates} with $p^\mathrm{C}>0.95$ that were identified to K-type Sco-Cen members.
In the last column we report whether these systems have been mentioned in previous literature.
\rev{A full version of the table that is including the parameter uncertainties is available online.}
}
\label{tbl:high_priority_companions_1}
\tiny
\def\arraystretch{1.2}
\centering
\begin{tabular}{@{}lllllllllllll@{}}
\hline\hline
\multicolumn{2}{c}{\textit{Gaia} EDR3 ID} & \multicolumn{2}{c}{Coordinates\tablefootmark{c}} & $\rho$\tablefootmark{d} & PA\tablefootmark{e} & \multicolumn{2}{c}{$\mu_{\alpha *}$\tablefootmark{f}} & \multicolumn{2}{c}{$\mu_{\delta}$} & \multicolumn{2}{c}{$G$} & References \\    
Prim.\tablefootmark{a} & Sec.\tablefootmark{b} & Prim. & Sec. & & & Prim. & Sec. & Prim. & Sec. & Prim. & Sec. & \\
& & (hh:mm:ss.s $\pm$dd:mm:ss.s) & (hh:mm:ss.s $\pm$dd:mm:ss.s) & (\arcsec) & (\degr) & (mas\,yr$^{-1}$) &  (mas\,yr$^{-1}$) & (mas\,yr$^{-1}$) &  (mas\,yr$^{-1}$) & (mag) & (mag) \\
\hline
5232514298297802880 & 5232514298301348864 & 10:31:37.1 -69:01:58.7 & 10:31:36.4 -69:01:49.1 & 10.3 & 338.4 & -19.27 & -19.09 & 8.46 & 8.66 & 11.6 & 19.9 &  \\
5240643988513309952 & 5240643988513310592 & 11:17:58.1 -64:02:33.4 & 11:17:51.8 -64:02:05.6 & 49.6 & 304.0 & -28.59 & -28.64 & -0.63 & -1.23 & 7.8 & 11.4 & B19,D19,G18 \\
5332922112460126976 & 5332922116810986880 & 11:51:50.4 -64:07:27.7 & 11:51:50.7 -64:07:27.4 & 1.6 & 78.8 & -22.80 & -22.99 & 4.70 & 8.47 & 11.8 & 16.6 &  \\
5343603288120259072 & 5343603288130858112 & 11:55:42.9 -56:37:31.7 & 11:55:42.8 -56:37:31.2 & 0.8 & 304.9 & -39.46 & -38.91 & -7.41 & -10.59 & 11.2 & 13.0 &  \\
6124952431621117056 & 6124952431621118848 & 12:05:12.5 -53:31:23.6 & 12:05:12.2 -53:31:16.7 & 7.4 & 339.2 & -33.79 & -34.33 & -9.84 & -10.52 & 12.4 & 16.5 & D19,G18 \\
6057456092878899072 & 6057456092897714816 & 12:07:42.3 -62:27:28.2 & 12:07:42.1 -62:27:29.3 & 1.6 & 230.1 & -36.37 & -37.43 & -8.11 & -10.30 & 10.5 & 13.2 &  \\
6071087597518919040 & 6071087597497876480 & 12:09:41.8 -58:54:45.2 & 12:09:40.2 -58:54:41.0 & 12.8 & 289.4 & -37.47 & -37.52 & -10.68 & -9.54 & 10.0 & 15.9 &  \\
6071087597518919040 & 6071087597518919808 & 12:09:41.8 -58:54:45.2 & 12:09:41.7 -58:54:41.8 & 3.5 & 350.7 & -37.47 & -34.48 & -10.68 & -8.39 & 10.0 & 13.1 &  \\
6129878999619301120 & 6129879755533544832 & 12:10:10.6 -48:55:47.8 & 12:10:09.2 -48:55:44.0 & 14.1 & 285.3 & -37.49 & -36.09 & -12.36 & -11.87 & 10.8 & 13.5 & D19,G18 \\
5860803662284763392 & 5860803696599969280 & 12:12:08.0 -65:54:55.1 & 12:12:13.0 -65:54:49.3 & 31.6 & 79.4 & -37.22 & -37.03 & -8.27 & -8.53 & 11.1 & 17.0 & G18 \\
6129584364863667840 & 6129584364863668096 & 12:12:11.1 -49:50:08.4 & 12:12:11.0 -49:50:02.0 & 6.4 & 352.1 & -31.04 & -30.20 & -9.80 & -10.20 & 11.0 & 14.9 & D19,G18 \\
6075815841096386816 & 6075816592695096576 & 12:12:35.7 -55:20:27.5 & 12:12:36.1 -55:20:00.5 & 27.3 & 8.1 & -34.08 & -33.77 & -11.10 & -9.86 & 10.2 & 12.6 & D19,G18 \\
6126585034583698944 & 6126585034581064192 & 12:14:34.0 -51:10:12.7 & 12:14:31.8 -51:10:15.9 & 21.2 & 261.4 & -34.14 & -33.66 & -11.78 & -10.70 & 10.3 & 12.2 & D19,G18 \\
6075548419252906368 & 6075548419252913152 & 12:14:50.7 -55:47:23.7 & 12:14:52.2 -55:47:03.8 & 24.0 & 33.9 & -35.76 & -35.21 & -10.96 & -11.21 & 8.3 & 9.6 & D19,G18 \\
5855198832989286400 & 5855198828676205696 & 12:16:40.1 -70:07:36.3 & 12:16:39.5 -70:07:35.8 & 3.2 & 279.2 & -36.10 & -36.39 & -8.53 & -7.80 & 10.4 & 13.8 & D19,G18 \\
6126648698878768384 & 6126648703176939392 & 12:19:59.4 -50:18:40.8 & 12:19:59.3 -50:18:37.7 & 3.1 & 355.5 & -25.80 & -25.56 & -9.31 & -9.28 & 12.3 & 13.9 &  \\
6077461328965244160 & 6077461328965240576 & 12:21:08.0 -52:12:22.9 & 12:21:06.8 -52:12:41.4 & 21.8 & 211.9 & -35.11 & -34.68 & -12.80 & -13.28 & 11.4 & 15.8 & D19,G18 \\
6127110188823860480 & 6127110188823860864 & 12:22:04.3 -48:41:25.1 & 12:22:03.9 -48:41:17.7 & 8.3 & 333.0 & -28.63 & -27.60 & -10.63 & -9.90 & 10.4 & 15.7 & D19,G18 \\
6078760195785278208 & 6078760195785506048 & 12:30:29.5 -52:22:27.2 & 12:30:30.5 -52:22:28.9 & 9.2 & 100.6 & -33.60 & -34.67 & -13.31 & -14.35 & 11.3 & 13.3 & D19,G18 \\
6073105991611649024 & 6073106193460538240 & 12:33:33.7 -57:14:06.8 & 12:33:32.1 -57:14:10.3 & 13.4 & 254.6 & -37.85 & -36.63 & -13.91 & -14.76 & 10.5 & 17.9 &  \\
6074563978391541888 & 6074563978391541504 & 12:36:58.9 -54:12:18.2 & 12:36:56.5 -54:12:17.5 & 20.7 & 272.0 & -35.02 & -35.13 & -13.13 & -13.17 & 10.1 & 13.0 & D19,G18 \\
6074374346993398144 & 6074374346993397760 & 12:39:14.0 -54:54:47.0 & 12:39:13.8 -54:54:49.1 & 2.6 & 216.1 & -35.60 & -34.44 & -14.18 & -16.07 & 11.5 & 14.3 &  \\
6078074169253752064 & 6078074237969848960 & 12:40:46.6 -52:11:04.8 & 12:40:46.1 -52:11:04.3 & 4.8 & 276.4 & -26.63 & -25.89 & -11.36 & -10.35 & 11.6 & 15.0 & D19,G18 \\
6079389185156706944 & 6079389253876183808 & 12:40:54.5 -50:31:55.4 & 12:40:53.3 -50:31:47.6 & 14.1 & 303.8 & -43.33 & -43.35 & -19.41 & -19.86 & 11.4 & 14.4 &  \\
6060864815105080064 & 6060864819427739904 & 12:42:00.5 -57:59:48.8 & 12:42:00.4 -57:59:49.0 & 1.3 & 261.0 & -35.42 & -32.62 & -14.79 & -13.70 & 12.8 & 13.6 &  \\
5862876928853591040 & 5862876928853590144 & 12:44:34.7 -63:31:46.5 & 12:44:35.8 -63:31:49.9 & 7.9 & 116.1 & -36.13 & -37.35 & -13.72 & -14.82 & 10.3 & 15.2 &  \\
6074796250199371648 & 6074796245904371968 & 12:45:48.8 -54:10:58.7 & 12:45:49.6 -54:10:41.4 & 18.8 & 22.9 & -34.84 & -34.46 & -13.91 & -13.38 & 11.0 & 13.6 & D19,G18 \\
6061376298490107520 & 6061376298480880512 & 12:48:48.1 -56:35:38.0 & 12:48:47.8 -56:35:39.2 & 2.8 & 245.8 & -29.55 & -28.67 & -12.13 & -12.81 & 10.1 & 16.1 &  \\
6061310052916532096 & 6061310052892185088 & 12:50:44.8 -56:54:48.3 & 12:50:44.9 -56:54:49.6 & 1.7 & 138.2 & -37.56 & -34.54 & -15.76 & -13.30 & 12.4 & 12.5 &  \\
6075310478057303936 & 6075310478050174592 & 12:50:51.4 -51:56:35.6 & 12:50:51.7 -51:56:35.1 & 3.2 & 81.1 & -34.36 & -34.25 & -16.59 & -15.89 & 11.5 & 14.6 & D19,G18 \\
6080364825623490688 & 6080364791268622336 & 13:06:40.1 -51:59:38.9 & 13:06:39.5 -51:59:44.7 & 8.1 & 224.3 & -32.16 & -32.35 & -17.24 & -17.78 & 10.3 & 14.2 & D19,G18 \\
6056115131031531264 & 6056115135337482496 & 13:07:13.0 -59:52:11.3 & 13:07:13.0 -59:52:09.4 & 1.9 & 354.9 & -30.46 & -29.90 & -12.80 & -16.47 & 9.9 & 13.9 &  \\
6067650970822132224 & 6067650970822131584 & 13:12:18.0 -54:38:54.2 & 13:12:18.6 -54:39:05.7 & 12.5 & 157.2 & -28.59 & -29.09 & -11.91 & -13.20 & 10.2 & 11.3 &  \\
6088027047281877120 & 6088027051573430400 & 13:13:07.1 -45:37:43.9 & 13:13:07.1 -45:37:44.6 & 0.7 & 157.5 & -28.26 & -29.62 & -16.57 & -16.34 & 11.3 & 12.7 &  \\
6083750638577673088 & 6083750638540951552 & 13:23:35.8 -47:18:46.9 & 13:23:35.7 -47:18:51.0 & 4.4 & 204.2 & -31.18 & -32.38 & -20.10 & -19.69 & 10.8 & 16.8 &  \\
6082617587508723200 & 6082617591804976512 & 13:27:05.9 -48:56:18.5 & 13:27:06.5 -48:56:18.2 & 5.8 & 86.6 & -38.26 & -38.05 & -25.86 & -22.45 & 10.4 & 13.5 & D19,G18 \\
5863747467220738432 & 5863747462890662144 & 13:33:54.7 -65:36:41.9 & 13:33:54.8 -65:36:40.6 & 1.3 & 8.3 & -34.02 & -35.47 & -20.78 & -19.88 & 10.7 & 14.6 &  \\
6113000942074857216 & 6113000946373209216 & 13:36:40.9 -40:43:36.3 & 13:36:41.1 -40:43:40.1 & 4.6 & 146.2 & -30.16 & -28.98 & -21.69 & -20.90 & 12.0 & 14.2 &  \\
6094572719180018432 & 6094572753539758976 & 13:45:41.9 -49:04:59.3 & 13:45:44.2 -49:04:50.2 & 24.1 & 68.0 & -24.23 & -23.82 & -17.62 & -18.36 & 9.1 & 10.3 &  \\
6094529696485204864 & 6094529696485204736 & 13:47:50.5 -49:02:05.8 & 13:47:51.4 -49:01:49.0 & 19.0 & 27.4 & -23.40 & -23.76 & -15.39 & -16.28 & 10.6 & 11.8 & D19,G18 \\
\hline
\end{tabular}
\tablefoot{
\tablefoottext{a}{Primary of the binary system (i.e. the heavier mass star of the pair).}
\tablefoottext{b}{Secondary of the binary system (i.e. the lower mass object of the pair).}
\tablefoottext{c}{Coordinates are given at \textit{Gaia} EDR3 epoch J2016.0.}
\tablefoottext{d}{$\rho$ denotes the projected separation of the binary.}
\tablefoottext{e}{PA denotes the position angle (east of north) of the secondary with respect to the primary.}
\tablefoottext{f}{$\mu_{\alpha *}$ and $\mu_\delta$ denote the proper motions in RA and Dec, respectively.}
}
\tablebib{
B19:~\citet{bohn2019};
D19:~\citet{damiani2019};
G18:~\citet{goldman2018}
}
\end{sidewaystable*}
\begin{sidewaystable*}
\setcounter{table}{\the\numexpr\value{table}-1\relax}
\caption
{
(continued).
}
\tiny
\def\arraystretch{1.2}
\centering
\begin{tabular}{@{}lllllllllllll@{}}
\hline\hline
\multicolumn{2}{c}{\textit{Gaia} EDR3 ID} & \multicolumn{2}{c}{Coordinates\tablefootmark{c}} & $\rho$\tablefootmark{d} & PA\tablefootmark{e} & \multicolumn{2}{c}{$\mu_{\alpha *}$\tablefootmark{f}} & \multicolumn{2}{c}{$\mu_{\delta}$} & \multicolumn{2}{c}{$G$} & References \\    
Prim.\tablefootmark{a} & Sec.\tablefootmark{b} & Prim. & Sec. & & & Prim. & Sec. & Prim. & Sec. & Prim. & Sec. & \\
& & (hh:mm:ss.s $\pm$dd:mm:ss.s) & (hh:mm:ss.s $\pm$dd:mm:ss.s) & (\arcsec) & (\degr) & (mas\,yr$^{-1}$) &  (mas\,yr$^{-1}$) & (mas\,yr$^{-1}$) &  (mas\,yr$^{-1}$) & (mag) & (mag) \\
\hline
5850443307764629376 & 5850443303440130816 & 13:54:07.3 -67:33:45.0 & 13:54:03.1 -67:33:30.7 & 28.0 & 300.7 & -29.23 & -30.18 & -5.06 & -4.66 & 10.8 & 12.4 &  \\
5850443307764629376 & 5850443303446183040 & 13:54:07.3 -67:33:45.0 & 13:54:03.4 -67:33:27.3 & 28.7 & 308.0 & -29.23 & -30.50 & -5.06 & -5.43 & 10.8 & 13.7 &  \\
6095161370216796928 & 6095162843384927744 & 13:55:25.5 -47:06:56.8 & 13:55:27.6 -47:07:04.8 & 23.4 & 110.0 & -27.42 & -27.88 & -22.30 & -22.49 & 10.8 & 15.1 & D19,G18 \\
6114536929757358592 & 6114536929755100544 & 13:56:29.6 -38:39:13.0 & 13:56:29.6 -38:39:13.9 & 1.0 & 204.4 & -30.41 & -29.84 & -18.68 & -23.20 & 11.0 & 12.0 &  \\
6109646679996671360 & 6109646679996673024 & 14:00:49.7 -42:36:57.3 & 14:00:50.2 -42:36:48.0 & 11.0 & 32.7 & -24.53 & -25.37 & -20.76 & -21.39 & 10.7 & 13.2 &  \\
6116680904413813504 & 6116680908709624448 & 14:21:30.5 -38:45:25.1 & 14:21:30.5 -38:45:26.0 & 1.1 & 150.7 & -20.59 & -19.77 & -18.68 & -21.01 & 12.3 & 12.4 &  \\
6092201450556437632 & 6092201450556434944 & 14:27:05.5 -47:14:22.1 & 14:27:04.4 -47:14:37.8 & 19.1 & 215.0 & -28.17 & -28.52 & -24.70 & -24.30 & 10.4 & 16.0 &  \\
5894194318558985984 & 5894194348576228992 & 14:37:50.2 -54:57:41.6 & 14:37:52.0 -54:56:44.0 & 59.7 & 15.0 & -24.67 & -25.06 & -25.60 & -25.97 & 10.3 & 12.4 & D19,G18 \\
5905886387736675840 & 5905886215936891008 & 14:41:35.0 -47:00:29.2 & 14:41:39.3 -47:00:15.2 & 46.5 & 72.4 & -27.29 & -27.62 & -26.57 & -26.00 & 9.8 & 15.1 &  \\
6101758336902846208 & 6101758332601667712 & 14:42:15.9 -41:00:18.9 & 14:42:15.6 -41:00:15.7 & 4.4 & 316.5 & -27.60 & -29.13 & -25.83 & -26.06 & 11.1 & 11.8 &  \\
6204234706098895104 & 6204234706096290688 & 14:58:45.7 -33:15:10.7 & 14:58:45.9 -33:15:11.2 & 2.8 & 100.3 & -13.82 & -12.22 & -19.70 & -18.21 & 11.8 & 11.9 &  \\
6203845650778424960 & 6203845655074564864 & 14:59:44.7 -34:25:47.2 & 14:59:44.7 -34:25:47.1 & 0.9 & 87.1 & -25.02 & -29.59 & -28.70 & -31.21 & 11.1 & 11.5 &  \\
6203959484592343936 & 6203959488887357824 & 15:02:26.0 -34:05:13.7 & 15:02:25.9 -34:05:12.6 & 1.8 & 310.4 & -18.09 & -17.62 & -19.78 & -19.91 & 11.9 & 15.0 &  \\
6003698873422956928 & 6003698907782693248 & 15:08:51.4 -43:03:22.9 & 15:08:54.7 -43:03:14.2 & 37.3 & 76.6 & -19.91 & -20.40 & -26.14 & -26.20 & 9.4 & 11.0 &  \\
6201456107071825536 & 6201456107071825792 & 15:11:04.5 -32:51:30.8 & 15:11:04.7 -32:51:27.8 & 4.5 & 48.8 & -20.49 & -22.22 & -23.31 & -22.08 & 11.5 & 13.6 &  \\
6210702308369949952 & 6210702415745896192 & 15:12:44.5 -31:16:48.6 & 15:12:46.1 -31:17:02.2 & 24.9 & 123.2 & -17.28 & -16.54 & -22.64 & -21.62 & 11.3 & 14.0 &  \\
5903894175757504128 & 5903894175757506176 & 15:13:58.1 -46:29:14.9 & 15:13:57.2 -46:29:06.5 & 12.8 & 311.2 & -17.96 & -18.11 & -23.52 & -24.19 & 12.1 & 13.3 &  \\
5886747291936159104 & 5886747291923626240 & 15:15:22.9 -54:41:09.3 & 15:15:22.8 -54:41:09.1 & 1.5 & 280.0 & -21.33 & -18.11 & -22.89 & -25.03 & 11.9 & 13.3 &  \\
6200310519037175040 & 6200310484677437184 & 15:17:10.8 -34:34:20.0 & 15:17:14.3 -34:33:57.6 & 48.9 & 62.8 & -21.88 & -21.72 & -24.59 & -24.01 & 10.9 & 12.9 &  \\
6200310514738629504 & 6200310519037175040 & 15:17:10.7 -34:34:37.8 & 15:17:10.8 -34:34:20.0 & 17.9 & 5.1 & -20.73 & -21.88 & -22.73 & -24.59 & 8.0 & 10.9 &  \\
6000349589207590016 & 6000349657912636928 & 15:18:01.2 -44:44:27.5 & 15:18:01.0 -44:44:27.7 & 2.4 & 265.1 & -27.72 & -28.31 & -31.99 & -34.30 & 11.4 & 15.7 &  \\
6004331775495542656 & 6004331779800414464 & 15:19:16.0 -40:56:08.0 & 15:19:15.9 -40:56:11.8 & 3.9 & 190.2 & -21.74 & -20.92 & -27.20 & -27.21 & 11.0 & 16.4 &  \\
6213042069112745856 & 6213042073410590464 & 15:21:52.4 -28:42:38.7 & 15:21:52.4 -28:42:40.2 & 1.6 & 177.0 & -19.74 & -17.66 & -27.32 & -25.19 & 11.3 & 14.8 &  \\
6208136391829015936 & 6208136396126956928 & 15:24:11.4 -30:30:58.6 & 15:24:13.0 -30:30:57.6 & 20.3 & 87.2 & -22.09 & -22.11 & -25.07 & -24.57 & 10.6 & 12.9 &  \\
6208136396126956928 & 6208136396122878976 & 15:24:13.0 -30:30:57.6 & 15:24:12.8 -30:30:55.9 & 3.5 & 300.1 & -22.11 & -21.99 & -24.57 & -25.79 & 12.9 & 15.2 &  \\
6013649964744749184 & 6013649969039127552 & 15:29:38.5 -35:46:51.9 & 15:29:38.4 -35:46:50.6 & 2.6 & 299.1 & -23.95 & -21.95 & -30.37 & -29.91 & 10.7 & 15.7 &  \\
6013480055841212416 & 6013480060137095296 & 15:29:47.2 -36:28:37.7 & 15:29:47.2 -36:28:39.4 & 1.7 & 192.7 & -14.75 & -12.66 & -19.66 & -20.17 & 11.4 & 16.3 &  \\
6209492506280547200 & 6209492506275602816 & 15:31:29.6 -30:21:54.4 & 15:31:29.6 -30:21:53.2 & 1.2 & 19.1 & -14.45 & -13.24 & -17.99 & -20.95 & 12.4 & 14.3 &  \\
6014539542668216832 & 6014539645747431680 & 15:34:23.1 -33:00:09.2 & 15:34:26.3 -32:59:50.1 & 43.9 & 64.2 & -20.36 & -20.45 & -27.60 & -27.25 & 12.1 & 15.2 &  \\
6208381582919629568 & 6208381587215253504 & 15:37:02.1 -31:36:40.3 & 15:37:01.8 -31:36:39.5 & 4.7 & 279.8 & -20.59 & -21.49 & -27.87 & -28.36 & 9.8 & 11.8 &  \\
6208381582919629568 & 6208381587220236160 & 15:37:02.1 -31:36:40.3 & 15:37:01.7 -31:36:38.5 & 5.9 & 287.5 & -20.59 & -19.47 & -27.87 & -27.25 & 9.8 & 11.9 &  \\
6013399894569703040 & 6013399830146943104 & 15:39:27.8 -34:46:17.6 & 15:39:28.3 -34:46:18.4 & 6.4 & 97.8 & -13.22 & -13.12 & -22.12 & -21.81 & 11.7 & 13.9 &  \\
6013398451460692992 & 6013351520350805632 & 15:39:46.4 -34:51:02.9 & 15:39:46.4 -34:51:04.0 & 1.1 & 182.6 & -15.35 & -13.82 & -22.28 & -20.73 & 11.9 & 13.9 &  \\
6234377340635038848 & 6234377718592160384 & 15:41:06.8 -26:56:26.7 & 15:41:07.2 -26:56:25.9 & 6.3 & 82.1 & -18.07 & -17.87 & -26.64 & -24.87 & 11.0 & 16.1 &  \\
6014696841553696768 & 6014696841553696896 & 15:45:12.8 -34:17:31.0 & 15:45:12.6 -34:17:29.7 & 2.8 & 297.2 & -14.24 & -15.43 & -21.78 & -20.27 & 9.8 & 14.7 &  \\
5885915442658792576 & 5885915442658792448 & 15:46:29.6 -52:17:24.2 & 15:46:29.5 -52:17:25.8 & 1.6 & 183.3 & -21.95 & -24.04 & -22.60 & -21.96 & 11.1 & 14.7 &  \\
6236477064249469312 & 6236477064249562624 & 15:51:45.3 -24:56:51.7 & 15:51:45.5 -24:56:50.1 & 2.4 & 47.4 & -9.15 & -9.99 & -21.77 & -23.42 & 12.0 & 16.8 &  \\
6039427503765943936 & 6039427503765943040 & 15:54:51.4 -31:54:46.8 & 15:54:51.9 -31:54:47.6 & 6.3 & 97.7 & -9.75 & -10.71 & -27.23 & -27.27 & 11.6 & 17.6 &  \\
6248340348034690688 & 6248340343736161664 & 15:58:20.5 -18:37:25.5 & 15:58:20.5 -18:37:19.9 & 5.6 & 351.8 & -17.08 & -17.33 & -23.43 & -24.20 & 9.9 & 15.5 &  \\
6247571617610119808 & 6247571583249729280 & 15:59:11.0 -18:50:44.6 & 15:59:11.0 -18:50:59.2 & 14.6 & 183.0 & -10.28 & -10.21 & -20.89 & -21.31 & 12.6 & 13.3 &  \\
5984404849554176896 & 5984404849554176256 & 16:01:10.7 -48:04:44.3 & 16:01:12.1 -48:04:49.3 & 15.3 & 109.1 & -20.74 & -20.54 & -30.67 & -30.45 & 11.4 & 17.3 &  \\
6235806259081172736 & 6235806323497438592 & 16:02:39.1 -25:42:08.4 & 16:02:38.1 -25:41:39.5 & 31.7 & 335.8 & -20.03 & -20.08 & -32.57 & -32.68 & 11.6 & 12.4 &  \\
6236273895118889472 & 6236273895118890112 & 16:02:51.2 -24:01:57.8 & 16:02:51.2 -24:01:50.7 & 7.2 & 353.2 & -11.80 & -12.39 & -23.99 & -24.18 & 11.9 & 16.3 &  \\
6243393817024157184 & 6243393817024156288 & 16:04:21.6 -21:30:28.9 & 16:04:21.0 -21:30:42.1 & 16.2 & 215.7 & -12.45 & -12.64 & -23.81 & -24.75 & 11.7 & 13.6 &  \\
6249074718717359744 & 6249074718725644288 & 16:06:23.5 -18:14:19.5 & 16:06:23.5 -18:14:18.9 & 0.6 & 328.8 & -8.81 & -8.97 & -20.67 & -21.26 & 12.1 & 12.2 &  \\
\hline
\end{tabular}
\end{sidewaystable*}
\begin{sidewaystable*}
\setcounter{table}{\the\numexpr\value{table}-1\relax}
\caption
{
(continued).
}
\tiny
\def\arraystretch{1.2}
\centering
\begin{tabular}{@{}lllllllllllll@{}}
\hline\hline
\multicolumn{2}{c}{\textit{Gaia} EDR3 ID} & \multicolumn{2}{c}{Coordinates\tablefootmark{c}} & $\rho$\tablefootmark{d} & PA\tablefootmark{e} & \multicolumn{2}{c}{$\mu_{\alpha *}$\tablefootmark{f}} & \multicolumn{2}{c}{$\mu_{\delta}$} & \multicolumn{2}{c}{$G$} & References \\    
Prim.\tablefootmark{a} & Sec.\tablefootmark{b} & Prim. & Sec. & & & Prim. & Sec. & Prim. & Sec. & Prim. & Sec. & \\
& & (hh:mm:ss.s $\pm$dd:mm:ss.s) & (hh:mm:ss.s $\pm$dd:mm:ss.s) & (\arcsec) & (\degr) & (mas\,yr$^{-1}$) &  (mas\,yr$^{-1}$) & (mas\,yr$^{-1}$) &  (mas\,yr$^{-1}$) & (mag) & (mag) \\
\hline
6042124910722287744 & 6042124915024285440 & 16:06:57.9 -27:43:10.1 & 16:06:57.9 -27:43:08.7 & 1.5 & 6.1 & -12.84 & -9.82 & -24.24 & -24.14 & 8.1 & 11.8 &  \\
5998266426996208384 & 5998266426996209408 & 16:07:33.7 -37:59:24.7 & 16:07:33.5 -37:59:21.4 & 3.6 & 339.2 & -17.53 & -16.34 & -28.33 & -29.33 & 12.4 & 14.7 &  \\
5983063479721274880 & 5983063479721275520 & 16:08:07.7 -50:41:56.2 & 16:08:07.1 -50:41:52.9 & 6.2 & 302.0 & -22.03 & -21.01 & -36.43 & -37.88 & 11.5 & 13.6 &  \\
6043486660173351552 & 6043486655875354112 & 16:08:43.4 -26:02:17.2 & 16:08:44.4 -26:02:14.2 & 13.5 & 77.4 & -10.72 & -10.86 & -25.18 & -24.95 & 10.0 & 18.7 &  \\
6245777283349430912 & 6245777283349431552 & 16:09:00.7 -19:08:53.1 & 16:09:00.0 -19:08:37.3 & 19.0 & 326.6 & -9.37 & -9.44 & -25.12 & -24.89 & 12.9 & 15.4 &  \\
6035794030167079808 & 6035794030168507264 & 16:10:11.7 -32:26:36.1 & 16:10:11.7 -32:26:36.9 & 1.3 & 131.9 & -9.65 & -9.13 & -23.03 & -23.02 & 11.8 & 12.1 &  \\
6245781097280740864 & 6245781131640479360 & 16:10:21.7 -19:04:07.0 & 16:10:21.8 -19:04:02.4 & 4.6 & 6.6 & -9.05 & -8.60 & -24.05 & -24.78 & 13.6 & 16.9 &  \\
6042235793900264448 & 6042235725180784768 & 16:10:26.5 -27:56:30.0 & 16:10:28.2 -27:56:40.1 & 24.6 & 114.3 & -9.61 & -8.61 & -34.54 & -36.19 & 13.1 & 13.2 &  \\
6243833724749589632 & 6243833724749589760 & 16:10:42.0 -21:01:32.4 & 16:10:41.8 -21:01:30.6 & 3.2 & 305.7 & -9.88 & -10.09 & -23.34 & -24.59 & 11.8 & 15.5 &  \\
6242176653347275136 & 6242176829446854656 & 16:12:55.3 -23:19:46.1 & 16:12:52.6 -23:19:56.4 & 38.3 & 254.4 & -9.55 & -8.48 & -23.71 & -23.94 & 9.1 & 11.8 &  \\
6022499010422921088 & 6022499006128860928 & 16:13:58.0 -36:18:13.9 & 16:13:58.0 -36:18:19.6 & 5.7 & 178.8 & -16.14 & -17.36 & -30.32 & -31.12 & 10.8 & 13.3 &  \\
6243089247998315264 & 6243089217937775360 & 16:14:00.3 -21:08:44.3 & 16:13:58.1 -21:09:17.6 & 45.8 & 223.3 & -9.90 & -9.60 & -20.96 & -20.56 & 13.0 & 15.5 &  \\
5935099415271699072 & 5935099415267651328 & 16:14:52.0 -50:26:19.0 & 16:14:52.2 -50:26:19.9 & 2.2 & 114.4 & -20.08 & -16.52 & -31.57 & -32.12 & 10.2 & 11.4 &  \\
6242598526515737728 & 6242598526515738112 & 16:15:34.6 -22:42:43.1 & 16:15:34.5 -22:42:41.3 & 1.9 & 339.3 & -7.90 & -9.54 & -25.97 & -26.33 & 12.3 & 13.3 &  \\
6042418858284146688 & 6042418862581174016 & 16:16:14.2 -26:43:15.5 & 16:16:14.1 -26:43:14.2 & 2.5 & 300.5 & -17.79 & -17.25 & -26.92 & -28.46 & 11.7 & 12.6 &  \\
6050056478369804288 & 6050056482664014976 & 16:18:37.2 -24:05:23.0 & 16:18:37.2 -24:05:18.5 & 4.5 & 0.1 & -11.86 & -10.28 & -20.04 & -21.53 & 11.0 & 11.5 &  \\
6048571072519663872 & 6048571076818643968 & 16:19:12.2 -25:50:38.5 & 16:19:12.2 -25:50:39.8 & 1.3 & 175.5 & -11.38 & -10.81 & -23.39 & -25.84 & 12.5 & 16.1 &  \\
6037784008760031872 & 6037784004457590528 & 16:26:57.6 -30:32:28.4 & 16:26:57.0 -30:32:23.9 & 9.3 & 298.8 & -20.31 & -18.66 & -32.99 & -34.28 & 11.1 & 11.6 &  \\
6037784008760031872 & 6037784004457597568 & 16:26:57.6 -30:32:28.4 & 16:26:57.5 -30:32:31.8 & 3.6 & 197.1 & -20.31 & -18.70 & -32.99 & -33.03 & 11.1 & 12.6 &  \\
6031823178275372800 & 6031823173966757120 & 16:30:37.9 -29:54:22.9 & 16:30:37.8 -29:54:18.4 & 5.0 & 333.2 & -17.35 & -20.53 & -31.34 & -30.66 & 10.6 & 14.9 &  \\
6045791575844270208 & 6045791953801392128 & 16:32:00.6 -25:30:29.2 & 16:32:01.6 -25:30:25.8 & 14.2 & 76.2 & -11.52 & -10.65 & -23.08 & -22.83 & 12.5 & 12.7 &  \\
6047289699098449920 & 6047289699098450944 & 16:34:53.1 -25:18:17.4 & 16:34:53.2 -25:18:07.3 & 10.2 & 6.0 & -13.21 & -12.96 & -23.18 & -24.05 & 10.0 & 14.8 &  \\
5967552634824947456 & 5967552634824945152 & 16:43:01.4 -44:05:27.9 & 16:43:02.3 -44:05:33.2 & 10.8 & 119.1 & -8.89 & -8.60 & -19.79 & -19.36 & 11.6 & 17.1 &  \\
6046749289143173504 & 6046749319188591104 & 16:45:26.1 -25:03:17.0 & 16:45:29.0 -25:02:48.0 & 48.1 & 52.8 & -4.04 & -4.00 & -23.47 & -23.49 & 10.6 & 14.4 &  \\
4130416623473262720 & 4130416726552477440 & 16:47:37.1 -20:14:27.3 & 16:47:38.4 -20:14:15.8 & 22.1 & 58.6 & -8.10 & -8.50 & -21.94 & -21.26 & 11.8 & 14.6 &  \\
\hline
\end{tabular}
\end{sidewaystable*}
For each system we designated the object with the brightest $G$ band magnitude to be the primary.
The \rev{six} remaining triple systems in this list are represented by two individual entries each and we derived all properties with respect to the primary.

\begin{figure}
\resizebox{\hsize}{!}{\includegraphics{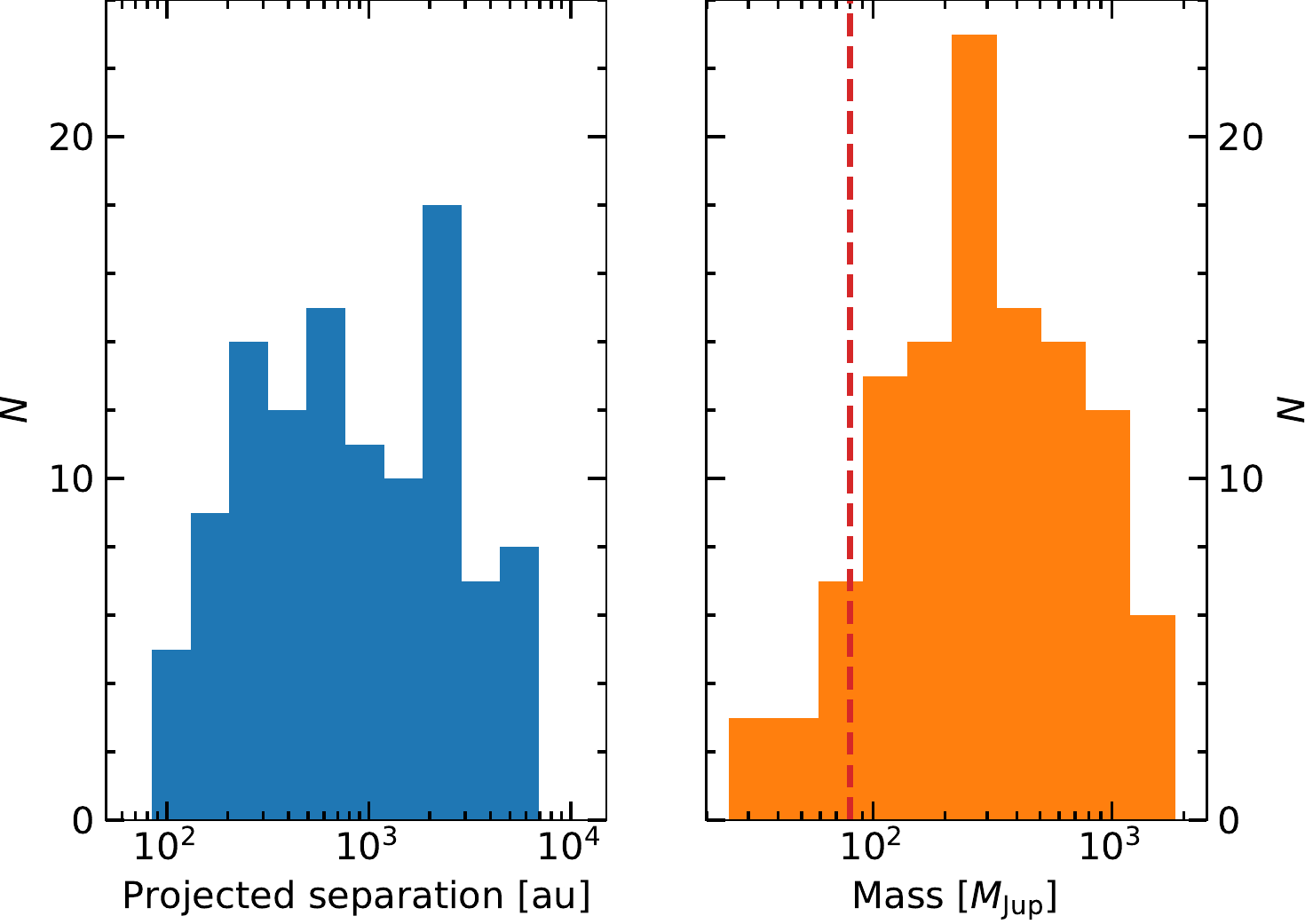}}
\caption{%
\rev{%
Properties of high confidence companions to K type Sco-Cen members.
\textit{Left panel}: Distribution of separations amongst our high confidence sample.
\textit{Right panel}: Mass distribution among this sample.
The red dashed line indicates the threshold of brown dwarf and stellar-mass companions at $80\,M_\mathrm{Jup}$.
}
}
\label{fig:high_confidence_comp_properties}
\end{figure}

\rev{%
The masses and separations of these high confidence binaries are presented in Fig.~\ref{fig:high_confidence_comp_properties}.
}
\rev{%
The majority of the companions exhibit projected separations that are smaller than 1000\,au.
We find a median separation of $763_{-539}^{+1933}$\,au and no companion is farther separated than 7\,000\,au.
The peak of the separation distribution is located at approximately 50\,au, which is in good agreement with general binary statistics \citep[e.g.,][]{duchene2013}.
For the masses of the companions we derived a median value of $277_{-172}^{+505}\,M_\mathrm{Jup}$, well in the M dwarf regime.
More massive and less massive objects are less frequent among the high confidence companions.
\revrev{As shown in Fig.~\ref{fig:separation_magnitude}, this statistical evaluation does not consider close-in, low-mass companions that are below the sensitivity of our detection method.
The derived median mass might thus be biased for these missing objects.
}
\revrev{Nevertheless, our results agree} well with the findings of \citet{duquennoy1991} and \citet{raghavan2010}, who established a peak in the companion mass ratio distributions to solar-type primaries at $q\approx0.3$ and $q\approx0.1$, respectively.
}

We identified \rev{ten} sub-stellar objects with masses below $80\,M_\mathrm{Jup}$ that are very likely to be comoving with $p^\mathrm{C}>0.95$.
\rev{The identifiers and mass estimates for these brown dwarf companions are listed in Table~\ref{tbl:bd_companions}.}
\begin{table}
\caption
{
\rev{%
Brown dwarf companions companions with $p^\mathrm{C}>0.95$.
}
}
\label{tbl:bd_companions}
\setlength{\tabcolsep}{12pt}
\def\arraystretch{1.2}
\centering
\begin{tabular}{@{}ll@{}}
\hline\hline
\textit{Gaia} EDR3 ID & $G$ band photometric mass \\    
 & ($M_\mathrm{Jup}$)\\
\hline
5232514298301348864 & $36.5\pm2.4$\\
6071087597497876480 & $76.0\pm16.7$ \\
5860803696599969280 & $66.2\pm4.5$ \\
6073106193460538240 & $46.5\pm3.3$ \\
6083750638540951552 & $48.7\pm11.1$ \\
6039427503765943040 & $33.8\pm3.6$ \\
5984404849554176256 & $73.8\pm4.4$ \\
6043486655875354112 & $24.8\pm1.0$ \\
6245781131640479360 & $43.1\pm5.8$ \\
6243833724749589760 & $78.9\pm12.5$ \\
\hline
\end{tabular}
\end{table}
\rev{%
Especially intriguing are Gaia EDR3 5232514298301348864, Gaia EDR3 6039427503765943040, and Gaia EDR3 6043486655875354112 that exhibit masses below $40\,M_\mathrm{Jup}$.
The latter is not too far away from the planetary mass regime with a photometric mass estimate of $24.8\pm1.0\,M_\mathrm{Jup}$.
}
Except for Gaia~EDR3 5860803696599969280, which was reported in \citet{goldman2018}, these are all newly identified brown dwarf members of Sco-Cen.
Their likely companionship to K-type stellar members of this association makes these especially intriguing laboratories not only for spectroscopic follow-up observations but also to study the dynamical evolution of multiple systems in a densely populated stellar birth environment.
Whereas the former will allow us to understand the atmospheric architectures and compositions of these sub-stellar objects, the latter will provide important implications for companion migration and ejection mechanisms caused by close stellar encounters.

\section{High-contrast imaging results}
\label{sec:hci_results}

\rev{%
In this section we present the results from our high-contrast imaging observations with SPHERE and NACO that were obtained within the scope of YSES.
Sect.~\ref{subsec:bd_companion_yses} contains a detailed analysis of a brown dwarf companion that we detected to TYC~8252-533-1.
We briefly assess the astrometry and photometry of other companions in the SPHERE field of view detected to other YSES targets from our input catalog in Sect.~\ref{subsec:sphere_stellar_companions}.
}

\subsection{A brown dwarf companion to TYC~8252-533-1}
\label{subsec:bd_companion_yses}

Our previous analysis revealed that the YSES target TYC~8252-533-1 (2MASS J13233587-4718467; Gaia EDR3 6083750638577673088) harbors a companion with a projected separation of 4\farcs4 and at a position angle of 204$^\circ$.
As \textit{Gaia} photometry suggests that the companion is of sub-stellar mass and likely a brown dwarf companion, we will refer to it as TYC~8252-533-1~B henceforth.
Due to the small angular separation of the binary, both objects were located within the SPHERE field of view during our regular YSES observations.
This object is thus an ideal test case to assess the quality of the parameters that were derived from the \textit{Gaia} catalog by comparison to independent photometry and astrometry acquired with SPHERE.
To characterize this companion independently from the \textit{Gaia} measurements, we acquired SPHERE dual-band measurements ranging from $Y$ to $K$ band and additional NACO $L'$ data as described in Sect.~\ref{subsec:methods_data_reduction}.
The final images for several SPHERE filters and our NACO data are presented in Appendix~\ref{sec:hci_observing_conditions}.

\subsubsection{Stellar parameters}
\label{subsubsec:stellar_parameters}

A summary of the basic stellar parameters of TYC~8252-533-1 is presented in Table~\ref{tbl:yses_companion_host_properties}.
\begin{table}
\caption{%
Properties of TYC~8252-533-1.
}
\label{tbl:yses_companion_host_properties}
\def\arraystretch{1.2}
\setlength{\tabcolsep}{8pt}
\small
\centering
\begin{tabular}{@{}lll@{}}
\hline \hline
Parameter & Value & Reference(s)\\ 
\hline
Main identifier & TYC~8252-533-1 & (1)\\
2MASS identifier & J13233587-4718467 & (2)\\
\textit{Gaia} \rev{EDR3} ID & 6083750638577673088 & (3) \\
Right Ascension\tablefootmark{a} & 13:23:35.8 & (3) \\
Declination\tablefootmark{a} & -47:18:46.9 & (3) \\
Spectral Type &  K3Ve & (4,5) \\
Mass [$M_\mathrm{\sun}$] & $1.2$ & (5) \\
$T_\mathrm{eff}$ [K] & $4,500\pm50$ & (6)\\
$\log\left(L_*/L_\mathrm{\sun}\right)$ & $-0.06\pm0.02$ & (6)\\
Age [Myr] & $5$ & (5)\\
Parallax [mas] & $7.71\pm0.09$ & (3) \\
Distance [pc] & $129.5\pm1.4$ & (3,7) \\
$\mu_{\alpha*}$ [mas / yr] & $-31.183\pm0.085$ & (3) \\
$\mu_{\delta}$ [mas / yr] & $-20.105\pm0.078$ & (3) \\
$B$ [mag]& $ 11.96$ & (1) \\
$V$ [mag] & $11.20$ & (1) \\
$G_\mathrm{BP}$ [mag] & $11.43$ & (3) \\
$G$ [mag] & $10.80$ & (3) \\
$G_\mathrm{RP}$ [mag] & $10.04$ & (3) \\
$J$ [mag] & $9.03$ & (2) \\
$H$ [mag] & $8.40$ & (2) \\
$K_\text{s}$ [mag] & $8.31$ & (2) \\
$W1$ [mag] & $8.17$ & (8) \\
$W2$ [mag] & $8.13$ & (8) \\
$W3$ [mag] & $7.94$ & (8) \\
$W4$ [mag] & $7.10$ & (8) \\
\hline
\end{tabular}
\tablefoot{
\tablefoottext{a}{Coordinates are given at \textit{Gaia} \rev{EDR3} epoch J2016.0.}
}
\tablebib{
(1)~\citet{hog2000}; 
(2)~\citet{cutri2012a}; 
(3)~\citet{gaia2020}; 
(4)~\citet{torres2006};
(5)~\citet{pecaut2016};
(6)~this work; 
(7)~\citet{bailer-jones2021};
(8)~\citet{cutri2012b}.
}
\end{table}
To convert the photometric contrasts measured in the SPHERE and NACO data into absolute fluxes, it was necessary to characterize the primary star first.
We analyzed its spectral energy distribution (SED) with VOSA \citep{bayo2008} using flux measurements from Tycho \citep{hog2000}, APASS \citep{henden2012}, \textit{Gaia} \rev{EDR3 \citep{gaia2020}}, DENIS \citep{denis_consortium2005}, 2MASS \citep{skrutskie2006,cutri2003}, and WISE \citep{cutri2012b}.
\citet{pecaut2016} measured an extinction of $A_V=0.16\,\mathrm{mag}$ and reported that the star is likely to host a debris disk based on the infrared excess that was observed in WISE $W3$ and $W4$ filters.
The STILISM reddening map \citep{lallement2014} provides a total visual extinction of $A_V=(0.05\pm0.05)\,\mathrm{mag}$ at the position and distance of our target.
We therefore allowed extinctions in the range $0\,\mathrm{mag}<A_V<0.2\,\mathrm{mag}$ and excluded the $W3$ and $W4$ measurements from the fit of the stellar SED.

As the stellar flux measurements of the primary might be affected by flux from the close companion we briefly assessed the extent of this potential contamination.
The fractional contribution from the secondary to the measured flux of the primary is maximized for the longest wavelengths of the analyzed SED\footnote{This statement is verified by the lowest magnitude contrast $\Delta$Mag between primary and secondary that is measured in the $L'$ band (see Table \ref{tbl:photometry_yses_bd}).}.
Equation (2) of \citet{bohn2020b} yields a contribution of 0.04\,mag to the stellar magnitude in $L'$ band if the binary was unresolved in the photometric data.
This contribution is significantly smaller for shorter wavelengths where the contrast between both objects is larger.
We therefore do not perform any correction of the flux measurements for the primary, as the contribution from the secondary is below 5\% and thus already considered in the larger uncertainties of the magnitude contrast values as presented in Table~\ref{tbl:photometry_yses_bd}, when deriving the fluxes of the companion.

The stellar distance was fixed to \rev{130\,pc} using the estimate from \citet{bailer-jones2021} based on \textit{Gaia} EDR3 parallaxes, and we assumed a solar-like metallicity of the primary.
We fitted a grid of BT-Settl models \citep{allard2012,baraffe2015} in a $\chi^2$ minimizing approach, and the best fit of the data was provided by a model with an effective temperature of $T_\mathrm{eff}=4500\pm50\,K$, a surface gravity of $\log\left(g\right)=2.6\pm0.3\,\mathrm{dex}$, a stellar luminosity of \rev{$\log\left(L_*/L_\sun\right)=-0.06\pm0.02$}, and a visual extinction parameter of $A_V=0.13\pm0.06\,\mathrm{mag}$.
These values are in good agreement with the properties that \citet{pecaut2016} derived for TYC~8252-533-1.

\subsubsection{Brown dwarf astrometry}
\label{subsubsec:yses_bd_astrometry}

\begin{figure}
\resizebox{\hsize}{!}{\includegraphics{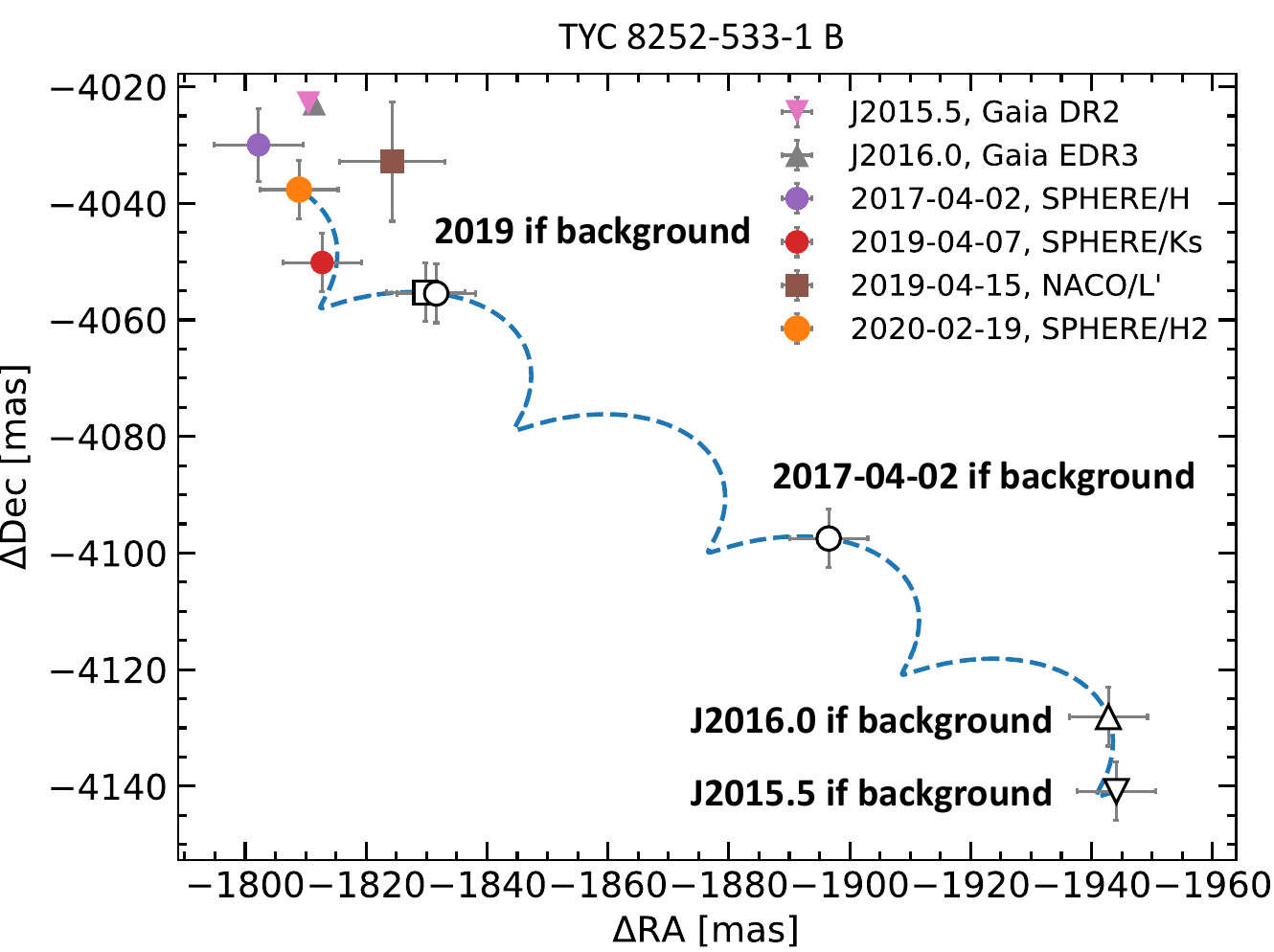}}
\caption{
Proper motion analysis of TYC~8252-533-1~B.
The colored markers represent the relative offsets in RA and Dec with respect to the primary that we measured in SPHERE and NACO imaging data or derived from the \textit{Gaia} DR2 \rev{and EDR3} catalogs.
The blue dashed line illustrates the trajectory of a static background object at infinity and the white markers along the curve represent the hypothetical positions of such an object at the epoch of the corresponding observation.
}
\label{fig:ppm_analysis_yses_companion}
\end{figure}

To assess whether our high-contrast data confirm a bound orbit of TYC~8252-533-1~B, we extracted the relative astrometry with respect to the primary star for all of our observations.
For SPHERE we used the general astrometric solution presented in \citet{maire2016} with a true north correction of $-1\fdg75\pm0\fdg08$ and plate scales of 12.255\,mas px$^{-1}$, 12.251\,mas px$^{-1}$, and 12.265\,mas px$^{-1}$ for $H2$, $H$, and $K_s$ band images, respectively.
The NACO data were calibrated with a plate scale of $27.193\pm0.059$\,mas px$^{-1}$ \citep{launhardt2020} but no true north correction was applied as discussed in \citet{bohn2020a}.
The derived separations and position angles are presented in Table~\ref{tbl:astrometry_yses_companion} and visualized in Fig.~\ref{fig:ppm_analysis_yses_companion}.
\begin{table}
\caption{%
Relative astrometry of TYC~8252-533-1~B with respect to the primary.
}
\label{tbl:astrometry_yses_companion}
\def\arraystretch{1.2}
\setlength{\tabcolsep}{8pt}
\centering
\begin{tabular}{@{}llll@{}}
\hline\hline
Epoch & Filter & Separation & PA\\ 
 & & (\arcsec) & (\degr)\\ 
\hline
J2015.5 & $G$ & $4.4115\pm0.0001$ & $204.230\pm0.002$ \\
J2016.0 & $G$ & $4.4119\pm0.0001$ & $204.241\pm0.001$ \\
2017-04-02 & $H$ & $4.414\pm0.005$ & $204.09\pm0.09$ \\
2019-04-07 & $K_s$ & $4.437\pm0.005$ & $204.11\pm0.09$ \\
2019-04-15 & $L'$ & $4.426\pm0.011$ & $204.34\pm0.11$ \\
2020-02-19 & $H2$ & $4.424\pm0.005$ & $204.13\pm0.09$ \\
\hline
\end{tabular}
\tablefoot{%
\rev{The $G$ band astrometric data originate from the $Gaia$ DR2 (epoch J2015.5) and EDR3 (epoch J2016.0) catalogs. 
These filters have different pass bands.}
}
\end{table}
The measurements clearly confirm that the path of TYC~8252-533-1~B is highly inconsistent with the simulated trajectory of a static background object.
Instead, the relative astrometric offsets with respect to the primary are in good agreement with the small relative velocity that is expected from an orbit with a projected angular separation of approximately \rev{570}\,au.
Even though there remain some systematic sources of uncertainties between the astrometric measurements that originate from different instruments or filters, the high-contrast imaging data clearly confirm the hypothesis, based on the \textit{Gaia} data, that TYC~8252-533-1~B is a bound companion to the solar-type primary TYC~8252-533-1.

\subsubsection{Brown dwarf photometry}
\label{subsubsec:yses_bd_photometry}

\begin{figure}
\resizebox{\hsize}{!}{\includegraphics{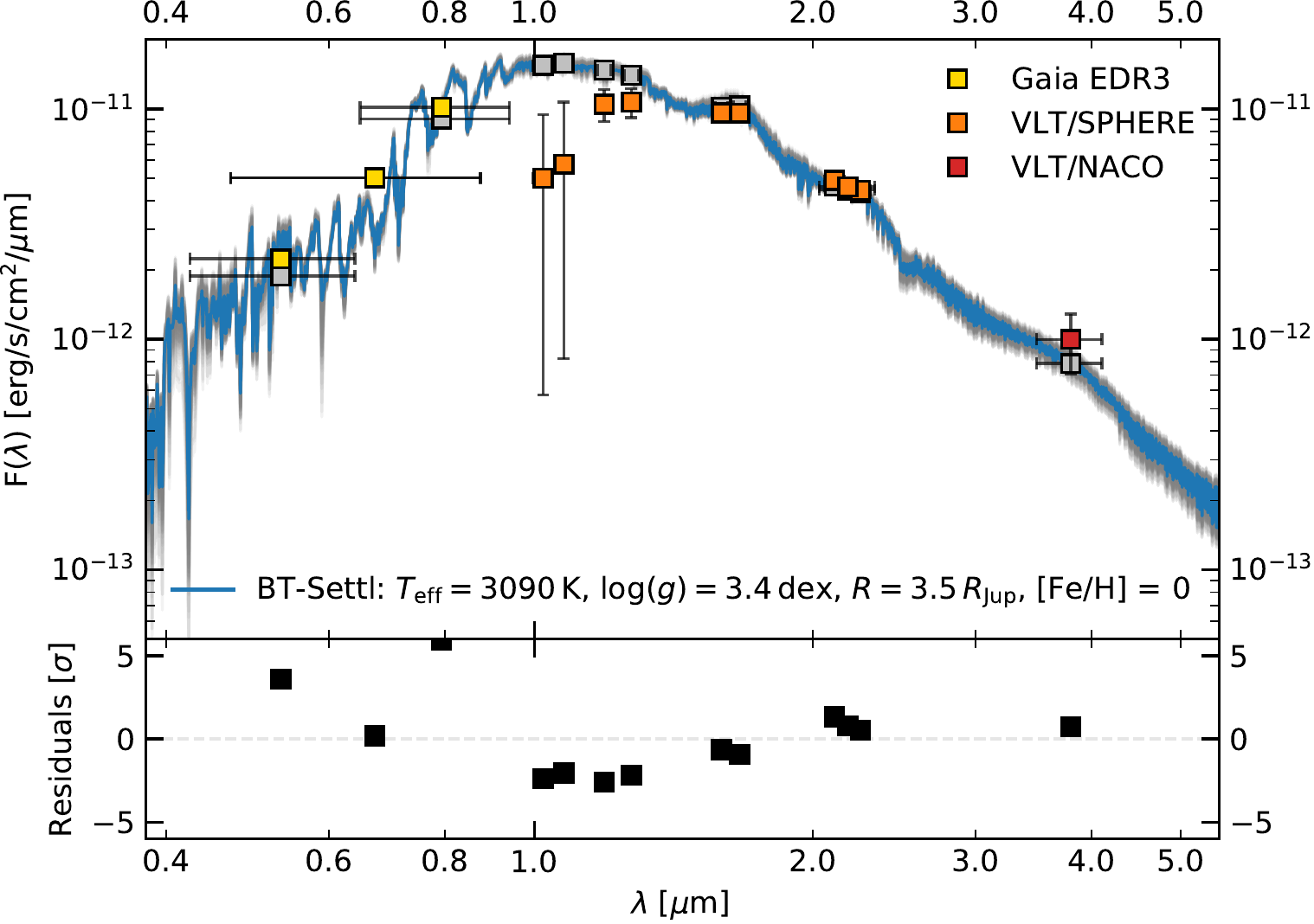}}
\caption{%
SED of TYC~8252-533-1~B.
The yellow, orange, and red squares indicate the photometry we measured for \textit{Gaia}, SPHERE, and NACO filters, respectively.
The bars in $x$ direction indicate the width of the filter transmission curves.
The blue curve presents the median model from our posterior distribution of the SED fit as shown in \rev{Fig.~\ref{fig:sed_mcmc_post_yses_companion} of} Appendix~\ref{sec:mcmc_posterior}.
We present the integrated flux of this median model in the applied filters with the grey squares and we visualize 100 randomly selected models from our posterior distribution as grey curves.
In the bottom panel we present the residuals of the fit.
}
\label{fig:sed_yses_companion}
\end{figure}

The de-reddened photometry of TYC~8252-533-1~B and its host star are presented in Table~\ref{tbl:photometry_yses_bd}.
\begin{table}
\caption{
De-reddened photometry of TYC~8252-533-1 and its brown dwarf companion.
}
\label{tbl:photometry_yses_bd}
\def\arraystretch{1.2}
\centering
\begin{tabular}{@{}llll@{}}
\hline\hline
Filter & Magnitude star & $\Delta$Mag & Flux companion\\ 
 & (mag) & (mag) & ($\mathrm{erg}\,\mathrm{s}^{-1}\,\mathrm{cm}^{-2}\,\mathrm{\mu m}^{-1}$)\\
\hline
$G_\mathrm{BP}$ & 11.43 & $6.73\pm0.05$ & $(0.22\pm0.01)\times10^{-11}$\\
$G$ & 10.80 & $5.94\pm0.01$ & $(0.503\pm0.002)\times10^{-11}$\\
$G_\mathrm{RP}$ & 10.04 & $5.20\pm0.02$ & $(1.02\pm0.02)\times10^{-11}$\\
\textit{Y2} & 9.34 & $5.84\pm0.96$ & $(0.50\pm    0.44)\times10^{-11}$\\
\textit{Y3} & 9.25 & $5.59\pm0.93$ & $(0.58\pm0.49)\times10^{-11}$\\
\textit{J2} & 9.06 & $4.79\pm0.17$ & $(1.05\pm0.16)\times10^{-11}$\\
\textit{J3} & 8.86 & $4.68\pm0.15$ & $(1.07\pm0.15)\times10^{-11}$\\
\textit{H2} & 8.34 & $4.47\pm0.09$ & $(0.96\pm0.08)\times10^{-11}$\\
\textit{H3} & 8.26 & $4.37\pm0.09$ & $(0.96\pm0.08)\times10^{-11}$\\
\textit{K1} & 8.22 & $4.25\pm0.05$ & $(0.49\pm0.02)\times10^{-11}$\\
\textit{K} & 8.22 & $4.20\pm0.05$ & $(0.44\pm0.02)\times10^{-11}$\\
\textit{K2} & 8.22 & $4.04\pm0.04$ & $(0.46\pm0.02)\times10^{-11}$\\
\textit{L'} & 8.16 & $3.62\pm0.31$ & $(0.10\pm0.03)\times10^{-11}$\\
\hline
\end{tabular}
\tablefoot{%
The $G$ band values originate directly from \textit{Gaia} \rev{EDR3} whereas the infrared measurements of the companion are based on our SPHERE and NACO data.
The stellar magnitudes at infrared wavelengths are derived from SED modeling as described in Sect.~\ref{subsubsec:stellar_parameters}.
}
\end{table}
The listed \textit{Gaia} photometry is directly obtained from the \rev{EDR3} catalog and the infrared photometry originates from our high-contrast imaging data.
We applied the previously determined extinction of $A_V=0.13$\,mag to correct the presented fluxes. 
As the SPHERE $H$ band data from 2017-04-02 were collected in poor atmospheric conditions and with a very unstable AO performance, we disregarded the photometry that was extracted from these observations in our further analysis.
We fitted the full optical and infrared SED of the companion utilizing the MCMC approach described in \citet{bohn2020c}.
We used a linearly interpolated grid of BT-Settl models \citep{allard2012,baraffe2015} with effective temperatures $T_\mathrm{eff}$ between 1500\,K and 4000\,K, surface gravity in the range of $0<\log\left(g\right)<6$, and solar metallicity.
We further allowed object radii $R$ from $0.5\,R_\mathrm{Jup}$ to $5\,R_\mathrm{Jup}$.
The MCMC sampler was implemented in the \texttt{emcee} framework \citep{foreman-mackey2013} and we used 100 walkers with 10,000 steps to sample the posterior distribution.
In accordance with the computed autocorrelation time of approximately 200 steps, the first 1000 samples of each chain were discarded as burn-in phase and we further continued using each twentieth step of the remaining chains.
This provided 45,000 samples for our posterior distribution in ($T_\mathrm{eff}$, $\log\left(g\right)$, $R$), which are visualized in \rev{Fig.~\ref{fig:sed_mcmc_post_yses_companion} of} Appendix~\ref{sec:mcmc_posterior}.
From these distributions we derived an effective temperature of $T_\mathrm{eff}=3092^{+186}_{-91}\,K$, a surface gravity of $\log\left(g\right)=3.41^{+1.07}_{-0.31}$\,dex, and a radius of $R=3.5^{+0.3}_{-0.4}\,R_\mathrm{Jup}$ for TYC~8252-533-1~B.
These values were obtained as the 95\,\% confidence intervals around the medians of the posterior distributions.
From these three parameters we derived the object luminosity as $\log\left(L_*/L_\sun\right)=-1.99^{+0.01}_{-0.02}$.
The results of this SED fit are presented in Fig.~\ref{fig:sed_yses_companion}.
Interestingly, the derived object radius is markedly larger than usual radii of field brown dwarfs of this mass, which are in the order of $1\,R_\mathrm{Jup}$ \citep[e.g.,][]{chabrier2009}.
This inflated photometric radius, however, is nothing unusual for sub-stellar companions at this young age, and similarly large values have been reported for various planetary-mass objects \citep[e.g.,][]{schmidt2008,bohn2019,stolker2020}.

To convert the results of this analysis to a mass estimate for the companion, we assumed a system age of $5\pm2$\,Myr \citep{pecaut2016}.
The kinematic parallax that \citet{pecaut2016} used for this target (7.83\,mas) is not far from the \textit{Gaia} \rev{EDR3} value of 7.71\,mas.
The difference in age due to the different adopted distances should therefore be considered in our assigned uncertainties.
Evaluation of BT-Settl isochrones at this age and with our derived luminosity yielded a mass of $52^{+17}_{-11}\,M_\mathrm{Jup}$ for TYC~8252-533-1~B, which makes it likely a brown dwarf companion to the solar-type primary.
\rev{This value is in good agreement with the photometric mass estimate of $48.7\pm11.1\,M_\mathrm{Jup}$ derived from the \textit{Gaia} $G$ band flux.}

\subsubsection{System architecture}
\label{subsubsec:yses_bd_companion_system_architecture}

Based on the observed infrared excess in the W3 and W4 filters the primary star likely hosts a debris disk \citep[][and Sect.~\ref{subsubsec:stellar_parameters}]{pecaut2016}.
Our direct imaging data do not reveal any scattered light flux of this postulated debris ring, which is most likely below the detection sensitivity of our total intensity observations.
Nevertheless, we assessed the relative location of this newly identified brown dwarf companion with respect to the circumstellar material.
As our photometric data cover only wavelengths shorter than 22\,$\mu$m (W4), we most likely do not sample the peak of the infrared excess emission;
hence we could only derive an upper limit for the temperature of the dust grains in the disk.
We assumed a single species of dust grains that is located at a radial separation $r_\mathrm{dust}$ from the central star, and modeled the thermal dust emission by a single blackbody with a dust temperature $T_\mathrm{dust}$.
As the peak of this emission is detected either in W4 or at longer wavelengths, we could constrain a maximum dust temperature of $T_\mathrm{dust}\leq130$\,K.
We converted this dust temperature to a minimum disk radius of $r_\mathrm{dust}\geq4$\,au, utilizing a temperature-radius relation as presented by \citet{backman1993} with
\begin{equation}
    \frac{r_\mathrm{dust}}{1\,\mathrm{au}} = \left(\frac{L_*}{L_\sun}\right)^{0.5} \left(\frac{278\,\mathrm{K}}{T_\mathrm{dust}}\right)^{2}\;,
\end{equation}
where $L_*$ denotes our previously determined stellar luminosity.
Even though this lower limit does not rule out that the debris disk around the primary might extend up to a separation in the order of the projected separation of \rev{570}\,au that we measure for TYC~8252-533-1~B, this scenario is expected to be extremely unlikely, as it would require dust temperatures as low as 10\,K.
In agreement with temperature and radius measurements from other debris disk studies that rely on either SED modeling \citep[e.g.,][]{moor2011} or spatially resolved imaging of these environments \citep[e.g.,][]{pawellek2014}, we conclude that TYC~8252-533-1~B is thus very likely located outside the debris disk of its primary star.

TYC~8252-533-1~B is a new data point to the catalog of debris-disk systems that are harboring giant exoplanets or brown dwarf companions such as \object{HR~8799} \citep{marois2008,marois2010}, \object{$\beta$~Pictoris} \citep{lagrange2009,lagrange2010}, \object{HD~95086} \citep{rameau2013}, \object{HD~106906} \citep{bailey2014}, \object{HR~3549} \citep{mawet2015}, \object{HR~2562} \citep{konopacky2016}, or \object{HR~206893} \citep{milli2017}.
While the giant companions HR~8799~bcde, $\beta$~Pic~b, HD~95086~b, HR~2562~B, and HR~206893~B are located inside the debris disks of their host stars, HD~106906~b, HR~3549~B, and TYC~8252-533-1~B are residing beyond these.
Whereas the objects in the former category might have formed via core accretion \citep{pollack1996,alibert2005,dodsonrobinson2009,lambrechts2012} or gravitational instabilities in the protoplanetary disk \citep{boss1997,rafikov2005,durisen2007,kratter2010,boss2011}, similar formation mechanisms are unlikely for the objects that are found outside the circumstellar disks of their hosts, as \textit{in situ} formation at these large separations is not supported by these mechanisms, and migration from within the current disk is unlikely without disrupting the circumstellar environment \citep{raymond2012}.
As concluded by \citet{bailey2014} for HD~106906, it is thus likely that TYC~8252-533-1~B rather formed via a star-like pathway by fragmentation processes in the collapsing protostellar cloud \citep{kroupa2001,chabrier2003}.
Future astrometric observations to constrain the orbital parameters of this brown dwarf and spectroscopic analysis of its atmospheric composition are required to confirm this suggested formation scenario.

\subsection{Stellar companions in the SPHERE field of view}
\label{subsec:sphere_stellar_companions}

\rev{%
Six additional companion candidates from the \textit{Gaia} preselection were also identified in the SPHERE observations that were collected within the scope of YSES.
Contrary to TYC~8252-533-1~B, most of these companions seem to exhibit masses that are in good agreement with stellar nature of these objects.
In this section we briefly assess the combined SPHERE and \textit{Gaia} astrometry and photometry of these objects.
The numerical values are presented in Table~\ref{tbl:yses_stellar_companions}.
We used 2MASS $JHK$ photometry of the systems to derive the absolute magnitudes for the companion candidates.
For 2MASS~J12195938-5018404, 2MASS~J12505143-5156353, and 2MASS~J12560830-6926539 the \textit{Gaia} companion candidates are spatially resolved by 2MASS, so they should not pollute the flux measurement of the primary.
This is not the case for 2MASS~J12391404-5454469, 2MASS~J13130714-4537438, and 2MASS~J13335481-6536414, for which the fluxes of both sources are blended.
Furthermore, we identified an additional companion candidate at a separation of approximately 0\farcs2 to 2MASS~J12560830-6926539, which is resolved neither by 2MASS nor \textit{Gaia}.
To obtain an unbiased estimate for the flux of the primary star, we corrected the 2MASS magnitudes of these unresolved pairs for the additional contribution, utilizing Eq.~(2) from \citet{bohn2020b}.
These revised values are reported in Table~\ref{tbl:yses_stellar_companions}.
As for the Gaia fluxes, the SPHERE photometric measurements were converted to object masses by BT-Settl models that were evaluated at the system age.
The corresponding imagery of these additional companions is shown in Fig.~\ref{fig:sphere_stellar_companions} in Appendix~\ref{sec:hci_observing_conditions}.
}
\begin{table*}
\caption{%
\rev{%
SPHERE astrometric and photometric measurements of stellar companions identified by the \textit{Gaia} search.
}
}
\label{tbl:yses_stellar_companions}
\def\arraystretch{1.2}
\centering
\begin{tabular}{@{}lllllllll@{}}
\hline\hline
Target & Date & Filter & Sep. & PA & $m_\star$\tablefootmark{a} & $\Delta m$ & $M_\mathrm{comp}$ & Phot. mass\\
(2MASS ID) & (yyyy-mm-dd) & & (mas) & (\degr) & (mag) & (mag) & (mag) & ($M_\mathrm{Jup}$)\\
\hline
12195938-5018404 & 2018-12-30 & $H$ & $3066\pm4$ & $355.5\pm0.1$ & 9.86 & $1.162\pm0.016$ & $5.107\pm0.016$ & $370.5^{+25.6}_{-38.7}$ \\
12195938-5018404 & 2018-12-30 & $K_s$ & $3067\pm4$ & $355.5\pm0.1$ & 9.65 & $1.086\pm0.072$ & $4.821\pm0.072$ & $380.4^{+34.4}_{-42.8}$ \\
12391404-5454469 & 2019-01-12 & $H$ & $2567\pm4$ & $215.9\pm0.1$ & 9.07 & $1.976\pm0.016$ & $5.887\pm0.016$ & $111.3^{+56.9}_{-40.6}$ \\
12391404-5454469 & 2019-01-12 & $K_s$ & $2568\pm3$ & $215.9\pm0.1$ & 8.96 & $1.835\pm0.006$ & $5.632\pm0.008$ & $106.1^{+59.9}_{-37.7}$ \\
12505143-5156353 & 2019-01-12 & $H$ & $3235\pm4$ & $81.0\pm0.1$ & 8.71 & $2.157\pm0.016$ & $5.675\pm0.016$ & $190.5^{+26.5}_{-51.3}$ \\
12505143-5156353 & 2019-01-12 & $K_s$ & $3240\pm4$ & $81.0\pm0.1$ & 8.63 & $2.061\pm0.002$ & $5.501\pm0.006$ & $180.1^{+24.1}_{-56.6}$ \\
12560830-6926539 & 2019-01-08 & $H$ & $5085\pm9$ & $41.8\pm0.1$ & 8.54 & $3.740\pm0.030$ & $7.291\pm0.053$ & $36.2^{+2.6}_{-6.9}$ \\
12560830-6926539 & 2019-01-08 & $K_s$ & $5086\pm6$ & $41.7\pm0.1$ & 8.30 & $3.388\pm0.377$ & $6.700\pm0.379$ & $39.9^{+12.5}_{-8.9}$ \\
13130714-4537438 & 2017-07-05 & $J$ & $707\pm3$ & $157.5\pm0.3$ & 9.59 & $0.800\pm0.003$ & $4.669\pm0.037$ & $312.4^{+181.0}_{-121.8}$ \\
13130714-4537438 & 2017-07-05 & $H$ & $706\pm3$ & $157.4\pm0.3$ & 8.99 & $0.659\pm0.002$ & $3.926\pm0.036$ & $318.3^{+176.5}_{-121.7}$ \\
13335481-6536414 & 2018-04-30 & $H$ & $1340\pm3$ & $8.2\pm0.2$ & 8.14 & $2.517\pm0.004$ & $5.599\pm0.005$ & $80.2^{+67.3}_{-3.2}$ \\
13335481-6536414 & 2018-04-30 & $K_s$ & $1341\pm3$ & $8.2\pm0.1$ & 7.91 & $2.529\pm0.003$ & $5.380\pm0.004$ & $76.7^{+61.9}_{-3.1}$ \\
13335481-6536414 & 2020-02-16 & $H$ & $1341\pm3$ & $8.4\pm0.1$ & 8.23 & $2.585\pm0.053$ & $5.661\pm0.053$ & $79.4^{+64.9}_{-5.1}$ \\
\hline
\end{tabular}
\tablefoot{%
\tablefoottext{a}{\rev{%
Stellar 2MASS magnitudes.
For 2MASS~J12391404-5454469, 2MASS~J12560830-6926539, 2MASS~J13130714-4537438, and 2MASS~J13335481-6536414 we corrected the 2MASS magnitudes for the contribution of close, unresolved companions.
}}
}
\end{table*}

\subsubsection{2MASS J12195938-5018404}
\label{subsubsec:results_2massj1219}

\begin{figure}
\resizebox{\hsize}{!}{\includegraphics{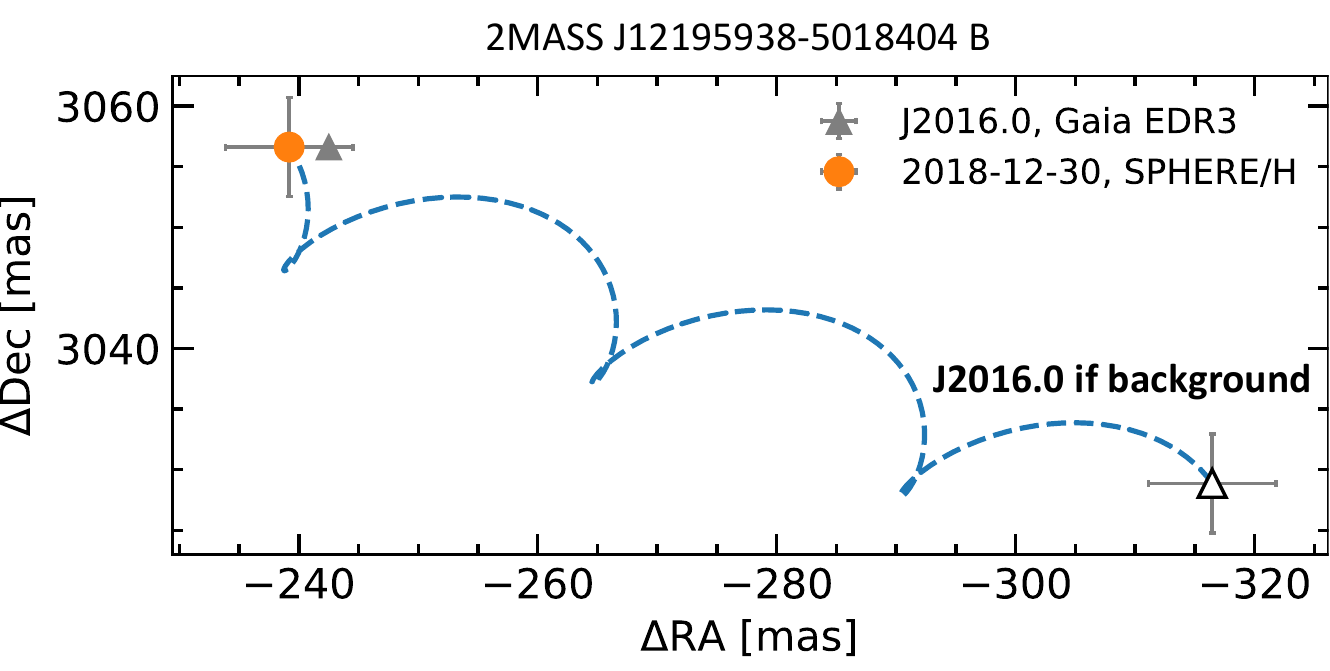}}
\caption{
\rev{%
Proper motion analysis of 2MASS~J12195938-5018404~B.
See Fig.~\ref{fig:ppm_analysis_yses_companion} for a detailed description of the plot elements.
}
}
\label{fig:2MASSJ1219_ppm_analysis}
\end{figure}

\rev{%
The candidate companion detected to 2MASS~J12195938-5018404 clearly seems to be stellar in nature.
Whereas the Gaia $G$ band photometry indicates an object mass of $464.7\pm29.3\,M_\mathrm{Jup}$, the SPHERE near infrared photometry favors slightly lower masses of approximately $375\,M_\mathrm{Jup}$.
Both measurements are consistent at 2$\sigma$ level and the slightly higher Gaia value might originate from uncorrected flux contribution of the primary star.
As visualized in Fig.~\ref{fig:2MASSJ1219_ppm_analysis}, SPHERE and Gaia astrometry for the companion candidate are highly consistent.
}
\rev{%
This is also supported by a companionship probability of $p^\mathrm{C}=100\,\%$ as derived from \textit{Gaia} astrometry alone.
We thus conclude that 2MASS~J12195938-5018404~B is a stellar secondary to its solar-mass primary.
The binary has a projected separation of approximately 470\,au.
}

\subsubsection{2MASS J12391404-5454469}
\label{subsubsec:results_2massj1239}

\begin{figure}
\resizebox{\hsize}{!}{\includegraphics{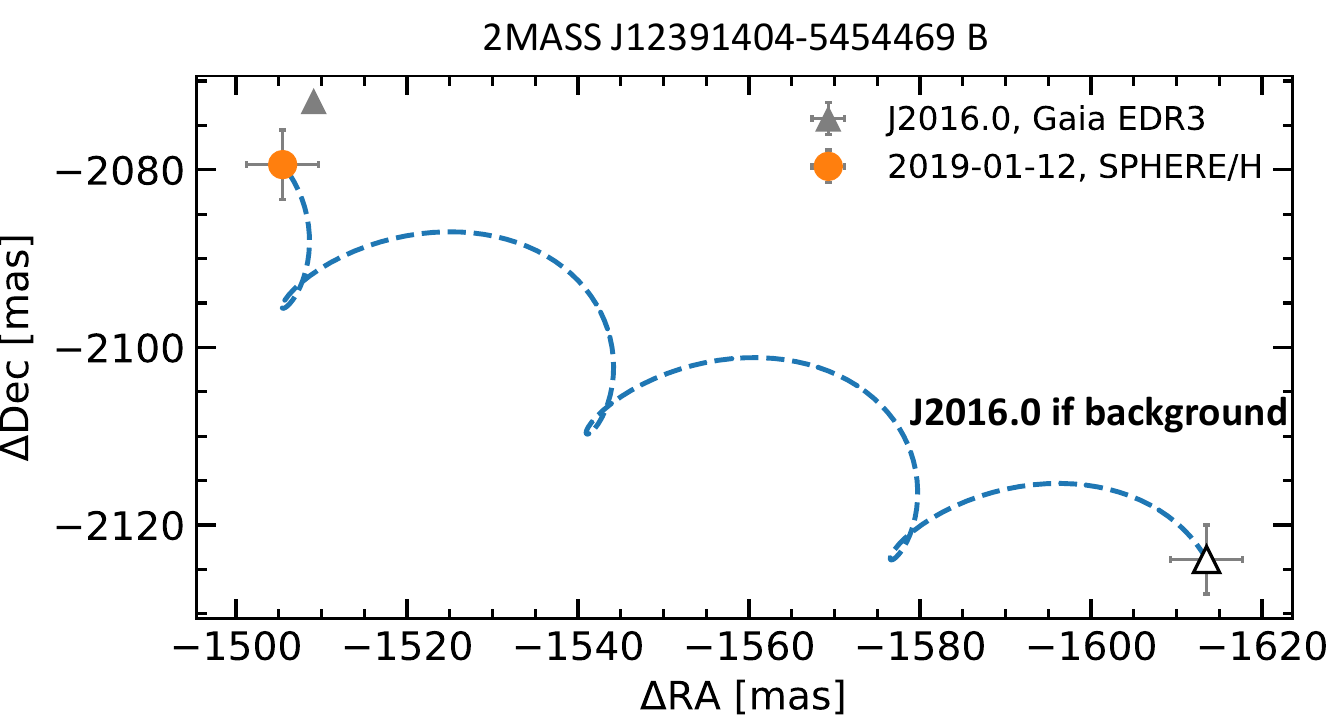}}
\caption{
\rev{%
Proper motion analysis of 2MASS~J12391404-5454469~B.
See Fig.~\ref{fig:ppm_analysis_yses_companion} for a detailed description of the plot elements.
}
}
\label{fig:2MASSJ1239_ppm_analysis}
\end{figure}

\rev{%
The companion candidate to 2MASS~J12391404-5454469 is supposedly of stellar mass.
The \textit{Gaia} photometry is consistent with a mass of $189.1\pm41.9\,M_\mathrm{Jup}$.
Again, the SPHERE data favor a smaller object mass of approximately $110\,M_\mathrm{Jup}$.
The proper motion analysis clearly supports that the companion candidate is gravitationally bound to the primary star, and has a projected separation of approximately 275\,au (see Fig.~\ref{fig:2MASSJ1239_ppm_analysis}).
}
\rev{%
This combined astrometric analysis is confirmed by $p^\mathrm{C}=100\,\%$ that was derived from \textit{Gaia} data alone.
We will thus refer to the object as 2MASS~J12391404-5454469~B henceforth.
}

\subsubsection{2MASS J12505143-5156353}
\label{subsubsec:results_2massj1250}

\begin{figure}
\resizebox{\hsize}{!}{\includegraphics{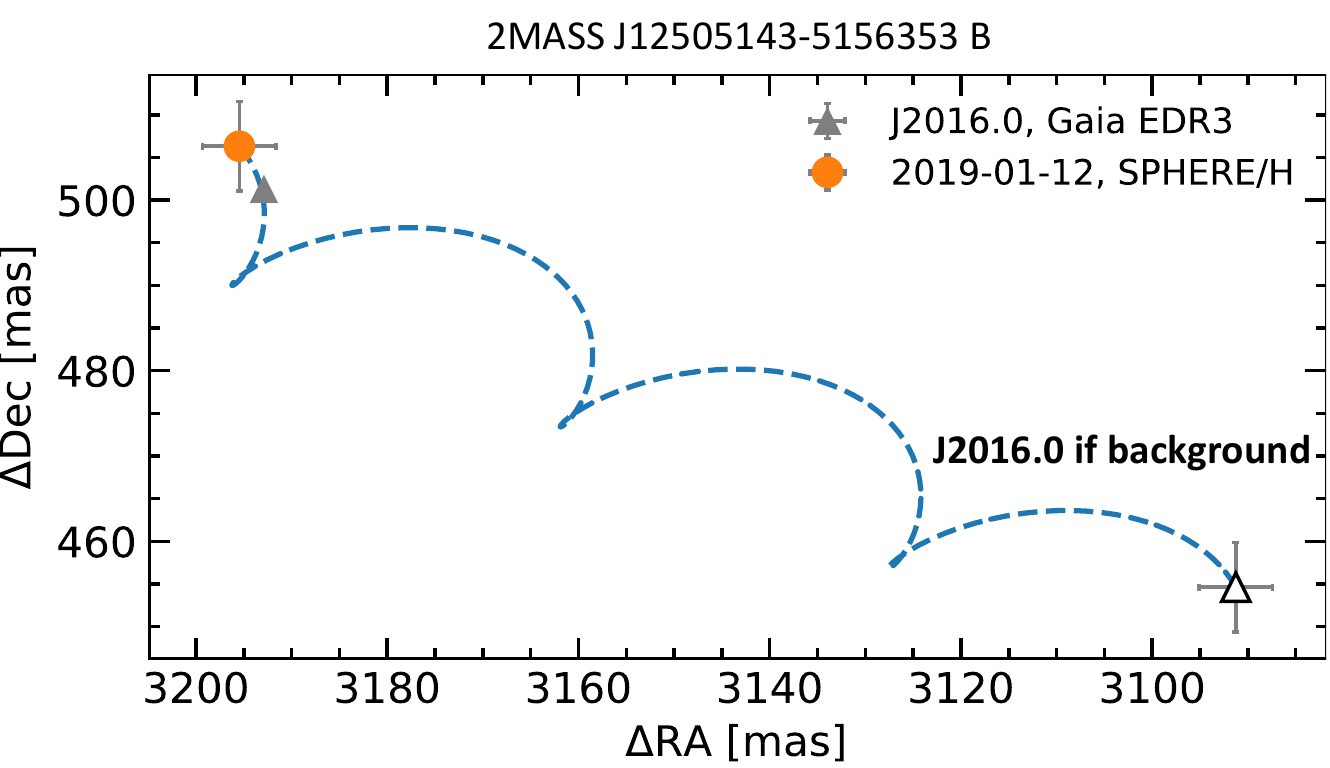}}
\caption{
\rev{%
Proper motion analysis of 2MASS~J12505143-5156353~B.
See Fig.~\ref{fig:ppm_analysis_yses_companion} for a detailed description of the plot elements.
}
}
\label{fig:2MASSJ1250_ppm_analysis}
\end{figure}

\rev{%
The three mass estimates derived from SPHERE $H$ and $K_s$ band and \textit{Gaia} $G$ band data agree particularly well for the candidate companion to 2MASS~J12505143-5156353.
As presented in Fig.~\ref{fig:2MASSJ1250_ppm_analysis}, the object is clearly comoving with the primary star.
}
\rev{%
The \textit{Gaia} companionship probability of $p^\mathrm{C}=100\,\%$ that we derived for this object strongly supports this conclusion.
It should therefore be named 2MASS~J12505143-5156353~B, a stellar companion with a mass of approximately $185\,M_\mathrm{Jup}$.
The projected separation between primary and secondary is approximately 350\,au.
}

\subsubsection{2MASS J12560830-6926539}
\label{subsubsec:results_2massj1256}

\begin{figure}
\resizebox{\hsize}{!}{\includegraphics{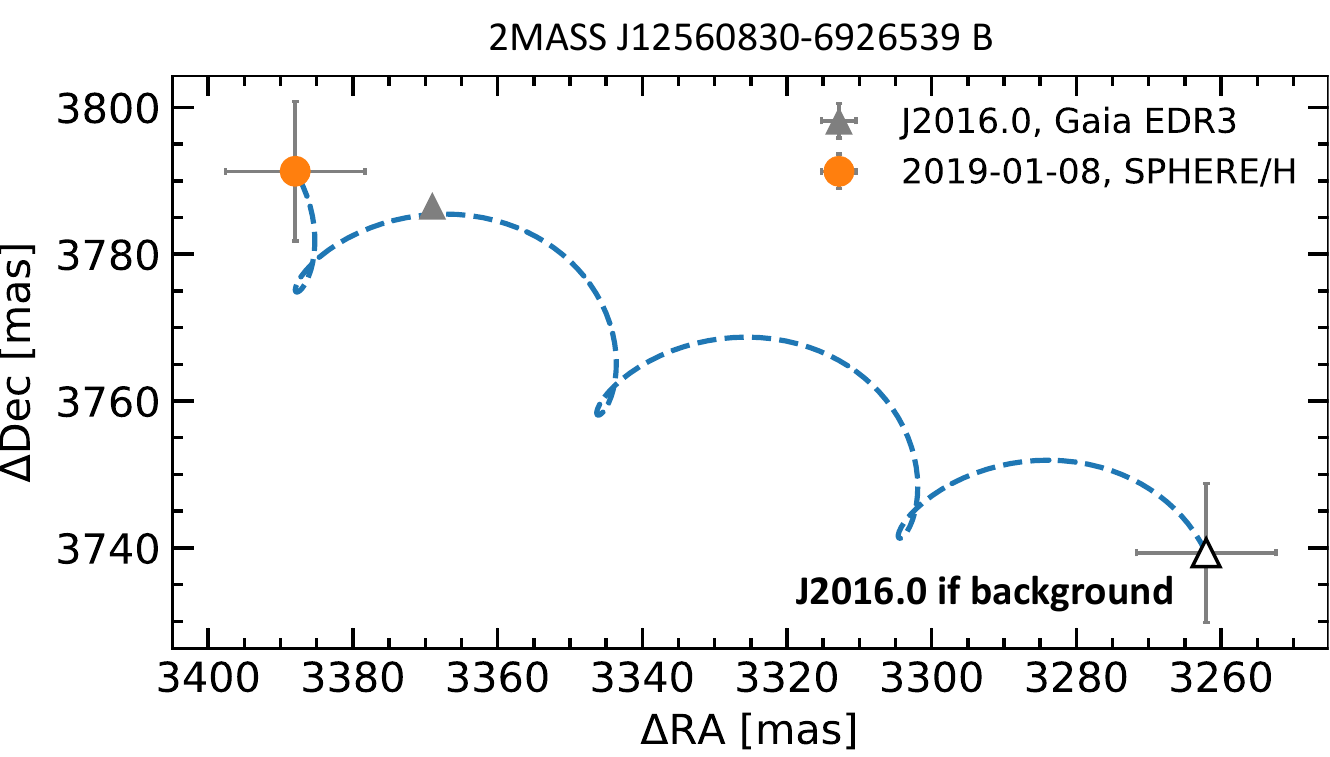}}
\caption{
\rev{%
Proper motion analysis of 2MASS~J12560830-6926539~B.
See Fig.~\ref{fig:ppm_analysis_yses_companion} for a detailed description of the plot elements.
}
}
\label{fig:2MASSJ1256_ppm_analysis}
\end{figure}

\rev{%
As presented in Fig.~\ref{fig:2MASSJ1256_ppm_analysis} the astrometry from \textit{Gaia} and SPHERE for the candidate companion to 2MASS J12560830-6926539 clearly disfavors a static background object.
}
\rev{
But the relative astrometric measurements for both epochs are too disjunct to allow a bound orbit within the measurement uncertainties.
This conclusion seems to be supported by a velocity ratio of $\nicefrac{v_\mathrm{proj}}{v_\mathrm{max}}=1.36\pm0.06$ and a \textit{Gaia} companionship probability of $p^\mathrm{C}=0\,\%$.
}

\rev{%
The reason for this inconsistency can be identified in the high-contrast imaging data collected on 2MASS~J12560830-6926539 (see Fig.~\ref{fig:sphere_stellar_companions} in Appendix~\ref{sec:hci_observing_conditions}).
The primary star is resolved into a visual binary with an angular separation of $0\farcs218\pm0\farcs006$.
The fainter of the two stars is detected at a position angle of 
$33\fdg8\pm1\fdg6$.
The binarity of the source was identified within the scope of the Search for associations containing young stars \citep[SACY;][]{torres2006,elliott2015} and confirmed by \citet{tokovinin2019}.
It also affects the accuracy of the provided \textit{Gaia} astrometry.
The \textit{Gaia} measurements exhibit an astrometric excess noise of 2.3\,mas (\texttt{astrometric\_excess\_noise} parameter) that is detected at a significance of approximately 6\,820 (\texttt{astrometric\_excess\_noise\_sig} parameter).
This significance is much larger than the threshold of 2 that is required for a certain corruption of the astrometric measurements \citep[see][]{lindegren2020b}.
This conclusion is also supported by a renormalised unit weight error (RUWE) of $\sim13$;
for well-behaved sources $\mathrm{RUWE}\approx1$ is expected.
As parallaxes and proper motions for the binary and the \textit{Gaia} companion candidate are highly consistent, it is very likely that these three objects form a gravitationally bound triple system.
Further astrometric monitoring will help to confirm this hypothesis.
}

\rev{
The companion candidate from \textit{Gaia} has been spectroscopically analyzed by \citet{riaz2006}.
The authors determined a spectral type of M1 and classified it as a stellar object.
This is in stark contrast to the photometric mass estimates that we derived from both \textit{Gaia} and SPHERE.
The \textit{Gaia} $G$ band photometry is consistent with a mass of $44.5\pm5.8\,M_\mathrm{Jup}$ and the SPHERE $H$ and $K_s$ band data indicate object masses of $36^{+3}_{-7}\,M_\mathrm{Jup}$ and $40^{+13}_{-9}\,M_\mathrm{Jup}$, respectively.
These values are consistent within their uncertainties and suggest that Gaia EDR3 5844909156504880128 (2MASS J12560892-6926503) is a brown dwarf rather than a stellar companion.
It is possible that \citet{riaz2006} misclassified the companion, perhaps due to contaminating flux from the primary star that was leaking into the slit with a width of 1\farcs5.
Further spectroscopic measurements will help to confirm the sub-stellar nature of this outer companion in a young triple system.
}

\subsubsection{2MASS J13130714-4537438}
\label{subsubsec:results_2massj1313}

\begin{figure}
\resizebox{\hsize}{!}{\includegraphics{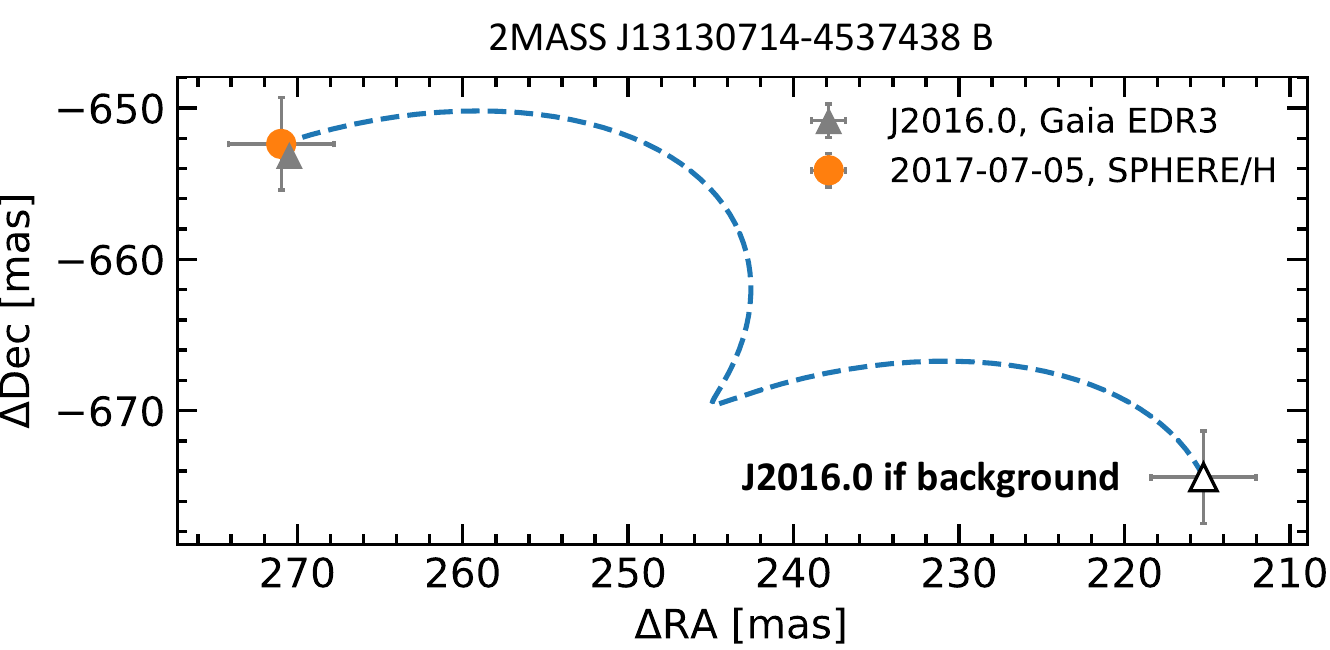}}
\caption{
\rev{%
Proper motion analysis of 2MASS~J13130714-4537438~B.
See Fig.~\ref{fig:ppm_analysis_yses_companion} for a detailed description of the plot elements.
}
}
\label{fig:2MASSJ1313_ppm_analysis}
\end{figure}

\rev{%
The \textit{Gaia} and SPHERE astrometric measurements for the companion candidate to 2MASS J13130714-4537438 are highly consistent and indicate that this object is gravitationally bound to its Sun-like primary (see Fig.~\ref{fig:2MASSJ1313_ppm_analysis}).
}
\rev{%
This conclusion was also drawn from \textit{Gaia} astrometry alone, for which we calculated $p^\mathrm{C}=100\,\%$.
We derived photometric mass estimates of $312.4^{+181.0}_{-121.8}\,M_\mathrm{Jup}$ and $318.3^{+176.5}_{-121.7}\,M_\mathrm{Jup}$ from the SPHERE $H$ and $K_s$ band data, respectively.
The corresponding \textit{Gaia} mass of $414.4\pm114.2\,M_\mathrm{Jup}$ is marginally higher, but well consistent within the uncertainties.
These mass errors are relatively large for this object.
Due the the young age of the system of $2_{-1}^{+2}$\,Myr, the object mass can vary significantly when propagating the age uncertainties.
2MASS~J13130714-4537438~B is likely a stellar-mass companion at a projected separation of approximately 100\,au.
}

\subsubsection{2MASS J13335481-6536414}
\label{subsubsec:results_2massj1333}

\begin{figure}
\resizebox{\hsize}{!}{\includegraphics{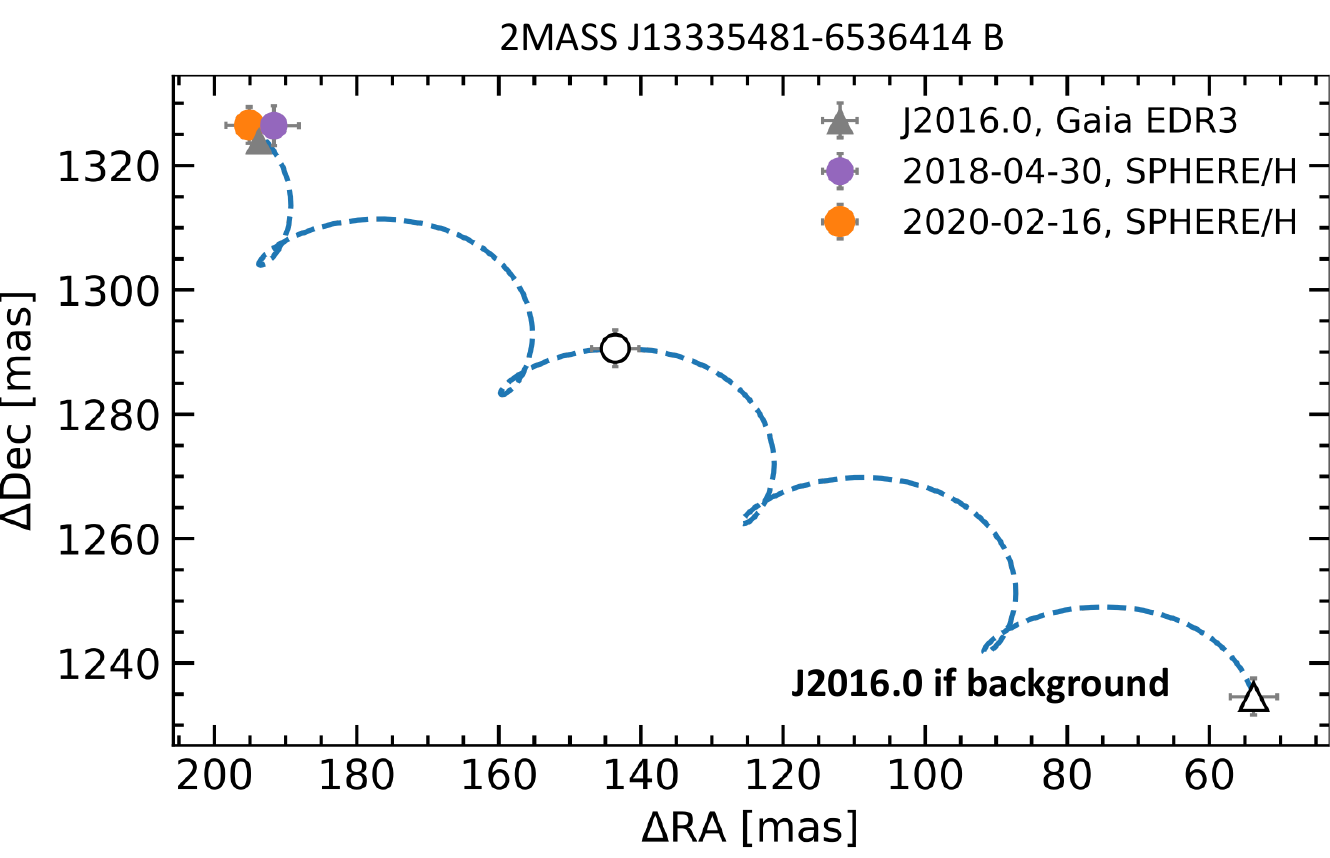}}
\caption{
\rev{%
Proper motion analysis of 2MASS~J13335481-6536414~B.
See Fig.~\ref{fig:ppm_analysis_yses_companion} for a detailed description of the plot elements.
}
}
\label{fig:2MASSJ1333_ppm_analysis}
\end{figure}

\rev{%
As visualized in Fig.~\ref{fig:2MASSJ1333_ppm_analysis}, the three astrometric data points collected with SPHERE and \textit{Gaia} are in very good agreement with each other.
}
\rev{%
This analysis and the \textit{Gaia} companionship probability of $p^\mathrm{C}=100\,\%$ strongly support that 2MASS~J13335481-6536414~B is a gravitationally bound companion with a projected separation of approximately 140\,au.
Both optical and near infrared photometric mass estimates for the secondary agree very well within their uncertainties.
Whereas the photometric mass estimate from \textit{Gaia} suggests that 2MASS~J13335481-6536414~B is a low-mass stellar object with a mass of $91.9\pm31.0\,M_\mathrm{Jup}$, the SPHERE data indicate that it might be of sub-stellar nature with a mass of approximately $78\,M_\mathrm{Jup}$.
But the upper boundaries of the SPHERE measurements also allow for stellar masses.
Future data will be required to discern whether 2MASS~J13335481-6536414~B is a high-mass brown dwarf or rather low-mass stellar companion.
}

\rev{%
An additional companion candidate to 2MASS~J13335481-6536414 at a projected separation of more than 1000\,au is detected by the \textit{Gaia} preselection algorithm.
For this object with the identifier Gaia EDR3 5863747467220738432 we derived $\nicefrac{v_\mathrm{proj}}{v_\mathrm{max}}=1.32\pm0.02$ and hence $p^\mathrm{C}=0\,\%$.
It is possible that the visual stellar binary of 2MASS~J13335481-6536414 AB is corrupting the astrometric measurements for the primary.
Further monitoring will be necessary to determine the orbital parameters of this binary.
These data will be helpful to assess whether the additional companion candidate is consistent with a bound orbit after all and was just rejected due to the velocity contribution from the inner binary.
}

\section{Discussion}
\label{sec:discussion}

\subsection{preselection criteria}
\label{subsec:discussion_preselection_criteria}

Even though we have detected several high confidence companions in \textit{Gaia} \rev{EDR3}, it is necessary to evaluate the quality of our preselection criteria as to whether \textit{bona fide} binaries might be neglected or too many unbound objects are identified.
Due to the nature of our companionship assessment, it is vital that our companion candidates must have a well constrained parallax measurement.
It is thus justified to dismiss any data without or with loosely constrained parallaxes.
Setting the maximum value of the parallax uncertainty to 0.5\,mas defines a rather conservative threshold and considers larger uncertainties due to imprecise astrometry for small separation binaries or objects close to \textit{Gaia}'s sensitivity limit.
Of course, true companions with small angular separations can exhibit even larger uncertainties and might thus be rejected from our preselected sample (these are the companions that reside close to or within the red, dashed exclusion triangle presented in Fig.~\ref{fig:separation_magnitude}).
Within the scope of this work, however, a rejection of these companions need not be considered as a drawback, because other techniques such as lucky imaging \citep[e.g.,][]{janson2012} speckle interferometry \citep[e.g.,][]{tokovinin2010} or AO-assisted high-contrast imaging \citep[e.g.,][]{wang2015,bonavita2021} are much stronger in detecting these close, visual binaries.

The \rev{applied 20\,\% deviation in parallax measurements} of the primary and companion candidate does not exclude any \textit{bona fide} binaries either;
again with the exception of very close pairs with corrupted astrometry.
Given the median parallaxes of our target stars \rev{($\varpi=7.55$\,mas, see Fig.~\ref{fig:sample_properties})}, this \rev{criterion} defines a potential radial separation of up to $\sim33$\,pc ($\approx7\times10^6$\,au), which is significantly larger than the average distance between stars in a young association \citep[e.g.,][]{kraus2008}.
\rev{%
The same cutoff threshold is applied for similar searches presented by \citet{fontanive2019} and \citet{fontanive2021}.
}
The \rev{10\,km\,s$^{-1}$} interval that we chose for the proper motions is justified in a similar manner.
Close companions show a non-negligible amount of orbital motion with respect to the primary star.
\rev{%
For a solar-mass primary, a BD companion of $40\,M_\mathrm{Jup}$ with a semi-major axis of 100\,au would exhibit an orbital velocity of 3\,km\,s$^{-1}$.
For a Sun-like secondary with a semi-major axis of 50\,au this velocity is even as large as 6\,km\,s$^{-1}$.
The 10\,km\,s$^{-1}$ interval is thus required to account for both potential scenarios and additional velocity uncertainties.
Both parallax and proper motion criteria applied in our preselection are rather conservative choices.
This ensures that no \textit{bona fide} binaries get neglected by the algorithm.
A refinement of the preselected sample as presented in Sect.~\ref{sec:results_analysis} is necessary to reject clearly unbound objects afterwards.
}

\begin{figure}
\resizebox{\hsize}{!}{\includegraphics{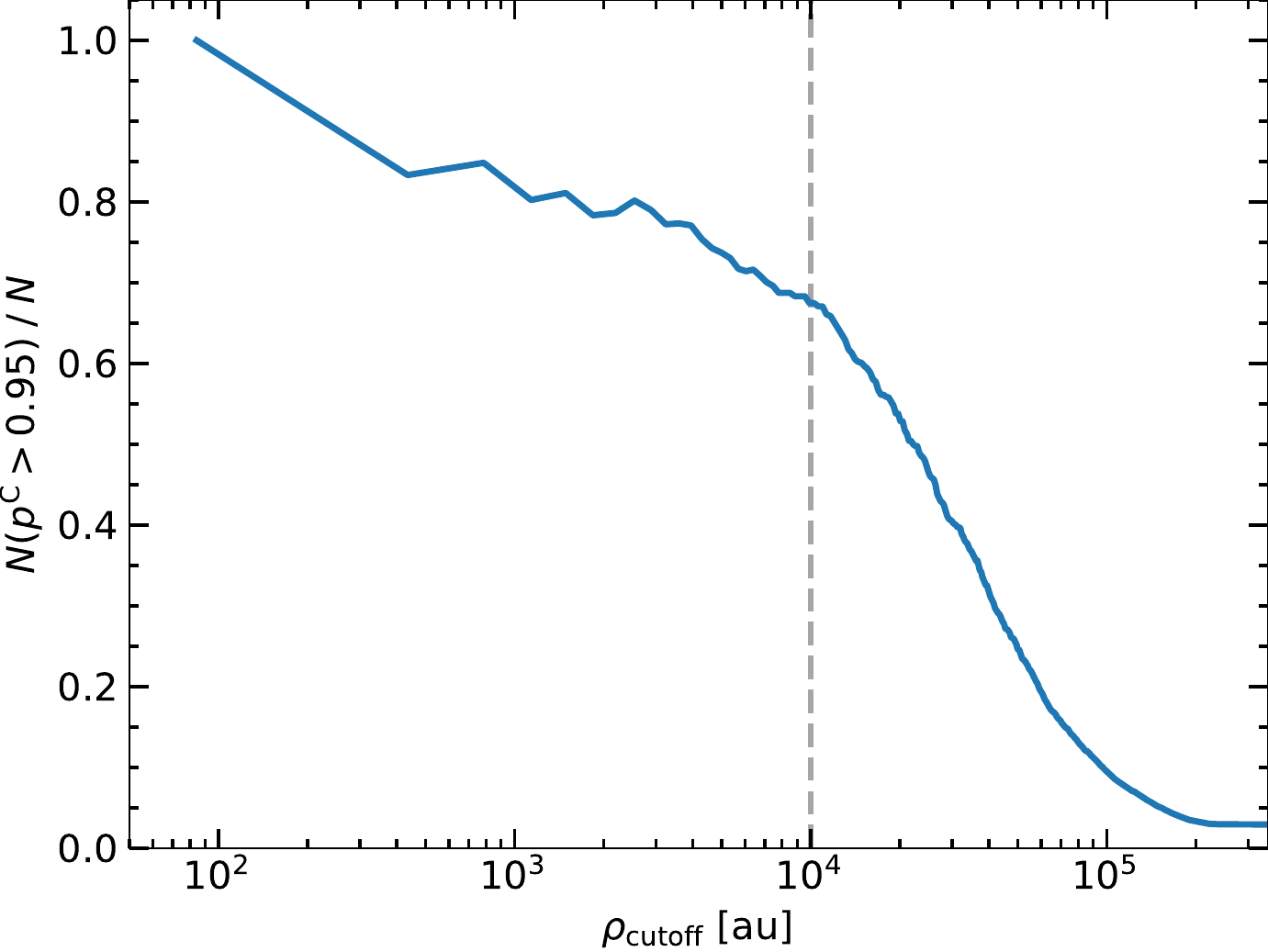}}
\caption{
Fraction of high confidence companions to the total number of identified companion candidates within a specified radius $\rho_\mathrm{cutoff}$.
We show the fractional amount of companions with $p^\mathrm{C}>0.95$ to the total number $N$ of preselected objects interior to our cutoff radius $\rho_\mathrm{cutoff}$.
The grey, dashed line marks the chosen cutoff radius of $\rho_\mathrm{cutoff}=10\,000$\,au.
}
\label{fig:p_bound_search_radius}
\end{figure}

The most important parameter for our selection is probably the cutoff radius \rev{of $\rho_\mathrm{cutoff}=10\,000$\,au} in projected separation.
\rev{The same threshold is also applied within the COCONUTS program that relies on a similar selection methodology \citep[][]{zhang_zj2020,zhang_zj2021}}.
Basic geometry demands that the larger one chooses this parameter, the more companions will be preselected by the algorithm, especially when applying it to a densely populated region on the sky.
It is expected, however, that the fraction of high confidence candidates with $p^\mathrm{C}>0.95$ among all preselected objects decreases with increasing $\rho_\mathrm{cutoff}$.
To assess this correlation, we repeated our full analysis with a cutoff radius of \rev{$\rho_\mathrm{cutoff}=200'000\,\mathrm{au}\approx$1\,pc}.
\rev{This agrees with the 20 times larger cutoff radius that was previously applied in our companionship probability assessment (see Sect.~\ref{subsec:companionship_assessment}).}
In Fig.~\ref{fig:p_bound_search_radius} we show the fraction of high confidence companions to the total number of potential companions identified within a radius of $\rho_\mathrm{cutoff}$.
A clear correlation between the chosen cutoff radius and the number of high confidence objects is visible and the ratio decreases continuously when increasing $\rho_\mathrm{cutoff}$.
For large separations the allowed velocities are in the order of the uncertainties we derive from the differential proper motions, so it becomes hard to assess whether these are actually bound or not.
Our chosen cutoff separation \rev{lies just before the steepest decrease of the fraction of high confidence companions}.
\rev{%
Increasing the cutoff separation would therefore also increase the number of preselected yet unbound companion candidates.
We thus argue that $\rho_\mathrm{cutoff}=10\,000$\,au is a good choice that allows for the detection of several wide-orbit companions, without diluting the preselected sample by an abundance of unbound contaminants.
This is supported by the non-detection of any high confidence companions with a projected angular separation that is larger than 7000\,au (see Table~\ref{tbl:high_priority_companions_1} and Fig.~\ref{fig:high_confidence_comp_properties}).
}

\subsection{Quality assessment of derived \textit{Gaia} properties}
\label{subsec:quality_gaia_properties}

As we detect \rev{seven companion candidates} in both SPHERE and \textit{Gaia} measurements, \rev{these systems} provide ideal test cases to assess the quality of the companion properties that we derived for the remaining sample members.
This analysis was focused on two important parameters that were extracted from our \textit{Gaia} analysis as presented in Sect.~\ref{sec:results_analysis}: (i) the co-movement hypothesis of the companion and (ii) the companion mass estimate.

\rev{
For all these companion candidates but the one detected to 2MASS J12560830-6926539 we derived a velocity ratio $v_\mathrm{proj}/v_\mathrm{max}$ that is smaller than unity.
This resulted in companionship probabilities $p^\mathrm{C}>0.95$, indicating that these are gravitationally bound companions.
These hypotheses are confirmed for all cases, when analyzing the relative astrometric offsets between the SPHERE and \textit{Gaia} epochs as presented in Sect.~\ref{sec:hci_results}.
SPHERE and \textit{Gaia} astrometric measurements are highly consistent and prove the claimed companionship status.
Even though the astrometric measurements from both instruments do not agree within their uncertainties for the companion candidate around 2MASS J12560830-6926539, this is no indication that our classification scheme does not work.
As detailed in Sect.~\ref{subsubsec:results_2massj1256}, the primary is a visual binary itself that is unresolved in the \textit{Gaia} catalog.
For that reason, the measured astrometry is corrupted and does not agree with our SPHERE data.
Previous work however strongly indicated that the companion candidate is indeed gravitationally bound, and all three objects form a triple system.
}

We conclude that the differential velocities directly measured in the \textit{Gaia} catalog are trustworthy, at least for companions that are as bright or brighter and at least as far separated from the primary as \rev{the benchmark targets with complementary SPHERE data}.
\rev{Due to the small angular separation of less than 5\farcs5, these are already quite challenging cases and for the majority of brighter and farther separated companions, a similar or better performance of the evaluation method can be expected.}
Further monitoring and reduction of the velocity uncertainties as presented in Fig.~\ref{fig:separation_escape_velocity} is required to confirm co-movement for all the companion candidates that were detected during our search.
Precise measurements of the RVs of the companions are necessary to obtain a complete three dimensional picture of the orbital dynamics and parameters.

\rev{
For all seven companions the photometric masses determined by either SPHERE or \textit{Gaia} are consistent within 2\,$\sigma$ intervals.
Especially the masses of the identified sub-stellar companions to TYC~8252-533-1, 2MASS J12560830-6926539, and 2MASS J13335481-6536414 are in exceptional agreement.
}
It is thus likely that the mass estimates we derived for all identified companions are good estimates for the actual object masses.
Given its rather close separation to the primary star and a $G$ band contrast of $\Delta G\approx6$\,mag, TYC~8252-533-1~B is one of the more challenging cases for extraction of \textit{Gaia} photometry (which is also visualized by the proximity of the object to the red exclusion zone that is highlighted in Fig.~\ref{fig:separation_magnitude}).
It is therefore likely that the extraction of these relevant \textit{Gaia} parameters could be performed at higher accuracy, leading to more trustworthy results for the objects that are farther away from the red-dashed triangle in Fig.~\ref{fig:separation_mass}.
Of course this is not true for companions that are closely located to \textit{Gaia}'s $G$ band magnitude limit of approximately 21\,mag, for which the small amount of received flux creates larger uncertainties in the measured parameters.

When inspecting the derived masses for the potential triple system comprising two of our targets from the input catalog (2MASS J15241147-3030582 and 2MASS J15241303-3030572) and a third lower-mass companion, it appears that our mass estimates for the stellar members of this trio are quite different from the input masses we adopted from \citet{pecaut2016}.
As listed in Table~\ref{tbl:companions_full_1}, we derived the same mass estimate of approximately $0.7\,M_\sun$ for both stars, which should be very distinct with $1.1\,M_\sun$ and $0.4\,M_\sun$ for 2MASS J15241147-3030582 and 2MASS J15241303-3030572, respectively.
This discrepancy is not caused by the marginally different parallaxes assumed for the corresponding primary, which only results in a magnitude difference of 0.05\,mag.
The main source of error stems from the two very different ages of 18\,Myr and 1\,Myr that \citet{pecaut2016} give for the two targets, which also results in the very different mass estimates of the third companion Gaia~EDR3 6208136396122878976.
Based on the same apparent $G$ band magnitude of this object, we derived masses of $0.23\,M_\sun$ and $0.08\,M_\sun$, respectively, with the higher mass corresponding to the older system age.
If the system is indeed a gravitationally bound stellar triple, which is strongly suggested by the derived values of $p^\mathrm{C}\approx1$, all three components should roughly be of the same age, which would also avoid the change in mass of more than 100\% for the lightest member of the system.
When trying to reproduce the stellar masses of the two targets from \citet{pecaut2016} based on the measured $G$ band photometry and BT-Settl isochrones, we find best agreement with a system age of approximately 4\,Myr, which provides masses of $1.13\,M_\sun$, $0.45\,M_\sun$, and $0.14\,M_\sun$ for 2MASS J15241147-3030582, 2MASS J15241303-3030572, and Gaia~EDR3 6208136396122878976, respectively.

We see similar age and mass discrepancies for the other multiple systems that contain stars from our input catalog as identified companion candidates.
This clearly shows that our derived object masses have to be used carefully as they are heavily affected by the underlying system age.
Further refinements of the stellar age such as updated \textit{Gaia} distances and coevality constraints for the identified multiple systems are thus required to properly constrain derived objects masses.
Further spectral coverage of the SEDs of the detected companions is also necessary, to determine their properties with much higher accuracy than our results, which are based on a single photometric data point.
As most of the identified objects are reasonably separated from the primary, spectroscopic follow-up measurements are easily possible and can in most cases even be performed without any AO-assisted instruments.

Further data releases of the \textit{Gaia} mission might also improve the parameters derived in this study, and it should be straightforward to update the presented results in the future.
The photometric uncertainties, especially of the very faint objects, will certainly be improved with more observational data points, even though the provided $G$ band magnitudes were already precise enough so that these only had a marginal influence on the precision of the derived quantities.
It is not clear if color measurements for all our companion candidates will be available in future releases as most of these incompletely characterized objects are very close to a source of equal or higher brightness, which will always be problematic for precise photometric measurements. 
Higher accuracy astrometry will certainly help to reject or confirm marginal cases from our current study and it will provide new candidate companions that were discarded by our present preselection, either due to non-existing proper motion measurements or too large astrometric uncertainties.

\subsection{Comparison to similar studies}
\label{subsec:comparision_similra_studies}

There are two main studies that have utilized \textit{Gaia} DR2 to reveal members of Sco-Cen.
\citet{goldman2018} identified additional \textit{Gaia} members in the LCC, which is the Sco-Cen sub-group that YSES is focusing on.
This moving group comprises several low-mass stellar and brown dwarf objects, whose membership had not been revealed before \textit{Gaia} DR2.
\citet{damiani2019} focus on the stellar population of Sco-Cen and they identify almost 11,000 pre-main sequence members by \textit{Gaia} astrometry and photometry.
To assess whether any of our candidates have been picked up by either of these studies, we cross-matched our identified companions with the output catalogs of both surveys.
We focused on the systems with a companionship probability $p^\mathrm{C}>0.95$ as presented in Table~\ref{tbl:high_priority_companions_1} and we found that \rev{20} of our detected companions were listed in both of the studied catalogs.
One additional source was identified only by \citet{goldman2018}.
Interestingly, only companions with a right ascension below 15h were picked up by either of these previous studies.
This is not surprising for the work of \citet{goldman2018}, as these authors were explicitly targeting the LCC, which does not extend to a right ascension greater than 15h.
However, none of these discussed surveys identified individual binary or multiple systems as part of their analysis.

\subsection{\rev{Completeness}}
\label{subsec:discussion_completeness}

\begin{figure}
\resizebox{\hsize}{!}{\includegraphics{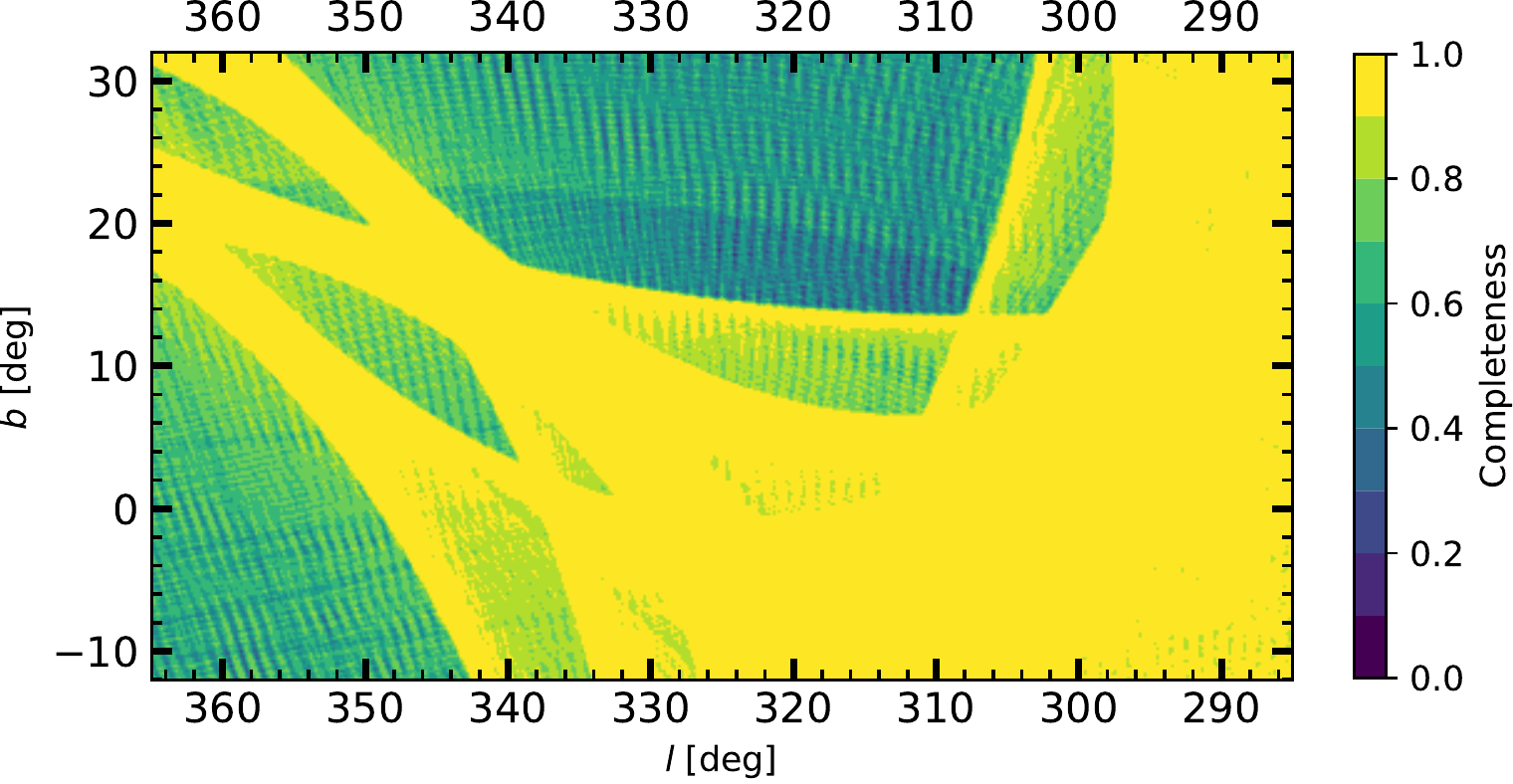}}
\caption{
\rev{
Completeness of \textit{Gaia} DR2 as a function of position on sky.
The completeness is calculated for objects as faint as $G=21$\,mag.
}
}
\label{fig:gaia_completeness}
\end{figure}

\rev{%
We assessed the completeness of our companion detection method based on the completeness of the \textit{Gaia} \revrev{DR2} catalog.
Using the \texttt{selectionfunctions} introduced by \citet{boubert2020a} and \citet{boubert2020b}, we derived the completeness as a function of the position on sky for objects down to the \textit{Gaia} magnitude threshold of $G=21$\,mag.
The resulting completeness map is presented in Fig.~\ref{fig:gaia_completeness}.
}
\rev{%
The map shows clear patterns that correspond to the scanning scheme of the \textit{Gaia} satellite.
For most of the coordinates of our Sco-Cen targets, the achieved completeness is higher than 80\,\% and the LCC, \revrev{which is covering an area of approximately $290\degr<l<310\degr$ and $-10\degr<b<20\degr$,}  has a completeness coverage of almost 100\,\%.
As the underlying code is relying on \textit{Gaia} DR2 we expect the actual completeness of Gaia EDR3 to be even higher.
}

\section{Conclusions}
\label{sec:conclusions}

We present a method to find comoving companions utilizing the \rev{early version of the third} data release of ESA's \textit{Gaia} mission.
We applied this preselection algorithm to a sample of 480 K-type pre-main sequence stars presented by \citet{pecaut2016} aiming to find new (sub-)stellar companions for them.
With our selection criteria we identify \rev{163} potential companions to \rev{142} members of our input catalog, including \rev{21} candidate triples systems.
We could show that the sensitivity of \textit{Gaia} to detect such companion candidates improves with increasing angular separation from the target star.
For our K-type stars we reach the background limit of $G\approx20$\,mag for separations larger than 10\arcsec.
Utilizing age estimates from \citet{pecaut2016}, we converted the measured photometry to mass estimates of the companion candidates using BT-Settl \citep{allard2012,baraffe2015} and MIST \citep{dotter2016,choi2016} models for (sub-)stellar objects.
The age and distance of our sample allows for detections of objects with masses as low as $20\,M_\mathrm{Jup}$.

To corroborate the companionship status of the preselected objects we performed tests based on the available photometry and relative astrometry.
Color-magnitude analysis, which was accessible for the majority of our candidate companions, is in agreement with evolutionary models of young, (sub-)stellar objects.
We further compared differential projected velocities $v_\mathrm{proj}$ of each potential binary to the maximum velocity that still permits bound orbits, $v_\mathrm{max}$.
Based on the posterior distributions of the calculated $v_\mathrm{proj}$ to $v_\mathrm{max}$ ratios \rev{and the statistical constraints from the stellar density around the target star} we derived likelihood \rev{estimates} of a companion to be bound, $p^\mathrm{C}$.
Only \rev{27}\% from our identified companion candidates exhibited a companionship probability of less than 10\%.
Instead, \rev{110} objects (\rev{67}\%) reside in the $p^\mathrm{C}>0.95$ regime, which we designate as our high confidence targets henceforth.
Amongst these we found \rev{ten candidate brown dwarf companions}:
\begin{itemize}
\item Gaia~EDR3 5232514298301348864, $M=36.5\pm2.4\,M_\mathrm{Jup}$,
\item Gaia~EDR3 6071087597497876480, $M=76.0\pm16.7\,M_\mathrm{Jup}$,
\item Gaia~EDR3 5860803696599969280, $M=66.2\pm4.5\,M_\mathrm{Jup}$,
\item Gaia~EDR3 6073106193460538240, $M=46.5\pm3.3\,M_\mathrm{Jup}$,
\item Gaia~EDR3 6083750638540951552, $M=48.7\pm11.1\,M_\mathrm{Jup}$,
\item Gaia~EDR3 6039427503765943040, $M=33.8\pm3.6\,M_\mathrm{Jup}$,
\item Gaia~EDR3 5984404849554176256, $M=73.8\pm4.4\,M_\mathrm{Jup}$,
\item Gaia~EDR3 6043486655875354112, $M=24.8\pm1.0\,M_\mathrm{Jup}$,
\item Gaia~EDR3 6245781131640479360, $M=43.1\pm5.8\,M_\mathrm{Jup}$,
\item Gaia~EDR3 6243833724749589760, $M=78.9\pm12.5\,M_\mathrm{Jup}$.
\end{itemize}
These \rev{ten} brown dwarf companions are suitable for future follow-up observations with ground- and space-based telescopes, aiming to understand their atmospheric architectures and dynamical histories.  

\rev{%
For seven of the detected candidate companions we used complementary high-contrast imaging data.
These were collected with the SPHERE and NACO instruments as part of YSES, a direct imaging search for sub-stellar companions around Sun-like stars \citep{bohn2020a}.
The astrometry that was extracted from this near infrared data was in excellent agreement with the \textit{Gaia} measurements and confirmed the companionship of six candidates.
The remaining target, 2MASS J12560830-6926539,  was resolved into a visual binary that was corrupting the astrometric measurements.
This system is likely to be a triple with a brown dwarf companion of about $40\,M_\mathrm{Jup}$.
}

Another interesting target that was observed with both instruments is TYC~8252-533-1.
This solar-type star hosts a sub-stellar companion at a projected separation of 4\farcs4 that was picked up by \textit{Gaia} and could be directly imaged within the scope of our survey.
Complementary data collected with SPHERE and NACO covered the spectral energy distribution of the companion from 1 to 4 microns and we derived an effective temperature of $T_\mathrm{eff}=3092^{+186}_{-91}\,K$, a surface gravity of $\log\left(g\right)=3.41^{+1.07}_{-0.31}$\,dex, a radius of $R=3.5^{+0.3}_{-0.4}\,R_\mathrm{Jup}$, and a luminosity of $\log\left(L_*/L_\sun\right)=-1.99^{+0.01}_{-0.02}$.
We converted the latter parameter to an object mass by evaluation of BT-Settl isochrones at the system age of $5\pm2$\,Myr, which provided a mass of $52^{+17}_{-11}\,M_\mathrm{Jup}$.
With a projected physical separation of \rev{570}\,au, this brown dwarf companion is likely to reside beyond the debris disk that is found around its host star based on an infrared excess in WISE W3 and W4 bands.
As postulated for the widely separated planetary-mass companion HD~106906~b that was detected outside the circumstellar disk around its primary \citep{bailey2014}, a star-like formation pathway for TYC~8252-533-1~B is favored over planetary formation channels such core accretion or gravitational instabilities in the protoplanetary disk.

\rev{
The photometric mass estimates independently derived from SPHERE and \textit{Gaia} data do also agree very well for the seven targets that were observed with both instruments.
}
We conclude that \textit{Gaia} can properly characterize all objects from our sample that are at lest equally bright and separated from their host stars.
Nevertheless, one should treat the derived mass estimates for the remaining objects with some care, as these are based on a single photometric measurement, and that this parameter is very sensitive to the system age, which needs to be determined better in some cases.

We have shown that our proposed method is a viable tool to identify comoving companions \rev{to Sco-Cen members} in \textit{Gaia} data.
It \rev{should be applicable to arbitrary samples} of input stars, but some justified age and primary mass estimates are required to derive secondary masses and their companionship probability.
The parameter space that is probed by this approach is highly complementary to the comparably small field of view of current high-contrast imaging instruments.
To obtain a detailed understanding of the occurrence rates of giant, sub-stellar companions and their underlying formation mechanisms, it is important not to neglect this population of widely separated objects.

\begin{acknowledgements}
\rev{
We would like to thank the anonymous referee for the detailed feedback they provided.
Their comments helped the authors to improve the quality of the manuscript.
}
We dedicate this article to the memory of France Allard, who passed away in October 2020.
Her groundbreaking work on low-mass stars, brown dwarfs, and exoplanets strongly shaped our current understanding of their atmospheric physics;
the models she (co-)developed -- that were also used in this work and that were frequently accessed by a vast number of researchers -- will remain as a crucial foundation to all future studies in this field and astronomy in general.

The research of A.J.B. leading to these results has received funding from the European Research Council under ERC Starting Grant agreement 678194 (FALCONER).
Part of this research was carried out at the Jet Propulsion Laboratory, California Institute of Technology, under a contract with the National Aeronautics and Space Administration (80NM0018D0004).
G.C. thanks the Swiss National Science Foundation for financial support under grant number 200021\_169131.

This research has used the SIMBAD database, operated at CDS, Strasbourg, France \citep{wenger2000}.

This work has used data from the European Space Agency (ESA) mission {\it Gaia} (\url{https://www.cosmos.esa.int/gaia}), processed by the {\it Gaia} Data Processing and Analysis Consortium (DPAC, \url{https://www.cosmos.esa.int/web/gaia/dpac/consortium}).

Funding for the DPAC has been provided by national institutions, in particular the institutions participating in the {\it Gaia} Multilateral Agreement.

This publication makes use of VOSA, developed under the Spanish Virtual Observatory project supported by the Spanish MINECO through grant AyA2017-84089. 

To achieve the scientific results presented in this article we made use of the \emph{Python} programming language\footnote{Python Software Foundation, \url{https://www.python.org/}}, especially the \emph{SciPy} \citep{virtanen2020}, \emph{NumPy} \citep{numpy}, \emph{Matplotlib} \citep{Matplotlib}, \emph{emcee} \citep{foreman-mackey2013}, \emph{scikit-image} \citep{scikit-image}, \emph{scikit-learn} \citep{scikit-learn}, \emph{photutils} \citep{photutils}, and \emph{astropy} \citep{astropy_1,astropy_2} packages.

\end{acknowledgements}

\bibliographystyle{aa} 
\bibliography{mybib} 

\begin{thebibliography}{104}
\expandafter\ifx\csname natexlab\endcsname\relax\def\natexlab#1{#1}\fi

\bibitem[{{Alcal{\'a}} {et~al.}(2017){Alcal{\'a}}, {Manara}, {Natta}, {Frasca},
  {Testi}, {Nisini}, {Stelzer}, {Williams}, {Antoniucci}, {Biazzo}, {Covino},
  {Esposito}, {Getman}, \& {Rigliaco}}]{alcala2017}
{Alcal{\'a}}, J.~M., {Manara}, C.~F., {Natta}, A., {et~al.} 2017, \aap, 600,
  A20

\bibitem[{{Alcal{\'a}} {et~al.}(2014){Alcal{\'a}}, {Natta}, {Manara}, {Spezzi},
  {Stelzer}, {Frasca}, {Biazzo}, {Covino}, {Randich}, {Rigliaco}, {Testi},
  {Comer{\'o}n}, {Cupani}, \& {D'Elia}}]{alcala2014}
{Alcal{\'a}}, J.~M., {Natta}, A., {Manara}, C.~F., {et~al.} 2014, \aap, 561, A2

\bibitem[{{Alibert} {et~al.}(2005){Alibert}, {Mordasini}, {Benz}, \&
  {Winisdoerffer}}]{alibert2005}
{Alibert}, Y., {Mordasini}, C., {Benz}, W., \& {Winisdoerffer}, C. 2005, \aap,
  434, 343

\bibitem[{{Allard} {et~al.}(2012){Allard}, {Homeier}, \&
  {Freytag}}]{allard2012}
{Allard}, F., {Homeier}, D., \& {Freytag}, B. 2012, Philosophical Transactions
  of the Royal Society of London Series A, 370, 2765

\bibitem[{{Astropy Collaboration} {et~al.}(2018){Astropy Collaboration},
  {Price-Whelan}, {Sip{\H o}cz}, {G{\"u}nther}, {Lim}, {Crawford}, {Conseil},
  {Shupe}, {Craig}, {Dencheva}, {Ginsburg}, {VanderPlas}, {Bradley},
  {P{\'e}rez-Su{\'a}rez}, {de Val-Borro}, {Aldcroft}, {Cruz}, {Robitaille},
  {Tollerud}, {Ardelean}, {Babej}, {Bach}, {Bachetti}, {Bakanov}, {Bamford},
  {Barentsen}, {Barmby}, {Baumbach}, {Berry}, {Biscani}, {Boquien}, {Bostroem},
  {Bouma}, {Brammer}, {Bray}, {Breytenbach}, {Buddelmeijer}, {Burke},
  {Calderone}, {Cano Rodr{\'{\i}}guez}, {Cara}, {Cardoso}, {Cheedella},
  {Copin}, {Corrales}, {Crichton}, {D'Avella}, {Deil}, {Depagne}, {Dietrich},
  {Donath}, {Droettboom}, {Earl}, {Erben}, {Fabbro}, {Ferreira}, {Finethy},
  {Fox}, {Garrison}, {Gibbons}, {Goldstein}, {Gommers}, {Greco}, {Greenfield},
  {Groener}, {Grollier}, {Hagen}, {Hirst}, {Homeier}, {Horton}, {Hosseinzadeh},
  {Hu}, {Hunkeler}, {Ivezi{\'c}}, {Jain}, {Jenness}, {Kanarek}, {Kendrew},
  {Kern}, {Kerzendorf}, {Khvalko}, {King}, {Kirkby}, {Kulkarni}, {Kumar},
  {Lee}, {Lenz}, {Littlefair}, {Ma}, {Macleod}, {Mastropietro}, {McCully},
  {Montagnac}, {Morris}, {Mueller}, {Mumford}, {Muna}, {Murphy}, {Nelson},
  {Nguyen}, {Ninan}, {N{\"o}the}, {Ogaz}, {Oh}, {Parejko}, {Parley}, {Pascual},
  {Patil}, {Patil}, {Plunkett}, {Prochaska}, {Rastogi}, {Reddy Janga},
  {Sabater}, {Sakurikar}, {Seifert}, {Sherbert}, {Sherwood-Taylor}, {Shih},
  {Sick}, {Silbiger}, {Singanamalla}, {Singer}, {Sladen}, {Sooley},
  {Sornarajah}, {Streicher}, {Teuben}, {Thomas}, {Tremblay}, {Turner},
  {Terr{\'o}n}, {van Kerkwijk}, {de la Vega}, {Watkins}, {Weaver}, {Whitmore},
  {Woillez}, {Zabalza}, \& {Astropy Contributors}}]{astropy_2}
{Astropy Collaboration}, {Price-Whelan}, A.~M., {Sip{\H o}cz}, B.~M., {et~al.}
  2018, \aj, 156, 123

\bibitem[{{Astropy Collaboration} {et~al.}(2013){Astropy Collaboration},
  {Robitaille}, {Tollerud}, {Greenfield}, {Droettboom}, {Bray}, {Aldcroft},
  {Davis}, {Ginsburg}, {Price-Whelan}, {Kerzendorf}, {Conley}, {Crighton},
  {Barbary}, {Muna}, {Ferguson}, {Grollier}, {Parikh}, {Nair}, {Unther},
  {Deil}, {Woillez}, {Conseil}, {Kramer}, {Turner}, {Singer}, {Fox}, {Weaver},
  {Zabalza}, {Edwards}, {Azalee Bostroem}, {Burke}, {Casey}, {Crawford},
  {Dencheva}, {Ely}, {Jenness}, {Labrie}, {Lim}, {Pierfederici}, {Pontzen},
  {Ptak}, {Refsdal}, {Servillat}, \& {Streicher}}]{astropy_1}
{Astropy Collaboration}, {Robitaille}, T.~P., {Tollerud}, E.~J., {et~al.} 2013,
  \aap, 558, A33

\bibitem[{{Backman} \& {Paresce}(1993)}]{backman1993}
{Backman}, D.~E. \& {Paresce}, F. 1993, in Protostars and Planets III, ed.
  E.~H. {Levy} \& J.~I. {Lunine}, 1253

\bibitem[{{Bailer-Jones} {et~al.}(2021){Bailer-Jones}, {Rybizki}, {Fouesneau},
  {Demleitner}, \& {Andrae}}]{bailer-jones2021}
{Bailer-Jones}, C.~A.~L., {Rybizki}, J., {Fouesneau}, M., {Demleitner}, M., \&
  {Andrae}, R. 2021, \aj, 161, 147

\bibitem[{{Bailey} {et~al.}(2014){Bailey}, {Meshkat}, {Reiter}, {Morzinski},
  {Males}, {Su}, {Hinz}, {Kenworthy}, {Stark}, {Mamajek}, {Briguglio}, {Close},
  {Follette}, {Puglisi}, {Rodigas}, {Weinberger}, \& {Xompero}}]{bailey2014}
{Bailey}, V., {Meshkat}, T., {Reiter}, M., {et~al.} 2014, \apjl, 780, L4

\bibitem[{{Baraffe} {et~al.}(2015){Baraffe}, {Homeier}, {Allard}, \&
  {Chabrier}}]{baraffe2015}
{Baraffe}, I., {Homeier}, D., {Allard}, F., \& {Chabrier}, G. 2015, \aap, 577,
  A42

\bibitem[{{Bayo} {et~al.}(2008){Bayo}, {Rodrigo}, {Barrado Y Navascu{\'e}s},
  {Solano}, {Guti{\'e}rrez}, {Morales-Calder{\'o}n}, \& {Allard}}]{bayo2008}
{Bayo}, A., {Rodrigo}, C., {Barrado Y Navascu{\'e}s}, D., {et~al.} 2008, \aap,
  492, 277

\bibitem[{{Beuzit} {et~al.}(2019){Beuzit}, {Vigan}, {Mouillet}, {Dohlen},
  {Gratton}, {Boccaletti}, {Sauvage}, {Schmid}, {Langlois}, {Petit},
  {Baruffolo}, {Feldt}, {Milli}, {Wahhaj}, {Abe}, {Anselmi}, {Antichi},
  {Barette}, {Baudrand}, {Baudoz}, {Bazzon}, {Bernardi}, {Blanchard}, {Brast},
  {Bruno}, {Buey}, {Carbillet}, {Carle}, {Cascone}, {Chapron}, {Charton},
  {Chauvin}, {Claudi}, {Costille}, {De Caprio}, {de Boer}, {Delboulb{\'e}},
  {Desidera}, {Dominik}, {Downing}, {Dupuis}, {Fabron}, {Fantinel}, {Farisato},
  {Feautrier}, {Fedrigo}, {Fusco}, {Gigan}, {Ginski}, {Girard}, {Giro},
  {Gisler}, {Gluck}, {Gry}, {Henning}, {Hubin}, {Hugot}, {Incorvaia}, {Jaquet},
  {Kasper}, {Lagadec}, {Lagrange}, {Le Coroller}, {Le Mignant}, {Le Ruyet},
  {Lessio}, {Lizon}, {Llored}, {Lundin}, {Madec}, {Magnard}, {Marteaud},
  {Martinez}, {Maurel}, {M{\'e}nard}, {Mesa}, {M{\"o}ller-Nilsson}, {Moulin},
  {Moutou}, {Orign{\'e}}, {Parisot}, {Pavlov}, {Perret}, {Pragt}, {Puget},
  {Rabou}, {Ramos}, {Reess}, {Rigal}, {Rochat}, {Roelfsema}, {Rousset}, {Roux},
  {Saisse}, {Salasnich}, {Santambrogio}, {Scuderi}, {Segransan}, {Sevin},
  {Siebenmorgen}, {Soenke}, {Stadler}, {Suarez}, {Tiph{\`e}ne}, {Turatto},
  {Udry}, {Vakili}, {Waters}, {Weber}, {Wildi}, {Zins}, \&
  {Zurlo}}]{beuzit2019}
{Beuzit}, J.~L., {Vigan}, A., {Mouillet}, D., {et~al.} 2019, \aap, 631, A155

\bibitem[{{Bohn} {et~al.}(2021){Bohn}, {Ginski}, {Kenworthy}, {Mamajek},
  {Pecaut}, {Mugrauer}, {Vogt}, {Adam}, {Meshkat}, {Reggiani}, \&
  {Snik}}]{bohn2021}
{Bohn}, A.~J., {Ginski}, C., {Kenworthy}, M.~A., {et~al.} 2021, \aap, 648, A73

\bibitem[{{Bohn} {et~al.}(2019){Bohn}, {Kenworthy}, {Ginski}, {Benisty}, {de
  Boer}, {Keller}, {Mamajek}, {Meshkat}, {Muro-Arena}, {Pecaut}, {Snik},
  {Wolff}, \& {Reggiani}}]{bohn2019}
{Bohn}, A.~J., {Kenworthy}, M.~A., {Ginski}, C., {et~al.} 2019, \aap, 624, A87

\bibitem[{{Bohn} {et~al.}(2020{\natexlab{a}}){Bohn}, {Kenworthy}, {Ginski},
  {Manara}, {Pecaut}, {de Boer}, {Keller}, {Mamajek}, {Meshkat}, {Reggiani},
  {Todorov}, \& {Snik}}]{bohn2020a}
{Bohn}, A.~J., {Kenworthy}, M.~A., {Ginski}, C., {et~al.} 2020{\natexlab{a}},
  \mnras, 492, 431

\bibitem[{{Bohn} {et~al.}(2020{\natexlab{b}}){Bohn}, {Kenworthy}, {Ginski},
  {Rieder}, {Mamajek}, {Meshkat}, {Pecaut}, {Reggiani}, {de Boer}, {Keller},
  {Snik}, \& {Southworth}}]{bohn2020c}
{Bohn}, A.~J., {Kenworthy}, M.~A., {Ginski}, C., {et~al.} 2020{\natexlab{b}},
  \apjl, 898, L16

\bibitem[{{Bohn} {et~al.}(2020{\natexlab{c}}){Bohn}, {Southworth}, {Ginski},
  {Kenworthy}, {Maxted}, \& {Evans}}]{bohn2020b}
{Bohn}, A.~J., {Southworth}, J., {Ginski}, C., {et~al.} 2020{\natexlab{c}},
  \aap, 635, A73

\bibitem[{{Bonavita} {et~al.}(2021){Bonavita}, {Gratton}, {Desidera},
  {Squicciarini}, {D'Orazi}, {Zurlo}, {Biller}, {Chauvin}, {Fontanive},
  {Janson}, {Messina}, {Menard}, {Meyer}, {Vigan}, {Avenhaus}, {Asensio
  Torres}, {Beuzit}, {Boccaletti}, {Bonnefoy}, {Brandner}, {Cantalloube},
  {Cheetham}, {Cudel}, {Daemgen}, {Delorme}, {Desgrange}, {Dominik}, {Engler},
  {Feautrier}, {Feldt}, {Galicher}, {Garufi}, {Gasparri}, {Ginski}, {Girard},
  {Grandjean}, {Hagelberg}, {Henning}, {Hunziker}, {Kasper}, {Keppler},
  {Lagadec}, {Lagrange}, {Langlois}, {Lannier}, {Lazzoni}, {Le Coroller},
  {Ligi}, {Lombart}, {Maire}, {Mazevet}, {Mesa}, {Mouillet}, {Moutou},
  {Muller}, {Peretti}, {Perrot}, {Petrus}, {Potier}, {Ramos}, {Rickman},
  {Rouan}, {Salter}, {Samland}, {Schmidt}, {Sissa}, {Stolker}, {Szulagyil},
  {Udry}, \& {Wildi}}]{bonavita2021}
{Bonavita}, M., {Gratton}, R., {Desidera}, S., {et~al.} 2021, arXiv e-prints,
  arXiv:2103.13706

\bibitem[{{Boss}(1997)}]{boss1997}
{Boss}, A.~P. 1997, Science, 276, 1836

\bibitem[{{Boss}(2011)}]{boss2011}
{Boss}, A.~P. 2011, \apj, 731, 74

\bibitem[{{Boubert} \& {Everall}(2020)}]{boubert2020b}
{Boubert}, D. \& {Everall}, A. 2020, \mnras, 497, 4246

\bibitem[{{Boubert} {et~al.}(2020){Boubert}, {Everall}, \&
  {Holl}}]{boubert2020a}
{Boubert}, D., {Everall}, A., \& {Holl}, B. 2020, \mnras, 497, 1826

\bibitem[{{Bradley} {et~al.}(2016){Bradley}, {Sipocz}, {Robitaille},
  {Tollerud}, {Deil}, {Vin{\'\i}cius}, {Barbary}, {G{\"u}nther}, {Bostroem},
  {Droettboom}, {Bray}, {Bratholm}, {Pickering}, {Craig}, {Pascual}, {Greco},
  {Donath}, {Kerzendorf}, {Littlefair}, {Barentsen}, {D'Eugenio}, \&
  {Weaver}}]{photutils}
{Bradley}, L., {Sipocz}, B., {Robitaille}, T., {et~al.} 2016, {Photutils:
  Photometry tools}

\bibitem[{{Chabrier}(2003)}]{chabrier2003}
{Chabrier}, G. 2003, \pasp, 115, 763

\bibitem[{{Chabrier} {et~al.}(2009){Chabrier}, {Baraffe}, {Leconte},
  {Gallardo}, \& {Barman}}]{chabrier2009}
{Chabrier}, G., {Baraffe}, I., {Leconte}, J., {Gallardo}, J., \& {Barman}, T.
  2009, in American Institute of Physics Conference Series, Vol. 1094, 15th
  Cambridge Workshop on Cool Stars, Stellar Systems, and the Sun, ed.
  E.~{Stempels}, 102--111

\bibitem[{{Chauvin} {et~al.}(2017){Chauvin}, {Desidera}, {Lagrange}, {Vigan},
  {Gratton}, {Langlois}, {Bonnefoy}, {Beuzit}, {Feldt}, {Mouillet}, {Meyer},
  {Cheetham}, {Biller}, {Boccaletti}, {D'Orazi}, {Galicher}, {Hagelberg},
  {Maire}, {Mesa}, {Olofsson}, {Samland}, {Schmidt}, {Sissa}, {Bonavita},
  {Charnay}, {Cudel}, {Daemgen}, {Delorme}, {Janin-Potiron}, {Janson},
  {Keppler}, {Le Coroller}, {Ligi}, {Marleau}, {Messina}, {Molli{\`e}re},
  {Mordasini}, {M{\"u}ller}, {Peretti}, {Perrot}, {Rodet}, {Rouan}, {Zurlo},
  {Dominik}, {Henning}, {Menard}, {Schmid}, {Turatto}, {Udry}, {Vakili}, {Abe},
  {Antichi}, {Baruffolo}, {Baudoz}, {Baudrand}, {Blanchard}, {Bazzon}, {Buey},
  {Carbillet}, {Carle}, {Charton}, {Cascone}, {Claudi}, {Costille}, {Deboulbe},
  {De Caprio}, {Dohlen}, {Fantinel}, {Feautrier}, {Fusco}, {Gigan}, {Giro},
  {Gisler}, {Gluck}, {Hubin}, {Hugot}, {Jaquet}, {Kasper}, {Madec}, {Magnard},
  {Martinez}, {Maurel}, {Le Mignant}, {M{\"o}ller-Nilsson}, {Llored}, {Moulin},
  {Orign{\'e}}, {Pavlov}, {Perret}, {Petit}, {Pragt}, {Puget}, {Rabou},
  {Ramos}, {Rigal}, {Rochat}, {Roelfsema}, {Rousset}, {Roux}, {Salasnich},
  {Sauvage}, {Sevin}, {Soenke}, {Stadler}, {Suarez}, {Weber}, {Wildi},
  {Antoniucci}, {Augereau}, {Baudino}, {Brandner}, {Engler}, {Girard}, {Gry},
  {Kral}, {Kopytova}, {Lagadec}, {Milli}, {Moutou}, {Schlieder},
  {Szul{\'a}gyi}, {Thalmann}, \& {Wahhaj}}]{chauvin2017b}
{Chauvin}, G., {Desidera}, S., {Lagrange}, A.~M., {et~al.} 2017, Astronomy and
  Astrophysics, 605, L9

\bibitem[{{Choi} {et~al.}(2016){Choi}, {Dotter}, {Conroy}, {Cantiello},
  {Paxton}, \& {Johnson}}]{choi2016}
{Choi}, J., {Dotter}, A., {Conroy}, C., {et~al.} 2016, \apj, 823, 102

\bibitem[{{Comer{\'o}n}(2008)}]{comeron2008}
{Comer{\'o}n}, F. 2008, {The Lupus Clouds}, ed. B.~{Reipurth}, Vol.~5, 295

\bibitem[{{Cropper} {et~al.}(2018){Cropper}, {Katz}, {Sartoretti}, {Prusti},
  {de Bruijne}, {Chassat}, {Charvet}, {Boyadjian}, {Perryman}, {Sarri}, {Gare},
  {Erdmann}, {Munari}, {Zwitter}, {Wilkinson}, {Arenou}, {Vallenari},
  {G{\'o}mez}, {Panuzzo}, {Seabroke}, {Allende Prieto}, {Benson}, {Marchal},
  {Huckle}, {Smith}, {Dolding}, {Jan{\ss}en}, {Viala}, {Blomme}, {Baker},
  {Boudreault}, {Crifo}, {Soubiran}, {Fr{\'e}mat}, {Jasniewicz}, {Guerrier},
  {Guy}, {Turon}, {Jean-Antoine-Piccolo}, {Th{\'e}venin}, {David}, {Gosset}, \&
  {Damerdji}}]{cropper2018}
{Cropper}, M., {Katz}, D., {Sartoretti}, P., {et~al.} 2018, \aap, 616, A5

\bibitem[{{Cutri} {et~al.}(2012{\natexlab{a}}){Cutri}, {Skrutskie}, {van Dyk},
  {Beichman}, {Carpenter}, {Chester}, {Cambresy}, {Evans}, {Fowler}, {Gizis},
  {Howard}, {Huchra}, {Jarrett}, {Kopan}, {Kirkpatrick}, {Light}, {Marsh},
  {McCallon}, {Schneider}, {Stiening}, {Sykes}, {Weinberg}, {Wheaton},
  {Wheelock}, \& {Zacharias}}]{cutri2012a}
{Cutri}, R.~M., {Skrutskie}, M.~F., {van Dyk}, S., {et~al.} 2012{\natexlab{a}},
  VizieR Online Data Catalog, II/281

\bibitem[{{Cutri} {et~al.}(2003){Cutri}, {Skrutskie}, {van Dyk}, {Beichman},
  {Carpenter}, {Chester}, {Cambresy}, {Evans}, {Fowler}, {Gizis}, {Howard},
  {Huchra}, {Jarrett}, {Kopan}, {Kirkpatrick}, {Light}, {Marsh}, {McCallon},
  {Schneider}, {Stiening}, {Sykes}, {Weinberg}, {Wheaton}, {Wheelock}, \&
  {Zacarias}}]{cutri2003}
{Cutri}, R.~M., {Skrutskie}, M.~F., {van Dyk}, S., {et~al.} 2003, VizieR Online
  Data Catalog, II/246

\bibitem[{{Cutri} {et~al.}(2012{\natexlab{b}})}]{cutri2012b}
{Cutri}, R.~M. {et~al.} 2012{\natexlab{b}}, VizieR Online Data Catalog, II/311

\bibitem[{{Damiani} {et~al.}(2019){Damiani}, {Prisinzano}, {Pillitteri},
  {Micela}, \& {Sciortino}}]{damiani2019}
{Damiani}, F., {Prisinzano}, L., {Pillitteri}, I., {Micela}, G., \&
  {Sciortino}, S. 2019, \aap, 623, A112

\bibitem[{{de Zeeuw} {et~al.}(1999){de Zeeuw}, {Hoogerwerf}, {de Bruijne},
  {Brown}, \& {Blaauw}}]{dezeeuw1999}
{de Zeeuw}, P.~T., {Hoogerwerf}, R., {de Bruijne}, J.~H.~J., {Brown}, A.~G.~A.,
  \& {Blaauw}, A. 1999, \aj, 117, 354

\bibitem[{{DENIS Consortium}(2005)}]{denis_consortium2005}
{DENIS Consortium}. 2005, VizieR Online Data Catalog, II/263

\bibitem[{{Dodson-Robinson} {et~al.}(2009){Dodson-Robinson}, {Veras}, {Ford},
  \& {Beichman}}]{dodsonrobinson2009}
{Dodson-Robinson}, S.~E., {Veras}, D., {Ford}, E.~B., \& {Beichman}, C.~A.
  2009, \apj, 707, 79

\bibitem[{{Dotter}(2016)}]{dotter2016}
{Dotter}, A. 2016, \apjs, 222, 8

\bibitem[{{Duch{\^e}ne} \& {Kraus}(2013)}]{duchene2013}
{Duch{\^e}ne}, G. \& {Kraus}, A. 2013, \araa, 51, 269

\bibitem[{{Duquennoy} \& {Mayor}(1991)}]{duquennoy1991}
{Duquennoy}, A. \& {Mayor}, M. 1991, \aap, 500, 337

\bibitem[{{Durisen} {et~al.}(2007){Durisen}, {Boss}, {Mayer}, {Nelson},
  {Quinn}, \& {Rice}}]{durisen2007}
{Durisen}, R.~H., {Boss}, A.~P., {Mayer}, L., {et~al.} 2007, in Protostars and
  Planets V, ed. B.~{Reipurth}, D.~{Jewitt}, \& K.~{Keil}, 607

\bibitem[{{Eggleton} \& {Tokovinin}(2008)}]{eggleton2008}
{Eggleton}, P.~P. \& {Tokovinin}, A.~A. 2008, \mnras, 389, 869

\bibitem[{{Elliott} {et~al.}(2015){Elliott}, {Hu{\'e}lamo}, {Bouy}, {Bayo},
  {Melo}, {Torres}, {Sterzik}, {Quast}, {Chauvin}, \& {Barrado}}]{elliott2015}
{Elliott}, P., {Hu{\'e}lamo}, N., {Bouy}, H., {et~al.} 2015, \aap, 580, A88

\bibitem[{{Fontanive} \& {Bardalez Gagliuffi}(2021)}]{fontanive2021}
{Fontanive}, C. \& {Bardalez Gagliuffi}, D. 2021, Frontiers in Astronomy and
  Space Sciences, 8, 16

\bibitem[{{Fontanive} {et~al.}(2019){Fontanive}, {Rice}, {Bonavita}, {Lopez},
  {Mu{\v{z}}i{\'c}}, {}, \& {Biller}}]{fontanive2019}
{Fontanive}, C., {Rice}, K., {Bonavita}, M., {et~al.} 2019, \mnras, 485, 4967

\bibitem[{{Foreman-Mackey} {et~al.}(2013){Foreman-Mackey}, {Hogg}, {Lang}, \&
  {Goodman}}]{foreman-mackey2013}
{Foreman-Mackey}, D., {Hogg}, D.~W., {Lang}, D., \& {Goodman}, J. 2013, \pasp,
  125, 306

\bibitem[{{Gaia Collaboration} {et~al.}(2018){Gaia Collaboration}, {Brown},
  {Vallenari}, {Prusti}, {de Bruijne}, {Babusiaux}, {Bailer-Jones}, {Biermann},
  {Evans}, {Eyer}, {Jansen}, {Jordi}, {Klioner}, {Lammers}, {Lindegren},
  {Luri}, {Mignard}, {Panem}, {Pourbaix}, {Randich}, {Sartoretti}, {Siddiqui},
  {Soubiran}, {van Leeuwen}, {Walton}, {Arenou}, {Bastian}, {Cropper},
  {Drimmel}, {Katz}, {Lattanzi}, {Bakker}, {Cacciari}, {Casta{\~n}eda},
  {Chaoul}, {Cheek}, {De Angeli}, {Fabricius}, {Guerra}, {Holl}, {Masana},
  {Messineo}, {Mowlavi}, {Nienartowicz}, {Panuzzo}, {Portell}, {Riello},
  {Seabroke}, {Tanga}, {Th{\'e}venin}, {Gracia-Abril}, {Comoretto},
  {Garcia-Reinaldos}, {Teyssier}, {Altmann}, {Andrae}, {Audard},
  {Bellas-Velidis}, {Benson}, {Berthier}, {Blomme}, {Burgess}, {Busso},
  {Carry}, {Cellino}, {Clementini}, {Clotet}, {Creevey}, {Davidson}, {De
  Ridder}, {Delchambre}, {Dell'Oro}, {Ducourant},
  {Fern{\'a}ndez-Hern{\'a}ndez}, {Fouesneau}, {Fr{\'e}mat}, {Galluccio},
  {Garc{\'\i}a-Torres}, {Gonz{\'a}lez-N{\'u}{\~n}ez}, {Gonz{\'a}lez-Vidal},
  {Gosset}, {Guy}, {Halbwachs}, {Hambly}, {Harrison}, {Hern{\'a}ndez},
  {Hestroffer}, {Hodgkin}, {Hutton}, {Jasniewicz}, {Jean-Antoine-Piccolo},
  {Jordan}, {Korn}, {Krone-Martins}, {Lanzafame}, {Lebzelter}, {L{\"o}ffler},
  {Manteiga}, {Marrese}, {Mart{\'\i}n-Fleitas}, {Moitinho}, {Mora}, {Muinonen},
  {Osinde}, {Pancino}, {Pauwels}, {Petit}, {Recio-Blanco}, {Richards},
  {Rimoldini}, {Robin}, {Sarro}, {Siopis}, {Smith}, {Sozzetti}, {S{\"u}veges},
  {Torra}, {van Reeven}, {Abbas}, {Abreu Aramburu}, {Accart}, {Aerts},
  {Altavilla}, {{\'A}lvarez}, {Alvarez}, {Alves}, {Anderson}, {Andrei},
  {Anglada Varela}, {Antiche}, {Antoja}, {Arcay}, {Astraatmadja}, {Bach},
  {Baker}, {Balaguer-N{\'u}{\~n}ez}, {Balm}, {Barache}, {Barata}, {Barbato},
  {Barblan}, {Barklem}, {Barrado}, {Barros}, {Barstow}, {Bartholom{\'e}
  Mu{\~n}oz}, {Bassilana}, {Becciani}, {Bellazzini}, {Berihuete}, {Bertone},
  {Bianchi}, {Bienaym{\'e}}, {Blanco-Cuaresma}, {Boch}, {Boeche}, {Bombrun},
  {Borrachero}, {Bossini}, {Bouquillon}, {Bourda}, {Bragaglia}, {Bramante},
  {Breddels}, {Bressan}, {Brouillet}, {Br{\"u}semeister}, {Brugaletta},
  {Bucciarelli}, {Burlacu}, {Busonero}, {Butkevich}, {Buzzi}, {Caffau},
  {Cancelliere}, {Cannizzaro}, {Cantat-Gaudin}, {Carballo}, {Carlucci},
  {Carrasco}, {Casamiquela}, {Castellani}, {Castro-Ginard}, {Charlot},
  {Chemin}, {Chiavassa}, {Cocozza}, {Costigan}, {Cowell}, {Crifo}, {Crosta},
  {Crowley}, {Cuypers}, {Dafonte}, {Damerdji}, {Dapergolas}, {David}, {David},
  {de Laverny}, {De Luise}, {De March}, {de Martino}, {de Souza}, {de Torres},
  {Debosscher}, {del Pozo}, {Delbo}, {Delgado}, {Delgado}, {Di Matteo},
  {Diakite}, {Diener}, {Distefano}, {Dolding}, {Drazinos}, {Dur{\'a}n},
  {Edvardsson}, {Enke}, {Eriksson}, {Esquej}, {Eynard Bontemps}, {Fabre},
  {Fabrizio}, {Faigler}, {Falc{\~a}o}, {Farr{\`a}s Casas}, {Federici},
  {Fedorets}, {Fernique}, {Figueras}, {Filippi}, {Findeisen}, {Fonti},
  {Fraile}, {Fraser}, {Fr{\'e}zouls}, {Gai}, {Galleti}, {Garabato},
  {Garc{\'\i}a-Sedano}, {Garofalo}, {Garralda}, {Gavel}, {Gavras}, {Gerssen},
  {Geyer}, {Giacobbe}, {Gilmore}, {Girona}, {Giuffrida}, {Glass}, {Gomes},
  {Granvik}, {Gueguen}, {Guerrier}, {Guiraud}, {Guti{\'e}rrez-S{\'a}nchez},
  {Haigron}, {Hatzidimitriou}, {Hauser}, {Haywood}, {Heiter}, {Helmi}, {Heu},
  {Hilger}, {Hobbs}, {Hofmann}, {Holland}, {Huckle}, {Hypki}, {Icardi},
  {Jan{\ss}en}, {Jevardat de Fombelle}, {Jonker}, {Juh{\'a}sz}, {Julbe},
  {Karampelas}, {Kewley}, {Klar}, {Kochoska}, {Kohley}, {Kolenberg},
  {Kontizas}, {Kontizas}, {Koposov}, {Kordopatis}, {Kostrzewa-Rutkowska},
  {Koubsky}, {Lambert}, {Lanza}, {Lasne}, {Lavigne}, {Le Fustec}, {Le
  Poncin-Lafitte}, {Lebreton}, {Leccia}, {Leclerc}, {Lecoeur-Taibi},
  {Lenhardt}, {Leroux}, {Liao}, {Licata}, {Lindstr{\o}m}, {Lister}, {Livanou},
  {Lobel}, {L{\'o}pez}, {Managau}, {Mann}, {Mantelet}, {Marchal}, {Marchant},
  {Marconi}, {Marinoni}, {Marschalk{\'o}}, {Marshall}, {Martino}, {Marton},
  {Mary}, {Massari}, {Matijevi{\v{c}}}, {Mazeh}, {McMillan}, {Messina},
  {Michalik}, {Millar}, {Molina}, {Molinaro}, {Moln{\'a}r}, {Montegriffo},
  {Mor}, {Morbidelli}, {Morel}, {Morris}, {Mulone}, {Muraveva}, {Musella},
  {Nelemans}, {Nicastro}, {Noval}, {O'Mullane}, {Ord{\'e}novic},
  {Ord{\'o}{\~n}ez-Blanco}, {Osborne}, {Pagani}, {Pagano}, {Pailler},
  {Palacin}, {Palaversa}, {Panahi}, {Pawlak}, {Piersimoni}, {Pineau}, {Plachy},
  {Plum}, {Poggio}, {Poujoulet}, {Pr{\v{s}}a}, {Pulone}, {Racero}, {Ragaini},
  {Rambaux}, {Ramos-Lerate}, {Regibo}, {Reyl{\'e}}, {Riclet}, {Ripepi}, {Riva},
  {Rivard}, {Rixon}, {Roegiers}, {Roelens}, {Romero-G{\'o}mez}, {Rowell},
  {Royer}, {Ruiz-Dern}, {Sadowski}, {Sagrist{\`a} Sell{\'e}s}, {Sahlmann},
  {Salgado}, {Salguero}, {Sanna}, {Santana-Ros}, {Sarasso}, {Savietto},
  {Schultheis}, {Sciacca}, {Segol}, {Segovia}, {S{\'e}gransan}, {Shih},
  {Siltala}, {Silva}, {Smart}, {Smith}, {Solano}, {Solitro}, {Sordo}, {Soria
  Nieto}, {Souchay}, {Spagna}, {Spoto}, {Stampa}, {Steele},
  {Steidelm{\"u}ller}, {Stephenson}, {Stoev}, {Suess}, {Surdej}, {Szabados},
  {Szegedi-Elek}, {Tapiador}, {Taris}, {Tauran}, {Taylor}, {Teixeira},
  {Terrett}, {Teyssand ier}, {Thuillot}, {Titarenko}, {Torra Clotet}, {Turon},
  {Ulla}, {Utrilla}, {Uzzi}, {Vaillant}, {Valentini}, {Valette}, {van Elteren},
  {Van Hemelryck}, {van Leeuwen}, {Vaschetto}, {Vecchiato}, {Veljanoski},
  {Viala}, {Vicente}, {Vogt}, {von Essen}, {Voss}, {Votruba}, {Voutsinas},
  {Walmsley}, {Weiler}, {Wertz}, {Wevers}, {Wyrzykowski}, {Yoldas},
  {{\v{Z}}erjal}, {Ziaeepour}, {Zorec}, {Zschocke}, {Zucker}, {Zurbach}, \&
  {Zwitter}}]{gaia2018}
{Gaia Collaboration}, {Brown}, A.~G.~A., {Vallenari}, A., {et~al.} 2018, \aap,
  616, A1

\bibitem[{{Gaia Collaboration} {et~al.}(2021){Gaia Collaboration}, {Brown},
  {Vallenari}, {Prusti}, {de Bruijne}, {Babusiaux}, {Biermann}, {Creevey},
  {Evans}, {Eyer}, {Hutton}, {Jansen}, {Jordi}, {Klioner}, {Lammers},
  {Lindegren}, {Luri}, {Mignard}, {Panem}, {Pourbaix}, {Randich}, {Sartoretti},
  {Soubiran}, {Walton}, {Arenou}, {Bailer-Jones}, {Bastian}, {Cropper},
  {Drimmel}, {Katz}, {Lattanzi}, {van Leeuwen}, {Bakker}, {Cacciari},
  {Casta{\~n}eda}, {De Angeli}, {Ducourant}, {Fabricius}, {Fouesneau},
  {Fr{\'e}mat}, {Guerra}, {Guerrier}, {Guiraud}, {Jean-Antoine Piccolo},
  {Masana}, {Messineo}, {Mowlavi}, {Nicolas}, {Nienartowicz}, {Pailler},
  {Panuzzo}, {Riclet}, {Roux}, {Seabroke}, {Sordo}, {Tanga}, {Th{\'e}venin},
  {Gracia-Abril}, {Portell}, {Teyssier}, {Altmann}, {Andrae}, {Bellas-Velidis},
  {Benson}, {Berthier}, {Blomme}, {Brugaletta}, {Burgess}, {Busso}, {Carry},
  {Cellino}, {Cheek}, {Clementini}, {Damerdji}, {Davidson}, {Delchambre},
  {Dell'Oro}, {Fern{\'a}ndez-Hern{\'a}ndez}, {Galluccio}, {Garc{\'\i}a-Lario},
  {Garcia-Reinaldos}, {Gonz{\'a}lez-N{\'u}{\~n}ez}, {Gosset}, {Haigron},
  {Halbwachs}, {Hambly}, {Harrison}, {Hatzidimitriou}, {Heiter},
  {Hern{\'a}ndez}, {Hestroffer}, {Hodgkin}, {Holl}, {Jan{\ss}en}, {Jevardat de
  Fombelle}, {Jordan}, {Krone-Martins}, {Lanzafame}, {L{\"o}ffler}, {Lorca},
  {Manteiga}, {Marchal}, {Marrese}, {Moitinho}, {Mora}, {Muinonen}, {Osborne},
  {Pancino}, {Pauwels}, {Petit}, {Recio-Blanco}, {Richards}, {Riello},
  {Rimoldini}, {Robin}, {Roegiers}, {Rybizki}, {Sarro}, {Siopis}, {Smith},
  {Sozzetti}, {Ulla}, {Utrilla}, {van Leeuwen}, {van Reeven}, {Abbas}, {Abreu
  Aramburu}, {Accart}, {Aerts}, {Aguado}, {Ajaj}, {Altavilla}, {{\'A}lvarez},
  {{\'A}lvarez Cid-Fuentes}, {Alves}, {Anderson}, {Anglada Varela}, {Antoja},
  {Audard}, {Baines}, {Baker}, {Balaguer-N{\'u}{\~n}ez}, {Balbinot}, {Balog},
  {Barache}, {Barbato}, {Barros}, {Barstow}, {Bartolom{\'e}}, {Bassilana},
  {Bauchet}, {Baudesson-Stella}, {Becciani}, {Bellazzini}, {Bernet}, {Bertone},
  {Bianchi}, {Blanco-Cuaresma}, {Boch}, {Bombrun}, {Bossini}, {Bouquillon},
  {Bragaglia}, {Bramante}, {Breedt}, {Bressan}, {Brouillet}, {Bucciarelli},
  {Burlacu}, {Busonero}, {Butkevich}, {Buzzi}, {Caffau}, {Cancelliere},
  {C{\'a}novas}, {Cantat-Gaudin}, {Carballo}, {Carlucci}, {Carnerero},
  {Carrasco}, {Casamiquela}, {Castellani}, {Castro-Ginard}, {Castro Sampol},
  {Chaoul}, {Charlot}, {Chemin}, {Chiavassa}, {Cioni}, {Comoretto}, {Cooper},
  {Cornez}, {Cowell}, {Crifo}, {Crosta}, {Crowley}, {Dafonte}, {Dapergolas},
  {David}, {David}, {de Laverny}, {De Luise}, {De March}, {De Ridder}, {de
  Souza}, {de Teodoro}, {de Torres}, {del Peloso}, {del Pozo}, {Delbo},
  {Delgado}, {Delgado}, {Delisle}, {Di Matteo}, {Diakite}, {Diener},
  {Distefano}, {Dolding}, {Eappachen}, {Edvardsson}, {Enke}, {Esquej}, {Fabre},
  {Fabrizio}, {Faigler}, {Fedorets}, {Fernique}, {Fienga}, {Figueras},
  {Fouron}, {Fragkoudi}, {Fraile}, {Franke}, {Gai}, {Garabato},
  {Garcia-Gutierrez}, {Garc{\'\i}a-Torres}, {Garofalo}, {Gavras}, {Gerlach},
  {Geyer}, {Giacobbe}, {Gilmore}, {Girona}, {Giuffrida}, {Gomel}, {Gomez},
  {Gonzalez-Santamaria}, {Gonz{\'a}lez-Vidal}, {Granvik},
  {Guti{\'e}rrez-S{\'a}nchez}, {Guy}, {Hauser}, {Haywood}, {Helmi}, {Hidalgo},
  {Hilger}, {H{\l}adczuk}, {Hobbs}, {Holland}, {Huckle}, {Jasniewicz},
  {Jonker}, {Juaristi Campillo}, {Julbe}, {Karbevska}, {Kervella}, {Khanna},
  {Kochoska}, {Kontizas}, {Kordopatis}, {Korn}, {Kostrzewa-Rutkowska},
  {Kruszy{\'n}ska}, {Lambert}, {Lanza}, {Lasne}, {Le Campion}, {Le Fustec},
  {Lebreton}, {Lebzelter}, {Leccia}, {Leclerc}, {Lecoeur-Taibi}, {Liao},
  {Licata}, {Lindstr{\o}m}, {Lister}, {Livanou}, {Lobel}, {Madrero Pardo},
  {Managau}, {Mann}, {Marchant}, {Marconi}, {Marcos Santos}, {Marinoni},
  {Marocco}, {Marshall}, {Martin Polo}, {Mart{\'\i}n-Fleitas}, {Masip},
  {Massari}, {Mastrobuono-Battisti}, {Mazeh}, {McMillan}, {Messina},
  {Michalik}, {Millar}, {Mints}, {Molina}, {Molinaro}, {Moln{\'a}r},
  {Montegriffo}, {Mor}, {Morbidelli}, {Morel}, {Morris}, {Mulone}, {Munoz},
  {Muraveva}, {Murphy}, {Musella}, {Noval}, {Ord{\'e}novic}, {Orr{\`u}},
  {Osinde}, {Pagani}, {Pagano}, {Palaversa}, {Palicio}, {Panahi}, {Pawlak},
  {Pe{\~n}alosa Esteller}, {Penttil{\"a}}, {Piersimoni}, {Pineau}, {Plachy},
  {Plum}, {Poggio}, {Poretti}, {Poujoulet}, {Pr{\v{s}}a}, {Pulone}, {Racero},
  {Ragaini}, {Rainer}, {Raiteri}, {Rambaux}, {Ramos}, {Ramos-Lerate}, {Re
  Fiorentin}, {Regibo}, {Reyl{\'e}}, {Ripepi}, {Riva}, {Rixon}, {Robichon},
  {Robin}, {Roelens}, {Rohrbasser}, {Romero-G{\'o}mez}, {Rowell}, {Royer},
  {Rybicki}, {Sadowski}, {Sagrist{\`a} Sell{\'e}s}, {Sahlmann}, {Salgado},
  {Salguero}, {Samaras}, {Sanchez Gimenez}, {Sanna}, {Santove{\~n}a},
  {Sarasso}, {Schultheis}, {Sciacca}, {Segol}, {Segovia}, {S{\'e}gransan},
  {Semeux}, {Shahaf}, {Siddiqui}, {Siebert}, {Siltala}, {Slezak}, {Smart},
  {Solano}, {Solitro}, {Souami}, {Souchay}, {Spagna}, {Spoto}, {Steele},
  {Steidelm{\"u}ller}, {Stephenson}, {S{\"u}veges}, {Szabados}, {Szegedi-Elek},
  {Taris}, {Tauran}, {Taylor}, {Teixeira}, {Thuillot}, {Tonello}, {Torra},
  {Torra}, {Turon}, {Unger}, {Vaillant}, {van Dillen}, {Vanel}, {Vecchiato},
  {Viala}, {Vicente}, {Voutsinas}, {Weiler}, {Wevers}, {Wyrzykowski}, {Yoldas},
  {Yvard}, {Zhao}, {Zorec}, {Zucker}, {Zurbach}, \& {Zwitter}}]{gaia2020}
{Gaia Collaboration}, {Brown}, A.~G.~A., {Vallenari}, A., {et~al.} 2021, \aap,
  649, A1

\bibitem[{{Gaia Collaboration} {et~al.}(2016){Gaia Collaboration}, {Prusti},
  {de Bruijne}, {Brown}, {Vallenari}, {Babusiaux}, {Bailer-Jones}, {Bastian},
  {Biermann}, {Evans}, {Eyer}, {Jansen}, {Jordi}, {Klioner}, {Lammers},
  {Lindegren}, {Luri}, {Mignard}, {Milligan}, {Panem}, {Poinsignon},
  {Pourbaix}, {Randich}, {Sarri}, {Sartoretti}, {Siddiqui}, {Soubiran},
  {Valette}, {van Leeuwen}, {Walton}, {Aerts}, {Arenou}, {Cropper}, {Drimmel},
  {H{\o}g}, {Katz}, {Lattanzi}, {O'Mullane}, {Grebel}, {Holland}, {Huc},
  {Passot}, {Bramante}, {Cacciari}, {Casta{\~n}eda}, {Chaoul}, {Cheek}, {De
  Angeli}, {Fabricius}, {Guerra}, {Hern{\'a}ndez}, {Jean-Antoine-Piccolo},
  {Masana}, {Messineo}, {Mowlavi}, {Nienartowicz}, {Ord{\'o}{\~n}ez-Blanco},
  {Panuzzo}, {Portell}, {Richards}, {Riello}, {Seabroke}, {Tanga},
  {Th{\'e}venin}, {Torra}, {Els}, {Gracia-Abril}, {Comoretto},
  {Garcia-Reinaldos}, {Lock}, {Mercier}, {Altmann}, {Andrae}, {Astraatmadja},
  {Bellas-Velidis}, {Benson}, {Berthier}, {Blomme}, {Busso}, {Carry},
  {Cellino}, {Clementini}, {Cowell}, {Creevey}, {Cuypers}, {Davidson}, {De
  Ridder}, {de Torres}, {Delchambre}, {Dell'Oro}, {Ducourant}, {Fr{\'e}mat},
  {Garc{\'\i}a-Torres}, {Gosset}, {Halbwachs}, {Hambly}, {Harrison}, {Hauser},
  {Hestroffer}, {Hodgkin}, {Huckle}, {Hutton}, {Jasniewicz}, {Jordan},
  {Kontizas}, {Korn}, {Lanzafame}, {Manteiga}, {Moitinho}, {Muinonen},
  {Osinde}, {Pancino}, {Pauwels}, {Petit}, {Recio-Blanco}, {Robin}, {Sarro},
  {Siopis}, {Smith}, {Smith}, {Sozzetti}, {Thuillot}, {van Reeven}, {Viala},
  {Abbas}, {Abreu Aramburu}, {Accart}, {Aguado}, {Allan}, {Allasia},
  {Altavilla}, {{\'A}lvarez}, {Alves}, {Anderson}, {Andrei}, {Anglada Varela},
  {Antiche}, {Antoja}, {Ant{\'o}n}, {Arcay}, {Atzei}, {Ayache}, {Bach},
  {Baker}, {Balaguer-N{\'u}{\~n}ez}, {Barache}, {Barata}, {Barbier}, {Barblan},
  {Baroni}, {Barrado y Navascu{\'e}s}, {Barros}, {Barstow}, {Becciani},
  {Bellazzini}, {Bellei}, {Bello Garc{\'\i}a}, {Belokurov}, {Bendjoya},
  {Berihuete}, {Bianchi}, {Bienaym{\'e}}, {Billebaud}, {Blagorodnova},
  {Blanco-Cuaresma}, {Boch}, {Bombrun}, {Borrachero}, {Bouquillon}, {Bourda},
  {Bouy}, {Bragaglia}, {Breddels}, {Brouillet}, {Br{\"u}semeister},
  {Bucciarelli}, {Budnik}, {Burgess}, {Burgon}, {Burlacu}, {Busonero}, {Buzzi},
  {Caffau}, {Cambras}, {Campbell}, {Cancelliere}, {Cantat-Gaudin}, {Carlucci},
  {Carrasco}, {Castellani}, {Charlot}, {Charnas}, {Charvet}, {Chassat},
  {Chiavassa}, {Clotet}, {Cocozza}, {Collins}, {Collins}, {Costigan}, {Crifo},
  {Cross}, {Crosta}, {Crowley}, {Dafonte}, {Damerdji}, {Dapergolas}, {David},
  {David}, {De Cat}, {de Felice}, {de Laverny}, {De Luise}, {De March}, {de
  Martino}, {de Souza}, {Debosscher}, {del Pozo}, {Delbo}, {Delgado},
  {Delgado}, {di Marco}, {Di Matteo}, {Diakite}, {Distefano}, {Dolding}, {Dos
  Anjos}, {Drazinos}, {Dur{\'a}n}, {Dzigan}, {Ecale}, {Edvardsson}, {Enke},
  {Erdmann}, {Escolar}, {Espina}, {Evans}, {Eynard Bontemps}, {Fabre},
  {Fabrizio}, {Faigler}, {Falc{\~a}o}, {Farr{\`a}s Casas}, {Faye}, {Federici},
  {Fedorets}, {Fern{\'a}ndez-Hern{\'a}ndez}, {Fernique}, {Fienga}, {Figueras},
  {Filippi}, {Findeisen}, {Fonti}, {Fouesneau}, {Fraile}, {Fraser}, {Fuchs},
  {Furnell}, {Gai}, {Galleti}, {Galluccio}, {Garabato}, {Garc{\'\i}a-Sedano},
  {Gar{\'e}}, {Garofalo}, {Garralda}, {Gavras}, {Gerssen}, {Geyer}, {Gilmore},
  {Girona}, {Giuffrida}, {Gomes}, {Gonz{\'a}lez-Marcos},
  {Gonz{\'a}lez-N{\'u}{\~n}ez}, {Gonz{\'a}lez-Vidal}, {Granvik}, {Guerrier},
  {Guillout}, {Guiraud}, {G{\'u}rpide}, {Guti{\'e}rrez-S{\'a}nchez}, {Guy},
  {Haigron}, {Hatzidimitriou}, {Haywood}, {Heiter}, {Helmi}, {Hobbs},
  {Hofmann}, {Holl}, {Holland }, {Hunt}, {Hypki}, {Icardi}, {Irwin}, {Jevardat
  de Fombelle}, {Jofr{\'e}}, {Jonker}, {Jorissen}, {Julbe}, {Karampelas},
  {Kochoska}, {Kohley}, {Kolenberg}, {Kontizas}, {Koposov}, {Kordopatis},
  {Koubsky}, {Kowalczyk}, {Krone-Martins}, {Kudryashova}, {Kull}, {Bachchan},
  {Lacoste-Seris}, {Lanza}, {Lavigne}, {Le Poncin-Lafitte}, {Lebreton},
  {Lebzelter}, {Leccia}, {Leclerc}, {Lecoeur-Taibi}, {Lemaitre}, {Lenhardt},
  {Leroux}, {Liao}, {Licata}, {Lindstr{\o}m}, {Lister}, {Livanou}, {Lobel},
  {L{\"o}ffler}, {L{\'o}pez}, {Lopez-Lozano}, {Lorenz}, {Loureiro},
  {MacDonald}, {Magalh{\~a}es Fernandes}, {Managau}, {Mann}, {Mantelet},
  {Marchal}, {Marchant}, {Marconi}, {Marie}, {Marinoni}, {Marrese},
  {Marschalk{\'o}}, {Marshall}, {Mart{\'\i}n-Fleitas}, {Martino}, {Mary},
  {Matijevi{\v{c}}}, {Mazeh}, {McMillan}, {Messina}, {Mestre}, {Michalik},
  {Millar}, {Miranda}, {Molina}, {Molinaro}, {Molinaro}, {Moln{\'a}r},
  {Moniez}, {Montegriffo}, {Monteiro}, {Mor}, {Mora}, {Morbidelli}, {Morel},
  {Morgenthaler}, {Morley}, {Morris}, {Mulone}, {Muraveva}, {Musella},
  {Narbonne}, {Nelemans}, {Nicastro}, {Noval}, {Ord{\'e}novic},
  {Ordieres-Mer{\'e}}, {Osborne}, {Pagani}, {Pagano}, {Pailler}, {Palacin},
  {Palaversa}, {Parsons}, {Paulsen}, {Pecoraro}, {Pedrosa}, {Pentik{\"a}inen},
  {Pereira}, {Pichon}, {Piersimoni}, {Pineau}, {Plachy}, {Plum}, {Poujoulet},
  {Pr{\v{s}}a}, {Pulone}, {Ragaini}, {Rago}, {Rambaux}, {Ramos-Lerate},
  {Ranalli}, {Rauw}, {Read}, {Regibo}, {Renk}, {Reyl{\'e}}, {Ribeiro},
  {Rimoldini}, {Ripepi}, {Riva}, {Rixon}, {Roelens}, {Romero-G{\'o}mez},
  {Rowell}, {Royer}, {Rudolph}, {Ruiz-Dern}, {Sadowski}, {Sagrist{\`a}
  Sell{\'e}s}, {Sahlmann}, {Salgado}, {Salguero}, {Sarasso}, {Savietto},
  {Schnorhk}, {Schultheis}, {Sciacca}, {Segol}, {Segovia}, {Segransan},
  {Serpell}, {Shih}, {Smareglia}, {Smart}, {Smith}, {Solano}, {Solitro},
  {Sordo}, {Soria Nieto}, {Souchay}, {Spagna}, {Spoto}, {Stampa}, {Steele},
  {Steidelm{\"u}ller}, {Stephenson}, {Stoev}, {Suess}, {S{\"u}veges}, {Surdej},
  {Szabados}, {Szegedi-Elek}, {Tapiador}, {Taris}, {Tauran}, {Taylor},
  {Teixeira}, {Terrett}, {Tingley}, {Trager}, {Turon}, {Ulla}, {Utrilla},
  {Valentini}, {van Elteren}, {Van Hemelryck}, {van Leeuwen}, {Varadi},
  {Vecchiato}, {Veljanoski}, {Via}, {Vicente}, {Vogt}, {Voss}, {Votruba},
  {Voutsinas}, {Walmsley}, {Weiler}, {Weingrill}, {Werner}, {Wevers},
  {Whitehead}, {Wyrzykowski}, {Yoldas}, {{\v{Z}}erjal}, {Zucker}, {Zurbach},
  {Zwitter}, {Alecu}, {Allen}, {Allende Prieto}, {Amorim},
  {Anglada-Escud{\'e}}, {Arsenijevic}, {Azaz}, {Balm}, {Beck}, {Bernstein},
  {Bigot}, {Bijaoui}, {Blasco}, {Bonfigli}, {Bono}, {Boudreault}, {Bressan},
  {Brown}, {Brunet}, {Bunclark}, {Buonanno}, {Butkevich}, {Carret}, {Carrion},
  {Chemin}, {Ch{\'e}reau}, {Corcione}, {Darmigny}, {de Boer}, {de Teodoro}, {de
  Zeeuw}, {Delle Luche}, {Domingues}, {Dubath}, {Fodor}, {Fr{\'e}zouls},
  {Fries}, {Fustes}, {Fyfe}, {Gallardo}, {Gallegos}, {Gardiol}, {Gebran},
  {Gomboc}, {G{\'o}mez}, {Grux}, {Gueguen}, {Heyrovsky}, {Hoar}, {Iannicola},
  {Isasi Parache}, {Janotto}, {Joliet}, {Jonckheere}, {Keil}, {Kim},
  {Klagyivik}, {Klar}, {Knude}, {Kochukhov}, {Kolka}, {Kos}, {Kutka}, {Lainey},
  {LeBouquin}, {Liu}, {Loreggia}, {Makarov}, {Marseille}, {Martayan},
  {Martinez-Rubi}, {Massart}, {Meynadier}, {Mignot}, {Munari}, {Nguyen},
  {Nordlander}, {Ocvirk}, {O'Flaherty}, {Olias Sanz}, {Ortiz}, {Osorio},
  {Oszkiewicz}, {Ouzounis}, {Palmer}, {Park}, {Pasquato}, {Peltzer}, {Peralta},
  {P{\'e}turaud}, {Pieniluoma}, {Pigozzi}, {Poels}, {Prat}, {Prod'homme},
  {Raison}, {Rebordao}, {Risquez}, {Rocca-Volmerange}, {Rosen}, {Ruiz-Fuertes},
  {Russo}, {Sembay}, {Serraller Vizcaino}, {Short}, {Siebert}, {Silva},
  {Sinachopoulos}, {Slezak}, {Soffel}, {Sosnowska}, {Strai{\v{z}}ys}, {ter
  Linden}, {Terrell}, {Theil}, {Tiede}, {Troisi}, {Tsalmantza}, {Tur},
  {Vaccari}, {Vachier}, {Valles}, {Van Hamme}, {Veltz}, {Virtanen}, {Wallut},
  {Wichmann}, {Wilkinson}, {Ziaeepour}, \& {Zschocke}}]{gaia2016}
{Gaia Collaboration}, {Prusti}, T., {de Bruijne}, J.~H.~J., {et~al.} 2016,
  \aap, 595, A1

\bibitem[{{Goldman} {et~al.}(2018){Goldman}, {R{\"o}ser}, {Schilbach},
  {Mo{\'o}r}, \& {Henning}}]{goldman2018}
{Goldman}, B., {R{\"o}ser}, S., {Schilbach}, E., {Mo{\'o}r}, A.~C., \&
  {Henning}, T. 2018, \apj, 868, 32

\bibitem[{{Haffert} {et~al.}(2019){Haffert}, {Bohn}, {de Boer}, {Snellen},
  {Brinchmann}, {Girard}, {Keller}, \& {Bacon}}]{haffert2019}
{Haffert}, S.~Y., {Bohn}, A.~J., {de Boer}, J., {et~al.} 2019, Nature
  Astronomy, 3, 749

\bibitem[{{Henden} {et~al.}(2012){Henden}, {Levine}, {Terrell}, {Smith}, \&
  {Welch}}]{henden2012}
{Henden}, A.~A., {Levine}, S.~E., {Terrell}, D., {Smith}, T.~C., \& {Welch}, D.
  2012, Journal of the American Association of Variable Star Observers
  (JAAVSO), 40, 430

\bibitem[{{H{\o}g} {et~al.}(2000){H{\o}g}, {Fabricius}, {Makarov}, {Urban},
  {Corbin}, {Wycoff}, {Bastian}, {Schwekendiek}, \& {Wicenec}}]{hog2000}
{H{\o}g}, E., {Fabricius}, C., {Makarov}, V.~V., {et~al.} 2000, \aap, 355, L27

\bibitem[{{Hunter}(2007)}]{Matplotlib}
{Hunter}, J.~D. 2007, Computing in Science and Engineering, 9, 90

\bibitem[{{Janson} {et~al.}(2012){Janson}, {Hormuth}, {Bergfors}, {Brand ner},
  {Hippler}, {Daemgen}, {Kudryavtseva}, {Schmalzl}, {Schnupp}, \&
  {Henning}}]{janson2012}
{Janson}, M., {Hormuth}, F., {Bergfors}, C., {et~al.} 2012, \apj, 754, 44

\bibitem[{{Keppler} {et~al.}(2018){Keppler}, {Benisty}, {M{\"u}ller},
  {Henning}, {van Boekel}, {Cantalloube}, {Ginski}, {van Holstein}, {Maire},
  {Pohl}, {Samland}, {Avenhaus}, {Baudino}, {Boccaletti}, {de Boer},
  {Bonnefoy}, {Chauvin}, {Desidera}, {Langlois}, {Lazzoni}, {Marleau},
  {Mordasini}, {Pawellek}, {Stolker}, {Vigan}, {Zurlo}, {Birnstiel},
  {Brandner}, {Feldt}, {Flock}, {Girard}, {Gratton}, {Hagelberg}, {Isella},
  {Janson}, {Juhasz}, {Kemmer}, {Kral}, {Lagrange}, {Launhardt}, {Matter},
  {M{\'e}nard}, {Milli}, {Molli{\`e}re}, {Olofsson}, {P{\'e}rez}, {Pinilla},
  {Pinte}, {Quanz}, {Schmidt}, {Udry}, {Wahhaj}, {Williams}, {Buenzli},
  {Cudel}, {Dominik}, {Galicher}, {Kasper}, {Lannier}, {Mesa}, {Mouillet},
  {Peretti}, {Perrot}, {Salter}, {Sissa}, {Wildi}, {Abe}, {Antichi},
  {Augereau}, {Baruffolo}, {Baudoz}, {Bazzon}, {Beuzit}, {Blanchard}, {Brems},
  {Buey}, {De Caprio}, {Carbillet}, {Carle}, {Cascone}, {Cheetham}, {Claudi},
  {Costille}, {Delboulb{\'e}}, {Dohlen}, {Fantinel}, {Feautrier}, {Fusco},
  {Giro}, {Gluck}, {Gry}, {Hubin}, {Hugot}, {Jaquet}, {Le Mignant}, {Llored},
  {Madec}, {Magnard}, {Martinez}, {Maurel}, {Meyer}, {M{\"o}ller-Nilsson},
  {Moulin}, {Mugnier}, {Orign{\'e}}, {Pavlov}, {Perret}, {Petit}, {Pragt},
  {Puget}, {Rabou}, {Ramos}, {Rigal}, {Rochat}, {Roelfsema}, {Rousset}, {Roux},
  {Salasnich}, {Sauvage}, {Sevin}, {Soenke}, {Stadler}, {Suarez}, {Turatto}, \&
  {Weber}}]{keppler2018}
{Keppler}, M., {Benisty}, M., {M{\"u}ller}, A., {et~al.} 2018, \aap, 617, A44

\bibitem[{{Konopacky} {et~al.}(2016){Konopacky}, {Rameau}, {Duch{\^e}ne},
  {Filippazzo}, {Giorla Godfrey}, {Marois}, {Nielsen}, {Pueyo}, {Rafikov},
  {Rice}, {Wang}, {Ammons}, {Bailey}, {Barman}, {Bulger}, {Bruzzone},
  {Chilcote}, {Cotten}, {Dawson}, {De Rosa}, {Doyon}, {Esposito}, {Fitzgerald},
  {Follette}, {Goodsell}, {Graham}, {Greenbaum}, {Hibon}, {Hung}, {Ingraham},
  {Kalas}, {Lafreni{\`e}re}, {Larkin}, {Macintosh}, {Maire}, {Marchis},
  {Marley}, {Matthews}, {Metchev}, {Millar-Blanchaer}, {Oppenheimer}, {Palmer},
  {Patience}, {Perrin}, {Poyneer}, {Rajan}, {Rantakyr{\"o}}, {Savransky},
  {Schneider}, {Sivaramakrishnan}, {Song}, {Soummer}, {Thomas}, {Wallace},
  {Ward-Duong}, {Wiktorowicz}, \& {Wolff}}]{konopacky2016}
{Konopacky}, Q.~M., {Rameau}, J., {Duch{\^e}ne}, G., {et~al.} 2016, \apjl, 829,
  L4

\bibitem[{{Kouwenhoven} {et~al.}(2010){Kouwenhoven}, {Goodwin}, {Parker},
  {Davies}, {Malmberg}, \& {Kroupa}}]{kouwenhoven2010}
{Kouwenhoven}, M.~B.~N., {Goodwin}, S.~P., {Parker}, R.~J., {et~al.} 2010,
  \mnras, 404, 1835

\bibitem[{{Kratter} {et~al.}(2010){Kratter}, {Murray-Clay}, \&
  {Youdin}}]{kratter2010}
{Kratter}, K.~M., {Murray-Clay}, R.~A., \& {Youdin}, A.~N. 2010, \apj, 710,
  1375

\bibitem[{{Kraus} \& {Hillenbrand}(2008)}]{kraus2008}
{Kraus}, A.~L. \& {Hillenbrand}, L.~A. 2008, \apjl, 686, L111

\bibitem[{{Kroupa}(2001)}]{kroupa2001}
{Kroupa}, P. 2001, \mnras, 322, 231

\bibitem[{{Lagrange} {et~al.}(2010){Lagrange}, {Bonnefoy}, {Chauvin}, {Apai},
  {Ehrenreich}, {Boccaletti}, {Gratadour}, {Rouan}, {Mouillet}, {Lacour}, \&
  {Kasper}}]{lagrange2010}
{Lagrange}, A.~M., {Bonnefoy}, M., {Chauvin}, G., {et~al.} 2010, Science, 329,
  57

\bibitem[{{Lagrange} {et~al.}(2009){Lagrange}, {Gratadour}, {Chauvin}, {Fusco},
  {Ehrenreich}, {Mouillet}, {Rousset}, {Rouan}, {Allard}, {Gendron}, {Charton},
  {Mugnier}, {Rabou}, {Montri}, \& {Lacombe}}]{lagrange2009}
{Lagrange}, A.~M., {Gratadour}, D., {Chauvin}, G., {et~al.} 2009, \aap, 493,
  L21

\bibitem[{{Lallement} {et~al.}(2014){Lallement}, {Vergely}, {Valette},
  {Puspitarini}, {Eyer}, \& {Casagrande}}]{lallement2014}
{Lallement}, R., {Vergely}, J.~L., {Valette}, B., {et~al.} 2014, \aap, 561, A91

\bibitem[{{Lambrechts} \& {Johansen}(2012)}]{lambrechts2012}
{Lambrechts}, M. \& {Johansen}, A. 2012, \aap, 544, A32

\bibitem[{{Launhardt} {et~al.}(2020){Launhardt}, {Henning}, {Quirrenbach},
  {S{\'e}gransan}, {Avenhaus}, {van Boekel}, {Brems}, {Cheetham}, {Cugno},
  {Girard}, {Godoy}, {Kennedy}, {Maire}, {Metchev}, {M{\"u}ller}, {Musso
  Barcucci}, {Olofsson}, {Pepe}, {Quanz}, {Queloz}, {Reffert}, {Rickman},
  {Ruh}, \& {Samland}}]{launhardt2020}
{Launhardt}, R., {Henning}, T., {Quirrenbach}, A., {et~al.} 2020, \aap, 635,
  A162

\bibitem[{{Lenzen} {et~al.}(2003){Lenzen}, {Hartung}, {Brandner}, {Finger},
  {Hubin}, {Lacombe}, {Lagrange}, {Lehnert}, {Moorwood}, \&
  {Mouillet}}]{lenzen2003}
{Lenzen}, R., {Hartung}, M., {Brandner}, W., {et~al.} 2003, in Society of
  Photo-Optical Instrumentation Engineers (SPIE) Conference Series, Vol. 4841,
  Instrument Design and Performance for Optical/Infrared Ground-based
  Telescopes, ed. M.~{Iye} \& A.~F.~M. {Moorwood}, 944--952

\bibitem[{{Lindegren} {et~al.}(2021{\natexlab{a}}){Lindegren}, {Bastian},
  {Biermann}, {Bombrun}, {de Torres}, {Gerlach}, {Geyer}, {Hern{\'a}ndez},
  {Hilger}, {Hobbs}, {Klioner}, {Lammers}, {McMillan}, {Ramos-Lerate},
  {Steidelm{\"u}ller}, {Stephenson}, \& {van Leeuwen}}]{lindegren2020a}
{Lindegren}, L., {Bastian}, U., {Biermann}, M., {et~al.} 2021{\natexlab{a}},
  \aap, 649, A4

\bibitem[{{Lindegren} {et~al.}(2021{\natexlab{b}}){Lindegren}, {Klioner},
  {Hern{\'a}ndez}, {Bombrun}, {Ramos-Lerate}, {Steidelm{\"u}ller}, {Bastian},
  {Biermann}, {de Torres}, {Gerlach}, {Geyer}, {Hilger}, {Hobbs}, {Lammers},
  {McMillan}, {Stephenson}, {Casta{\~n}eda}, {Davidson}, {Fabricius},
  {Gracia-Abril}, {Portell}, {Rowell}, {Teyssier}, {Torra}, {Bartolom{\'e}},
  {Clotet}, {Garralda}, {Gonz{\'a}lez-Vidal}, {Torra}, {Abbas}, {Altmann},
  {Anglada Varela}, {Balaguer-N{\'u}{\~n}ez}, {Balog}, {Barache}, {Becciani},
  {Bernet}, {Bertone}, {Bianchi}, {Bouquillon}, {Brown}, {Bucciarelli},
  {Busonero}, {Butkevich}, {Buzzi}, {Cancelliere}, {Carlucci}, {Charlot},
  {Cioni}, {Crosta}, {Crowley}, {del Peloso}, {del Pozo}, {Drimmel}, {Esquej},
  {Fienga}, {Fraile}, {Gai}, {Garcia-Reinaldos}, {Guerra}, {Hambly}, {Hauser},
  {Jan{\ss}en}, {Jordan}, {Kostrzewa-Rutkowska}, {Lattanzi}, {Liao}, {Licata},
  {Lister}, {L{\"o}ffler}, {Marchant}, {Masip}, {Mignard}, {Mints}, {Molina},
  {Mora}, {Morbidelli}, {Murphy}, {Pagani}, {Panuzzo}, {Pe{\~n}alosa Esteller},
  {Poggio}, {Re Fiorentin}, {Riva}, {Sagrist{\`a} Sell{\'e}s}, {Sanchez
  Gimenez}, {Sarasso}, {Sciacca}, {Siddiqui}, {Smart}, {Souami}, {Spagna},
  {Steele}, {Taris}, {Utrilla}, {van Reeven}, \& {Vecchiato}}]{lindegren2020b}
{Lindegren}, L., {Klioner}, S.~A., {Hern{\'a}ndez}, J., {et~al.}
  2021{\natexlab{b}}, \aap, 649, A2

\bibitem[{{Macintosh} {et~al.}(2015){Macintosh}, {Graham}, {Barman}, {De Rosa},
  {Konopacky}, {Marley}, {Marois}, {Nielsen}, {Pueyo}, {Rajan}, {Rameau},
  {Saumon}, {Wang}, {Patience}, {Ammons}, {Arriaga}, {Artigau}, {Beckwith},
  {Brewster}, {Bruzzone}, {Bulger}, {Burningham}, {Burrows}, {Chen}, {Chiang},
  {Chilcote}, {Dawson}, {Dong}, {Doyon}, {Draper}, {Duch{\^e}ne}, {Esposito},
  {Fabrycky}, {Fitzgerald}, {Follette}, {Fortney}, {Gerard}, {Goodsell},
  {Greenbaum}, {Hibon}, {Hinkley}, {Cotten}, {Hung}, {Ingraham},
  {Johnson-Groh}, {Kalas}, {Lafreniere}, {Larkin}, {Lee}, {Line}, {Long},
  {Maire}, {Marchis}, {Matthews}, {Max}, {Metchev}, {Millar-Blanchaer},
  {Mittal}, {Morley}, {Morzinski}, {Murray-Clay}, {Oppenheimer}, {Palmer},
  {Patel}, {Perrin}, {Poyneer}, {Rafikov}, {Rantakyr{\"o}}, {Rice}, {Rojo},
  {Rudy}, {Ruffio}, {Ruiz}, {Sadakuni}, {Saddlemyer}, {Salama}, {Savransky},
  {Schneider}, {Sivaramakrishnan}, {Song}, {Soummer}, {Thomas}, {Vasisht},
  {Wallace}, {Ward- Duong}, {Wiktorowicz}, {Wolff}, \&
  {Zuckerman}}]{macintosh2015}
{Macintosh}, B., {Graham}, J.~R., {Barman}, T., {et~al.} 2015, Science, 350, 64

\bibitem[{{Macintosh} {et~al.}(2014){Macintosh}, {Graham}, {Ingraham},
  {Konopacky}, {Marois}, {Perrin}, {Poyneer}, {Bauman}, {Barman}, {Burrows},
  {Cardwell}, {Chilcote}, {De Rosa}, {Dillon}, {Doyon}, {Dunn}, {Erikson},
  {Fitzgerald}, {Gavel}, {Goodsell}, {Hartung}, {Hibon}, {Kalas}, {Larkin},
  {Maire}, {Marchis}, {Marley}, {McBride}, {Millar-Blanchaer}, {Morzinski},
  {Norton}, {Oppenheimer}, {Palmer}, {Patience}, {Pueyo}, {Rantakyro},
  {Sadakuni}, {Saddlemyer}, {Savransky}, {Serio}, {Soummer},
  {Sivaramakrishnan}, {Song}, {Thomas}, {Wallace}, {Wiktorowicz}, \&
  {Wolff}}]{Macintosh2014}
{Macintosh}, B., {Graham}, J.~R., {Ingraham}, P., {et~al.} 2014, Proceedings of
  the National Academy of Science, 111, 12661

\bibitem[{{Maire} {et~al.}(2016){Maire}, {Langlois}, {Dohlen}, {Lagrange},
  {Gratton}, {Chauvin}, {Desidera}, {Girard}, {Milli}, {Vigan}, {Zins},
  {Delorme}, {Beuzit}, {Claudi}, {Feldt}, {Mouillet}, {Puget}, {Turatto}, \&
  {Wildi}}]{maire2016}
{Maire}, A.-L., {Langlois}, M., {Dohlen}, K., {et~al.} 2016, in Ground-based
  and Airborne Instrumentation for Astronomy VI, Vol. 9908, 990834

\bibitem[{{Malmberg} {et~al.}(2011){Malmberg}, {Davies}, \&
  {Heggie}}]{malmberg2011}
{Malmberg}, D., {Davies}, M.~B., \& {Heggie}, D.~C. 2011, \mnras, 411, 859

\bibitem[{{Marois} {et~al.}(2006){Marois}, {Lafreni{\`e}re}, {Doyon},
  {Macintosh}, \& {Nadeau}}]{marois2006}
{Marois}, C., {Lafreni{\`e}re}, D., {Doyon}, R., {Macintosh}, B., \& {Nadeau},
  D. 2006, \apj, 641, 556

\bibitem[{{Marois} {et~al.}(2008){Marois}, {Macintosh}, {Barman}, {Zuckerman},
  {Song}, {Patience}, {Lafreni{\`e}re}, \& {Doyon}}]{marois2008}
{Marois}, C., {Macintosh}, B., {Barman}, T., {et~al.} 2008, Science, 322, 1348

\bibitem[{{Marois} {et~al.}(2010){Marois}, {Zuckerman}, {Konopacky},
  {Macintosh}, \& {Barman}}]{marois2010}
{Marois}, C., {Zuckerman}, B., {Konopacky}, Q.~M., {Macintosh}, B., \&
  {Barman}, T. 2010, \nat, 468, 1080

\bibitem[{{Mawet} {et~al.}(2015){Mawet}, {David}, {Bottom}, {Hinkley},
  {Stapelfeldt}, {Padgett}, {Mennesson}, {Serabyn}, {Morales}, \&
  {Kuhn}}]{mawet2015}
{Mawet}, D., {David}, T., {Bottom}, M., {et~al.} 2015, \apj, 811, 103

\bibitem[{{Mayor} \& {Mazeh}(1987)}]{mayor1987}
{Mayor}, M. \& {Mazeh}, T. 1987, \aap, 171, 157

\bibitem[{{Milli} {et~al.}(2017){Milli}, {Hibon}, {Christiaens}, {Choquet},
  {Bonnefoy}, {Kennedy}, {Wyatt}, {Absil}, {G{\'o}mez Gonz{\'a}lez}, {del
  Burgo}, {Matr{\`a}}, {Augereau}, {Boccaletti}, {Delacroix}, {Ertel}, {Dent},
  {Forsberg}, {Fusco}, {Girard}, {Habraken}, {Huby}, {Karlsson}, {Lagrange},
  {Mawet}, {Mouillet}, {Perrin}, {Pinte}, {Pueyo}, {Reyes}, {Soummer},
  {Surdej}, {Tarricq}, \& {Wahhaj}}]{milli2017}
{Milli}, J., {Hibon}, P., {Christiaens}, V., {et~al.} 2017, \aap, 597, L2

\bibitem[{{Mo{\'o}r} {et~al.}(2011){Mo{\'o}r}, {Pascucci}, {K{\'o}sp{\'a}l},
  {{\'A}brah{\'a}m}, {Csengeri}, {Kiss}, {Apai}, {Grady}, {Henning}, {Kiss},
  {Bayliss}, {Juh{\'a}sz}, {Kov{\'a}cs}, \& {Szalai}}]{moor2011}
{Mo{\'o}r}, A., {Pascucci}, I., {K{\'o}sp{\'a}l}, {\'A}., {et~al.} 2011, \apjs,
  193, 4

\bibitem[{Oliphant(2006)}]{numpy}
Oliphant, T.~E. 2006, A guide to NumPy, Vol.~1 (Trelgol Publishing USA)

\bibitem[{{Pawellek} {et~al.}(2014){Pawellek}, {Krivov}, {Marshall},
  {Montesinos}, {{\'A}brah{\'a}m}, {Mo{\'o}r}, {Bryden}, \&
  {Eiroa}}]{pawellek2014}
{Pawellek}, N., {Krivov}, A.~V., {Marshall}, J.~P., {et~al.} 2014, \apj, 792,
  65

\bibitem[{{Pecaut} \& {Mamajek}(2016)}]{pecaut2016}
{Pecaut}, M.~J. \& {Mamajek}, E.~E. 2016, \mnras, 461, 794

\bibitem[{{Pedregosa} {et~al.}(2012){Pedregosa}, {Varoquaux}, {Gramfort},
  {Michel}, {Thirion}, {Grisel}, {Blondel}, {M{\"u}ller}, {Nothman}, {Louppe},
  {Prettenhofer}, {Weiss}, {Dubourg}, {Vanderplas}, {Passos}, {Cournapeau},
  {Brucher}, {Perrot}, \& {Duchesnay}}]{scikit-learn}
{Pedregosa}, F., {Varoquaux}, G., {Gramfort}, A., {et~al.} 2012, arXiv
  e-prints, arXiv:1201.0490

\bibitem[{{Pollack} {et~al.}(1996){Pollack}, {Hubickyj}, {Bodenheimer},
  {Lissauer}, {Podolak}, \& {Greenzweig}}]{pollack1996}
{Pollack}, J.~B., {Hubickyj}, O., {Bodenheimer}, P., {et~al.} 1996, \icarus,
  124, 62

\bibitem[{{Rafikov}(2005)}]{rafikov2005}
{Rafikov}, R.~R. 2005, \apjl, 621, L69

\bibitem[{{Raghavan} {et~al.}(2010){Raghavan}, {McAlister}, {Henry}, {Latham},
  {Marcy}, {Mason}, {Gies}, {White}, \& {ten Brummelaar}}]{raghavan2010}
{Raghavan}, D., {McAlister}, H.~A., {Henry}, T.~J., {et~al.} 2010, \apjs, 190,
  1

\bibitem[{{Rameau} {et~al.}(2013){Rameau}, {Chauvin}, {Lagrange}, {Boccaletti},
  {Quanz}, {Bonnefoy}, {Girard}, {Delorme}, {Desidera}, {Klahr}, {Mordasini},
  {Dumas}, \& {Bonavita}}]{rameau2013}
{Rameau}, J., {Chauvin}, G., {Lagrange}, A.~M., {et~al.} 2013, The
  Astrophysical Journal, 772, L15

\bibitem[{{Raymond} {et~al.}(2012){Raymond}, {Armitage}, {Moro-Mart{\'\i}n},
  {Booth}, {Wyatt}, {Armstrong}, {Mand ell}, {Selsis}, \& {West}}]{raymond2012}
{Raymond}, S.~N., {Armitage}, P.~J., {Moro-Mart{\'\i}n}, A., {et~al.} 2012,
  \aap, 541, A11

\bibitem[{{Riaz} {et~al.}(2006){Riaz}, {Gizis}, \& {Harvin}}]{riaz2006}
{Riaz}, B., {Gizis}, J.~E., \& {Harvin}, J. 2006, \aj, 132, 866

\bibitem[{{Riello} {et~al.}(2021){Riello}, {De Angeli}, {Evans}, {Montegriffo},
  {Carrasco}, {Busso}, {Palaversa}, {Burgess}, {Diener}, {Davidson}, {Rowell},
  {Fabricius}, {Jordi}, {Bellazzini}, {Pancino}, {Harrison}, {Cacciari}, {van
  Leeuwen}, {Hambly}, {Hodgkin}, {Osborne}, {Altavilla}, {Barstow}, {Brown},
  {Castellani}, {Cowell}, {De Luise}, {Gilmore}, {Giuffrida}, {Hidalgo},
  {Holland}, {Marinoni}, {Pagani}, {Piersimoni}, {Pulone}, {Ragaini}, {Rainer},
  {Richards}, {Sanna}, {Walton}, {Weiler}, \& {Yoldas}}]{riello2020}
{Riello}, M., {De Angeli}, F., {Evans}, D.~W., {et~al.} 2021, \aap, 649, A3

\bibitem[{{Rousset} {et~al.}(2003){Rousset}, {Lacombe}, {Puget}, {Hubin},
  {Gendron}, {Fusco}, {Arsenault}, {Charton}, {Feautrier}, \&
  {Gigan}}]{rousset2003}
{Rousset}, G., {Lacombe}, F., {Puget}, P., {et~al.} 2003, in Society of
  Photo-Optical Instrumentation Engineers (SPIE) Conference Series, Vol. 4839,
  Adaptive Optical System Technologies II, ed. P.~L. {Wizinowich} \&
  D.~{Bonaccini}, 140--149

\bibitem[{{Schmidt} {et~al.}(2008){Schmidt}, {Neuh{\"a}user}, {Seifahrt},
  {Vogt}, {Bedalov}, {Helling}, {Witte}, \& {Hauschildt}}]{schmidt2008}
{Schmidt}, T.~O.~B., {Neuh{\"a}user}, R., {Seifahrt}, A., {et~al.} 2008, \aap,
  491, 311

\bibitem[{{Skrutskie} {et~al.}(2006){Skrutskie}, {Cutri}, {Stiening},
  {Weinberg}, {Schneider}, {Carpenter}, {Beichman}, {Capps}, {Chester},
  {Elias}, {Huchra}, {Liebert}, {Lonsdale}, {Monet}, {Price}, {Seitzer},
  {Jarrett}, {Kirkpatrick}, {Gizis}, {Howard}, {Evans}, {Fowler}, {Fullmer},
  {Hurt}, {Light}, {Kopan}, {Marsh}, {McCallon}, {Tam}, {Van Dyk}, \&
  {Wheelock}}]{skrutskie2006}
{Skrutskie}, M.~F., {Cutri}, R.~M., {Stiening}, R., {et~al.} 2006, \aj, 131,
  1163

\bibitem[{{Stolker} {et~al.}(2019){Stolker}, {Bonse}, {Quanz}, {Amara},
  {Cugno}, {Bohn}, \& {Boehle}}]{stolker2019}
{Stolker}, T., {Bonse}, M.~J., {Quanz}, S.~P., {et~al.} 2019, \aap, 621, A59

\bibitem[{{Stolker} {et~al.}(2020){Stolker}, {Marleau}, {Cugno},
  {Molli{\`e}re}, {Quanz}, {Todorov}, \& {K{\"u}hn}}]{stolker2020}
{Stolker}, T., {Marleau}, G.~D., {Cugno}, G., {et~al.} 2020, \aap, 644, A13

\bibitem[{{Tokovinin} {et~al.}(2010){Tokovinin}, {Mason}, \&
  {Hartkopf}}]{tokovinin2010}
{Tokovinin}, A., {Mason}, B.~D., \& {Hartkopf}, W.~I. 2010, \aj, 139, 743

\bibitem[{{Tokovinin} {et~al.}(2019){Tokovinin}, {Mason}, {Mendez}, {Horch}, \&
  {Brice{\~n}o}}]{tokovinin2019}
{Tokovinin}, A., {Mason}, B.~D., {Mendez}, R.~A., {Horch}, E.~P., \&
  {Brice{\~n}o}, C. 2019, \aj, 158, 48

\bibitem[{{Torres} {et~al.}(2006){Torres}, {Quast}, {da Silva}, {de La Reza},
  {Melo}, \& {Sterzik}}]{torres2006}
{Torres}, C.~A.~O., {Quast}, G.~R., {da Silva}, L., {et~al.} 2006, \aap, 460,
  695

\bibitem[{Van~der Walt {et~al.}(2014)Van~der Walt, Sch{\"o}nberger,
  Nunez-Iglesias, Boulogne, Warner, Yager, Gouillart, \& Yu}]{scikit-image}
Van~der Walt, S., Sch{\"o}nberger, J.~L., Nunez-Iglesias, J., {et~al.} 2014,
  PeerJ, 2, e453

\bibitem[{{Virtanen} {et~al.}(2020){Virtanen}, {Gommers}, {Oliphant},
  {Haberland}, {Reddy}, {Cournapeau}, {Burovski}, {Peterson}, {Weckesser},
  {Bright}, {van der Walt}, {Brett}, {Wilson}, {Millman}, {Mayorov}, {Nelson},
  {Jones}, {Kern}, {Larson}, {Carey}, {Polat}, {Feng}, {Moore}, {Vand erPlas},
  {Laxalde}, {Perktold}, {Cimrman}, {Henriksen}, {Quintero}, {Harris},
  {Archibald}, {Ribeiro}, {Pedregosa}, {van Mulbregt}, \& {SciPy 1. 0
  Contributors}}]{virtanen2020}
{Virtanen}, P., {Gommers}, R., {Oliphant}, T.~E., {et~al.} 2020, Nature
  Methods, 17, 261

\bibitem[{{Wang} {et~al.}(2015){Wang}, {Fischer}, {Horch}, \& {Xie}}]{wang2015}
{Wang}, J., {Fischer}, D.~A., {Horch}, E.~P., \& {Xie}, J.-W. 2015, \apj, 806,
  248

\bibitem[{{Wenger} {et~al.}(2000){Wenger}, {Ochsenbein}, {Egret}, {Dubois},
  {Bonnarel}, {Borde}, {Genova}, {Jasniewicz}, {Lalo{\"e}}, {Lesteven}, \&
  {Monier}}]{wenger2000}
{Wenger}, M., {Ochsenbein}, F., {Egret}, D., {et~al.} 2000, \aaps, 143, 9

\bibitem[{{Zhang} {et~al.}(2021){Zhang}, {Liu}, {Claytor}, {Best}, {Dupuy}, \&
  {Siverd}}]{zhang_zj2021}
{Zhang}, Z., {Liu}, M.~C., {Claytor}, Z.~R., {et~al.} 2021, \apjl, 916, L11

\bibitem[{{Zhang} {et~al.}(2020){Zhang}, {Liu}, {Hermes}, {Magnier}, {Marley},
  {Tremblay}, {Tucker}, {Do}, {Payne}, \& {Shappee}}]{zhang_zj2020}
{Zhang}, Z., {Liu}, M.~C., {Hermes}, J.~J., {et~al.} 2020, \apj, 891, 171

\end{thebibliography}

\begin{appendix}

\section{High-contrast imaging observations}
\label{sec:hci_observing_conditions}
\begin{figure*}
\centering
\includegraphics[width=0.8\textwidth]{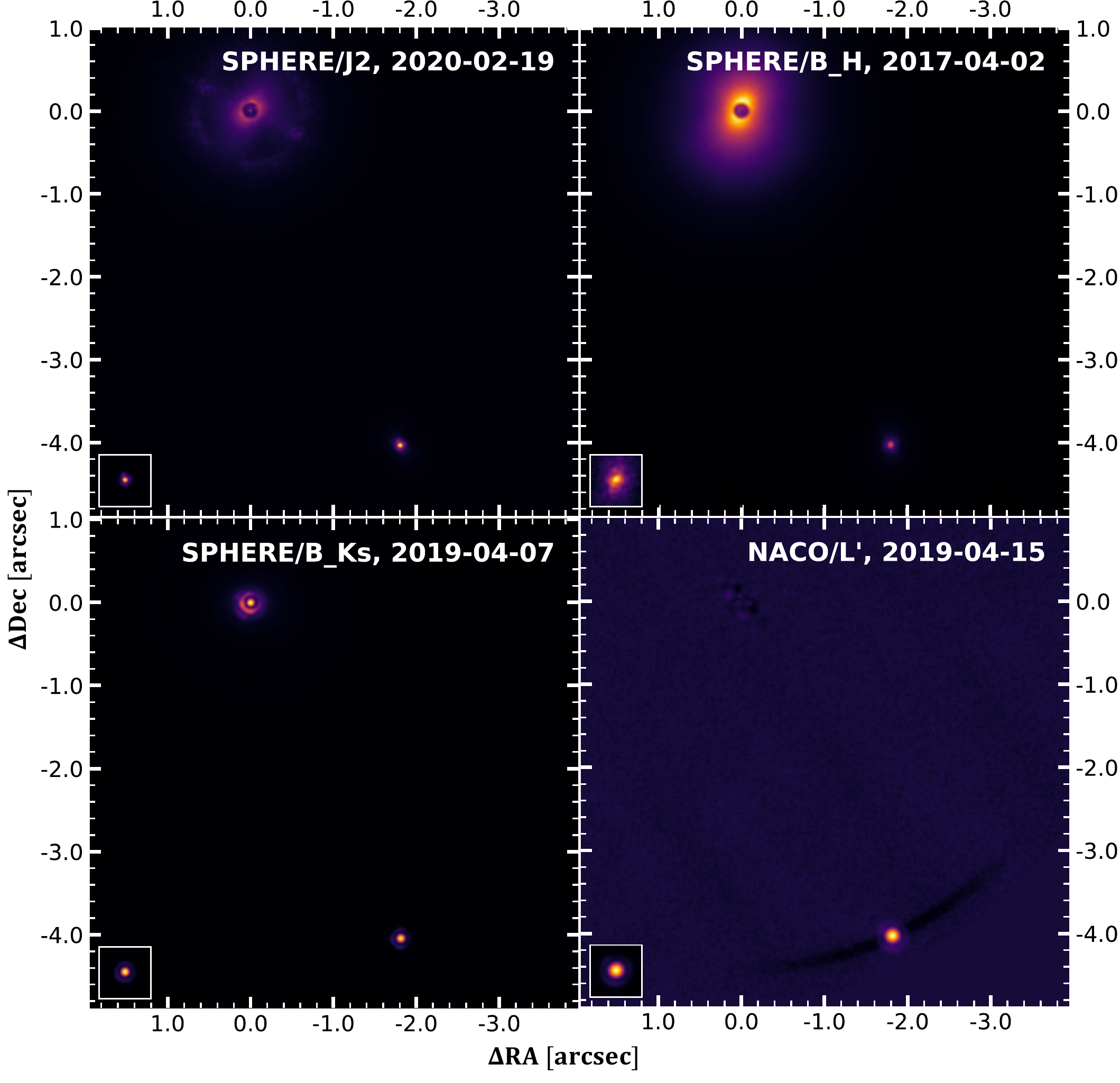}
\caption{
Reduced SPHERE and NACO data of TYC~8252-533-1 and its brown dwarf companion.
For each of the presented filters we show the median of the full observing sequence.
The images are derotated such that North points up and East towards the left.
No PSF subtraction is applied for the SPHERE data -- the star's intensity is attenuated by a coronagraphic mask, and it is located at the origin of the coordinate system that is representing the differential offsets in RA and Dec.
No coronagraph was used for the NACO observing sequence and the stellar PSF is removed by classical ADI \citep{marois2006} to obtain the presented image;
we did not perform any PSF subtraction for the extraction of the companion's photometry and astrometry.
The companion is detected at a separation of $\sim4\farcs4$ in south-western direction from the primary.
In the lower left of each panel we present the non-coronagraphic stellar PSF as a reference.
The images are shown with an arbitrary logarithmic color scale.
}
\label{fig:tyc8252_sphere_image}
\end{figure*}

\begin{figure*}
\centering
\includegraphics[width=0.8\textwidth]{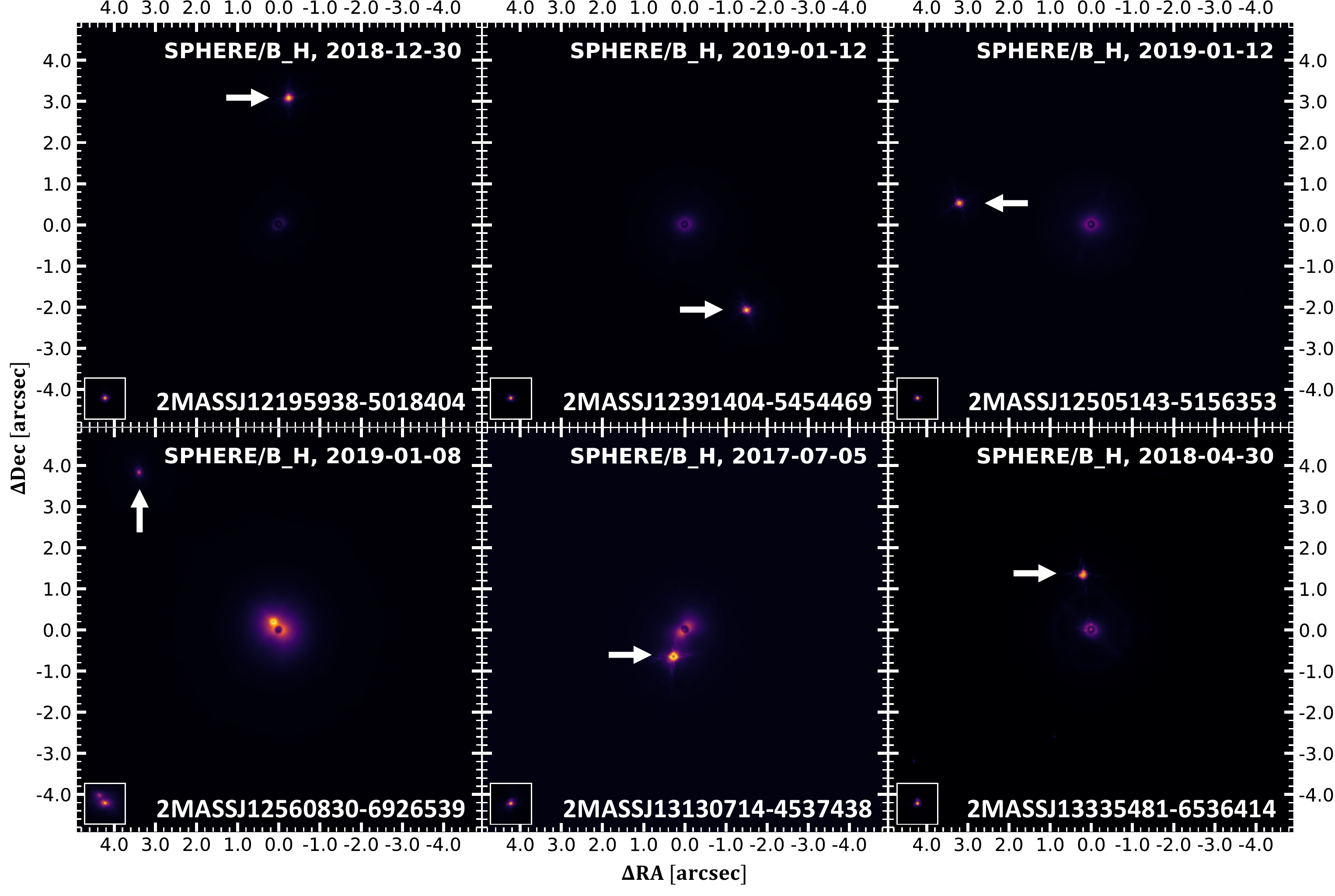}
\caption{
\rev{%
(Sub-)stellar companions identified in SPHERE data.
We present the de-rotated and median combined coronagraphic images for the epochs and filters as indicated in the upper right of each panel.
The star is located behind the coronagraphic mask at the center of each panel.
The companions that were also identified by our \textit{Gaia} selection algorithm are highlighted by the white arrows.
As discussed in Sect.~\ref{subsubsec:results_2massj1256} there is an additional, very close companion to 2MASS~J12560830-6926539.
With a separation of merely 0\farcs2 this stellar companion is not identified by our \textit{Gaia} selection method.
In the lower left of each panel we present the non-coronagraphic flux PSFs of each target.
In all frames north points up and east towards the left.
}
}
\label{fig:sphere_stellar_companions}
\end{figure*}

We present the setup of our SPHERE and NACO observations and the atmospheric conditions during the data acquisition in Table~\ref{tbl:yses_observations}.
The final images of our SPHERE and NACO data reductions on TYC~8252-533-1 are presented in Fig.~\ref{fig:tyc8252_sphere_image}.
\rev{%
In Fig.~\ref{fig:sphere_stellar_companions} we show the remaining stellar companions to YSES targets as observed with SPHERE.}
\begin{table*}
\caption{
High-contrast imaging observations \rev{of YSES targets}.
}
\label{tbl:yses_observations}
\def\arraystretch{1.2}
\setlength{\tabcolsep}{8pt}
\centering
\begin{tabular}{@{}llllllllll@{}}
\hline\hline
Target & Observation date & Instrument & Mode\tablefootmark{a} & Filter & NEXP$\times$NDIT$\times$DIT\tablefootmark{b} & $\Delta\pi$\tablefootmark{c} & $\langle X\rangle$\tablefootmark{d} & $\langle\omega\rangle$\tablefootmark{e} & $\langle\tau_0\rangle$\tablefootmark{f} \\    
(2MASS ID) & (yyyy-mm-dd) & & & & (1$\times$1$\times$s) & (\degr) & & (\arcsec) & (ms)\\
\hline
12195938-5018404 & 2018-12-30 & SPHERE & CI & $H$ & 4$\times$1$\times$32 & 0.4 & 1.6 & 0.5 & 8.0 \\
12195938-5018404 & 2018-12-30 & SPHERE & CI & $K_s$ & 4$\times$1$\times$32 & 0.4 & 1.6 & 0.6 & 9.0 \\
12391404-5454469 & 2019-01-12 & SPHERE & CI & $H$ & 4$\times$1$\times$32 & 0.5 & 1.4 & 1.0 & 4.3 \\
12391404-5454469 & 2019-01-12 & SPHERE & CI & $K_s$ & 4$\times$1$\times$32 & 0.5 & 1.4 & 0.9 & 4.2 \\
12505143-5156353 & 2019-01-12 & SPHERE & CI & $H$ & 4$\times$1$\times$32 & 0.5 & 1.3 & 1.1 & 3.8 \\
12505143-5156353 & 2019-01-12 & SPHERE & CI & $K_s$ & 4$\times$1$\times$32 & 0.5 & 1.3 & 1.0 & 4.1\\
12560830-6926539 & 2019-01-08 & SPHERE & CI & $H$ & 4$\times$1$\times$32 & 0.5 & 1.5 & 1.2 & 2.2 \\
12560830-6926539 & 2019-01-08 & SPHERE & CI & $K_s$ & 4$\times$1$\times$32 & 0.5 & 1.5 & 1.1 & 2.0 \\
13130714-4537438 & 2017-07-05 & SPHERE & CI & $J$ & 4$\times$2$\times$32 & 0.3 & 1.8 & 0.7 & 4.1 \\
13130714-4537438 & 2017-07-05 & SPHERE & CI & $H$ & 4$\times$1$\times$32 & 0.8 & 1.8 & 0.7 & 3.7 \\
13233587-4718467 & 2017-04-02 & SPHERE & CI & $H$ & 4$\times$1$\times$32 & 0.7 & 1.2 & 1.7 & 1.4 \\
13233587-4718467 & 2019-04-07 & SPHERE & CI & $K_s$ & 60$\times$2$\times$32 & 40.5 & 1.1 & 0.5 & 8.9 \\
13233587-4718467 & 2019-04-15 & NACO & CI & $L'$ & 63$\times$126$\times$0.2 & 24.9 & 1.1 & 0.5 & 3.9 \\
13233587-4718467 & 2020-02-19 & SPHERE & DBI & $Y23$ & 4$\times$2$\times$32 & 2.6 & 1.1 & 0.8 & 5.6 \\
13233587-4718467 & 2020-02-19 & SPHERE & DBI & $J23$ & 4$\times$2$\times$32 & 2.5 & 1.1 & 0.8 & 5.3 \\
13233587-4718467 & 2020-02-19 & SPHERE & DBI & $H23$ & 4$\times$2$\times$32 & 2.4 & 1.1 & 0.6 & 6.8 \\
13233587-4718467 & 2020-02-19 & SPHERE & DBI & $K12$ & 4$\times$2$\times$32 & 2.3 & 1.1 & 0.6 & 6.5 \\
13335481-6536414 & 2018-04-30 & SPHERE & CI & $H$ & 4$\times$2$\times$32 & 1.4 & 1.3 & 0.5 & 8.4 \\
13335481-6536414 & 2018-04-30 & SPHERE & CI & $K_s$ & 4$\times$2$\times$32 & 1.4 & 1.3 & 0.5 & 8.0 \\
13335481-6536414 & 2018-04-30 & SPHERE & DPI & $H$ & 16$\times$2$\times$32 & 6.4 & 1.3 & 0.5 & 10.3 \\
\hline
\end{tabular}
\tablefoot{
\tablefoottext{a}{
The observation mode is either classical imaging (CI) with a broadband filter, dual-band imaging (DBI) with two intermediate band filters simultaneously, \rev{or dual-polarization imaging (DPI) with orthogonal linear polarization filters in the optical path.}
}
\tablefoottext{b}{
NEXP describes the number of exposures, NDIT is the number of subintegrations per exposure and DIT is the detector integration time of an individual subintegration.
}
\tablefoottext{c}{
$\Delta\pi$ describes the amount of field rotation during the observation, if it is carried out in pupil-stabilized mode (only valid for CI observations).
}
\tablefoottext{d}{$\langle X\rangle$ denotes the average airmass during the observation.}
\tablefoottext{e}{$\langle\omega\rangle$ denotes the average seeing conditions during the observation.}
\tablefoottext{f}{$\langle\tau_0\rangle$ denotes the average coherence time during the observation.}
}
\end{table*}

\section{Identified companions}
\label{sec:identified_companions}

We present all identified \rev{candidate} companions in Table~\ref{tbl:companions_full_1}.

\begin{sidewaystable*}
\caption{
Identifiers and properties of all identified \rev{candidate} companions to K type Sco-Cen members from \citet{pecaut2016}.
}
\label{tbl:companions_full_1}
\tiny
\def\arraystretch{1.2}
\centering
\begin{tabular}{@{}llllllllllllllll@{}}
\hline \hline
\multicolumn{5}{c}{Target} & & \multicolumn{10}{c}{Companion} \\
\cline{1-5} 
\cline{7-16}
2MASS ID & \textit{Gaia} EDR3 ID & $\varpi$ & Age & Mass & & \textit{Gaia} EDR3 ID & \multicolumn{2}{c}{$\rho$} & $\Delta\mu_{\alpha *}$ & $\Delta\mu_{\delta}$ & $G$ & $G_\mathrm{BP}-G_\mathrm{RP}$ & Mass & $v_\mathrm{proj}/v_\mathrm{max}$ & $p^\mathrm{C}$\\
\cline{8-9}
 & & (mas) & (Myr) & ($M_\sun$) & & & (\arcsec) & (au) & (mas\,yr$^{-1}$) & (mas\,yr$^{-1}$) & (mag) & (mag) & ($M_\mathrm{Jup}$) & & (\%)\\
\hline
10313710-6901587 & \object{5232514298297802880} & 5.97 & 21 & 0.9 &  & \object{5232514298301348864} & 10.3 & 1725 & $-0.18\pm0.43$ & $-0.20\pm0.35$ & 19.87 & 2.50 & $36.5\pm2.4$ & $0.42^{+0.26}_{-0.21}$ & $98.23$ \\
10552886-6629147 & \object{5238609720201102080} & 7.57 & 15 & 1.1 &  & \object{5238609720201092736} & 20.8 & 2740 & $-2.29\pm0.05$ & $2.46\pm0.05$ & 16.40 & 3.03 & $120.2\pm9.0$ & $2.37^{+0.04}_{-0.04}$ & $0.00$ \\
11175186-6402056 & \object{5240643988513310592} & 7.20 & 11 & 1.2 &  & \object{5240643988513309952} & 49.6 & 6899 & $-0.05\pm0.03$ & $-0.60\pm0.02$ & 7.84 & 0.22 & $1841.2\pm54.6$ & $0.45^{+0.02}_{-0.02}$ & $99.75$ \\
11272881-3952572 & \object{5383924166203787648} & 7.52 & 6 & 1.2 &  & \object{5383924166203787904} & 23.0 & 3061 & $-0.69\pm0.20$ & $-3.20\pm0.18$ & 7.32 & 0.09 & $1985.8\pm69.7$ & $1.55^{+0.08}_{-0.09}$ & $0.00$ \\
11515049-6407278 & \object{5332922112460126976} & 6.70 & 24 & 1.0 &  & \object{5332922116810986880} & 1.6 & 243 & $0.19\pm0.17$ & $-3.77\pm0.16$ & 16.63 & - & $152.7\pm4.9$ & $0.92^{+0.04}_{-0.04}$ & $96.95$ \\
11554295-5637314 & \object{5343603288120259072} & 9.94 & 2 & 0.5 &  & \object{5343603180734334592} & 68.2 & 6865 & $-1.24\pm0.03$ & $2.07\pm0.03$ & 11.53 & 1.68 & $509.6\pm129.9$ & $2.28^{+0.03}_{-0.03}$ & $0.00$ \\
11554295-5637314 & \object{5343603288120259072} & 9.94 & 2 & 0.5 &  & \object{5343603288130858112} & 0.8 & 84 & $-0.55\pm0.23$ & $3.18\pm0.25$ & 12.98 & - & $265.2\pm77.3$ & $0.39^{+0.03}_{-0.03}$ & $100.00$ \\
12051254-5331233 & \object{6124952431621117056} & 8.67 & 17 & 0.9 &  & \object{6124952431621118848} & 7.4 & 853 & $0.54\pm0.04$ & $0.68\pm0.05$ & 16.47 & 3.46 & $100.6\pm7.2$ & $0.33^{+0.02}_{-0.02}$ & $100.00$ \\
12074236-6227282 & \object{6057456092878899072} & 8.98 & 3 & 1.2 &  & \object{6057456092897714816} & 1.6 & 181 & $1.06\pm0.06$ & $2.19\pm0.06$ & 13.23 & - & $306.5\pm89.6$ & $0.34^{+0.01}_{-0.01}$ & $100.00$ \\
12094184-5854450 & \object{6071087597518919040} & 9.50 & 6 & 1.3 &  & \object{6071087597497876480} & 12.8 & 1352 & $0.05\pm0.09$ & $-1.14\pm0.16$ & 15.89 & - & $76.0\pm16.7$ & $0.43^{+0.06}_{-0.06}$ & $100.00$ \\
12094184-5854450 & \object{6071087597518919040} & 9.50 & 6 & 1.3 &  & \object{6071087597518919808} & 3.5 & 365 & $-2.99\pm0.06$ & $-2.29\pm0.05$ & 13.10 & 2.20 & $395.5\pm47.3$ & $0.66^{+0.01}_{-0.01}$ & $99.75$ \\
12101065-4855476 & \object{6129878999619301120} & 9.72 & 12 & 1.0 &  & \object{6129879755533544832} & 14.1 & 1448 & $-1.40\pm0.21$ & $-0.50\pm0.20$ & 13.55 & 2.42 & $388.1\pm26.9$ & $0.56^{+0.08}_{-0.08}$ & $100.00$ \\
12120804-6554549 & \object{5860803662284763392} & 9.21 & 14 & 1.0 &  & \object{5860803696599969280} & 31.6 & 3435 & $-0.19\pm0.07$ & $0.26\pm0.07$ & 16.97 & 3.47 & $66.2\pm4.5$ & $0.23^{+0.05}_{-0.05}$ & $99.50$ \\
12121119-4950081 & \object{6129584364863667840} & 7.82 & 17 & 1.0 &  & \object{6129584364863668096} & 6.4 & 824 & $-0.83\pm0.03$ & $0.40\pm0.03$ & 14.92 & 2.80 & $270.4\pm12.8$ & $0.34^{+0.01}_{-0.01}$ & $100.00$ \\
12123577-5520273 & \object{6075815841096386816} & 8.91 & 7 & 1.2 &  & \object{6075816592695096576} & 27.3 & 3060 & $-0.31\pm0.03$ & $-1.23\pm0.03$ & 12.57 & 2.14 & $558.1\pm55.9$ & $0.67^{+0.02}_{-0.02}$ & $99.50$ \\
12123577-5520273 & \object{6075815841096386816} & 8.91 & 7 & 1.2 &  & \object{6075816596995760640} & 24.7 & 2772 & $2.04\pm0.09$ & $-0.90\pm0.08$ & 15.47 & 3.07 & $126.5\pm20.9$ & $1.29^{+0.05}_{-0.05}$ & $0.00$ \\
12143410-5110124 & \object{6126585034583698944} & 8.45 & 14 & 1.1 &  & \object{6126585034581064192} & 21.2 & 2505 & $-0.48\pm0.02$ & $-1.08\pm0.02$ & 12.23 & 1.82 & $794.3\pm17.4$ & $0.58^{+0.01}_{-0.01}$ & $100.00$ \\
12145229-5547037 & \object{6075548419252913152} & 8.78 & 10 & 1.4 &  & \object{6075548419252906368} & 24.0 & 2731 & $0.55\pm0.02$ & $-0.24\pm0.02$ & 8.33 & 0.64 & $1536.8\pm82.2$ & $0.24^{+0.01}_{-0.01}$ & $100.00$ \\
12164023-7007361 & \object{5855198832989286400} & 8.96 & 8 & 1.1 &  & \object{5855198828676205696} & 3.2 & 358 & $0.29\pm0.08$ & $-0.73\pm0.08$ & 13.78 & 2.36 & $336.1\pm30.8$ & $0.16^{+0.02}_{-0.02}$ & $100.00$ \\
12174048-5000266 & \object{6126715464151137792} & 7.17 & 14 & 0.7 &  & \object{6126715464149955968} & 10.8 & 1509 & $-0.86\pm0.11$ & $2.02\pm0.10$ & 14.29 & 2.78 & $389.5\pm23.1$ & $1.29^{+0.06}_{-0.06}$ & $0.01$ \\
12195938-5018404 & \object{6126648698878768384} & 6.56 & 10 & 0.9 &  & \object{6126648703176939392} & 3.1 & 468 & $-0.23\pm0.02$ & $-0.03\pm0.02$ & 13.93 & 2.24 & $464.7\pm29.3$ & $0.08^{+0.01}_{-0.01}$ & $100.00$ \\
12210808-5212226 & \object{6077461328965244160} & 9.09 & 8 & 1.0 &  & \object{6077461328965240576} & 21.8 & 2400 & $-0.43\pm0.04$ & $0.48\pm0.04$ & 15.78 & 3.61 & $104.1\pm16.4$ & $0.37^{+0.02}_{-0.02}$ & $100.00$ \\
12220430-4841248 & \object{6127110188823860480} & 7.13 & 5 & 1.3 &  & \object{6127110188823860864} & 8.3 & 1165 & $-1.02\pm0.05$ & $-0.73\pm0.04$ & 15.72 & 3.29 & $124.8\pm34.7$ & $0.57^{+0.02}_{-0.02}$ & $99.75$ \\
12302957-5222269 & \object{6078760195785278208} & 8.66 & 17 & 1.0 &  & \object{6078760195785506048} & 9.2 & 1061 & $1.07\pm0.03$ & $1.04\pm0.03$ & 13.35 & 2.75 & $538.6\pm26.5$ & $0.51^{+0.01}_{-0.01}$ & $100.00$ \\
12333381-5714066 & \object{6073105991611649024} & 9.82 & 18 & 1.0 &  & \object{6073106193460538240} & 13.4 & 1368 & $-1.22\pm0.10$ & $0.85\pm0.13$ & 17.93 & 4.17 & $46.5\pm3.3$ & $0.62^{+0.05}_{-0.05}$ & $100.00$ \\
12365895-5412178 & \object{6074563978391541888} & 8.66 & 5 & 1.3 &  & \object{6074563978391541504} & 20.7 & 2394 & $0.11\pm0.01$ & $0.04\pm0.02$ & 12.99 & 2.20 & $423.5\pm61.7$ & $0.06^{+0.01}_{-0.01}$ & $100.00$ \\
12373737-5143113 & \object{6078883925202611968} & 8.83 & 5 & 0.7 &  & \object{6078885402671365504} & 34.6 & 3918 & $-5.43\pm0.10$ & $2.32\pm0.08$ & 15.13 & 3.13 & $136.1\pm38.8$ & $5.18^{+0.08}_{-0.08}$ & $0.00$ \\
12391404-5454469 & \object{6074374346993398144} & 9.31 & 4 & 0.8 &  & \object{6074374346993397760} & 2.6 & 275 & $-1.16\pm0.10$ & $1.89\pm0.11$ & 14.30 & 2.46 & $189.1\pm41.9$ & $0.45^{+0.02}_{-0.02}$ & $100.00$ \\
12404664-5211046 & \object{6078074169253752064} & 6.88 & 14 & 1.1 &  & \object{6078074237969848960} & 4.8 & 693 & $-0.74\pm0.08$ & $-1.02\pm0.07$ & 15.03 & 2.75 & $277.2\pm15.9$ & $0.46^{+0.03}_{-0.03}$ & $100.00$ \\
12405458-5031550 & \object{6079389185156706944} & 11.30 & 17 & 0.8 &  & \object{6079389253876183808} & 14.1 & 1249 & $0.02\pm0.03$ & $0.45\pm0.03$ & 14.43 & 2.88 & $226.9\pm13.2$ & $0.16^{+0.01}_{-0.01}$ & $100.00$ \\
12420050-5759486 & \object{6060864815105080064} & 8.65 & 3 & 0.5 &  & \object{6060864819427739904} & 1.3 & 152 & $-2.80\pm0.28$ & $-1.09\pm0.51$ & 13.64 & - & $263.1\pm72.9$ & $0.56^{+0.06}_{-0.06}$ & $100.00$ \\
12443482-6331463 & \object{5862876928853591040} & 8.95 & 7 & 1.2 &  & \object{5862876928853590144} & 7.9 & 879 & $1.21\pm0.09$ & $1.10\pm0.11$ & 15.24 & 3.23 & $146.0\pm21.2$ & $0.53^{+0.03}_{-0.03}$ & $100.00$ \\
12454884-5410583 & \object{6074796250199371648} & 9.05 & 18 & 1.0 &  & \object{6074796245904371968} & 18.8 & 2074 & $-0.38\pm0.02$ & $-0.53\pm0.02$ & 13.61 & 2.48 & $465.6\pm23.8$ & $0.31^{+0.01}_{-0.01}$ & $100.00$ \\
12474824-5431308 & \object{6073980172067600640} & 8.47 & 1 & 0.5 &  & \object{6073980172067600000} & 2.6 & 311 & $-6.05\pm0.13$ & $2.86\pm0.13$ & 14.08 & 2.76 & $141.9\pm38.9$ & $1.97^{+0.04}_{-0.04}$ & $0.00$ \\
12474824-5431308 & \object{6073980172067600640} & 8.47 & 1 & 0.5 &  & \object{6073980240787079680} & 16.1 & 1904 & $-4.89\pm0.13$ & $-0.03\pm0.13$ & 14.05 & 3.21 & $144.2\pm40.0$ & $3.55^{+0.09}_{-0.10}$ & $0.00$ \\
12484818-5635378 & \object{6061376298490107520} & 7.73 & 18 & 1.2 &  & \object{6061376298480880512} & 2.8 & 367 & $-0.87\pm0.07$ & $0.69\pm0.11$ & 16.07 & 1.58 & $156.2\pm8.4$ & $0.27^{+0.02}_{-0.02}$ & $100.00$ \\
12504491-5654485 & \object{6061310052916532096} & 9.25 & 1 & 0.4 &  & \object{6061310052892185088} & 1.7 & 187 & $-3.02\pm0.06$ & $-2.46\pm0.07$ & 12.48 & 2.20 & $273.1\pm73.9$ & $0.80^{+0.01}_{-0.01}$ & $100.00$ \\
12505143-5156353 & \object{6075310478057303936} & 9.18 & 6 & 0.9 &  & \object{6075310478050174592} & 3.2 & 352 & $-0.11\pm0.03$ & $-0.70\pm0.03$ & 14.64 & 2.78 & $184.0\pm21.6$ & $0.16^{+0.01}_{-0.01}$ & $100.00$ \\
\hline
\end{tabular}
\tablefoot{
With $\varpi$ we denote the parallax of the target from our sample, $\rho$ denotes the projected separation between the identified companion and our primary target, and $\Delta\mu_{\alpha*}$ and $\Delta\mu_{\delta}$ refer to the RA and Dec differences in proper motion between them.
The ratio of $v_\mathrm{proj}$ and $v_\mathrm{max}$ is calculated as described in equation~\eqref{eqn:v_proj_v_max} (we present the 68\% confidence interval around the median of the posterior distribution) and $p^{c}$ is the upper limit we derive for the probability that the system is indeed gravitationally bound.
All astrometric and photometric measurements are adopted or derived from \textit{Gaia} EDR3 data products \citep{gaia2020}.
With one exception, the age and mass estimates for the sources are adopted from \citet{pecaut2016} and we assume that the age of the primary is also valid for the identified companion.
The age of the SZ~65 (2MASS J15392776-3446171, Gaia EDR3 6013399894569703040) system is based on Lupus membership \citep{comeron2008,alcala2014} and the mass estimate is taken from \citet{alcala2017}.
The masses are calculated with BT-Settl isochrones \citep{allard2012,baraffe2015}. 
}
\end{sidewaystable*}

\begin{sidewaystable*}
\setcounter{table}{\the\numexpr\value{table}-1\relax}
\caption
{
(continued).
}
\tiny
\def\arraystretch{1.2}
\centering
\begin{tabular}{@{}llllllllllllllll@{}}
\hline \hline
\multicolumn{5}{c}{Target} & & \multicolumn{10}{c}{Companion} \\
\cline{1-5} 
\cline{7-16}
2MASS ID & \textit{Gaia} EDR3 ID & $\varpi$ & Age & Mass & & \textit{Gaia} EDR3 ID & \multicolumn{2}{c}{$\rho$} & $\Delta\mu_{\alpha *}$ & $\Delta\mu_{\delta}$ & $G$ & $G_\mathrm{BP}-G_\mathrm{RP}$ & Mass & $v_\mathrm{proj}/v_\mathrm{max}$ & $p^\mathrm{C}$\\
\cline{8-9}
 & & (mas) & (Myr) & ($M_\sun$) & & & (\arcsec) & (au) & (mas\,yr$^{-1}$) & (mas\,yr$^{-1}$) & (mag) & (mag) & ($M_\mathrm{Jup}$) & & (\%)\\
\hline
12560830-6926539 & \object{5844909156504879360} & 10.07 & 3 & 0.7 &  & \object{5844909156504880128} & 5.1 & 503 & $-4.46\pm0.20$ & $-1.41\pm0.23$ & 16.17 & 3.41 & $44.5\pm5.8$ & $1.36^{+0.06}_{-0.06}$ & $0.00$ \\
13064012-5159386 & \object{6080364825623490688} & 8.75 & 10 & 1.2 &  & \object{6080364791268622336} & 8.1 & 929 & $0.19\pm0.02$ & $0.55\pm0.02$ & 14.23 & 2.67 & $295.2\pm17.4$ & $0.19^{+0.01}_{-0.01}$ & $100.00$ \\
13071310-5952108 & \object{6056115131031531264} & 8.60 & 3 & 1.2 &  & \object{6056115135337482496} & 1.9 & 221 & $-0.56\pm0.10$ & $3.67\pm0.12$ & 13.91 & - & $229.4\pm65.0$ & $0.61^{+0.02}_{-0.02}$ & $99.75$ \\
13071310-5952108 & \object{6056115131031531264} & 8.60 & 3 & 1.2 &  & \object{6056115169731484288} & 13.7 & 1597 & $1.56\pm0.09$ & $3.81\pm0.12$ & 15.46 & 3.39 & $72.6\pm24.0$ & $1.91^{+0.05}_{-0.05}$ & $0.00$ \\
13121859-5439054 & \object{6067650970822131584} & 7.81 & 26 & 1.0 &  & \object{6067650970822132224} & 12.5 & 1602 & $-0.50\pm0.02$ & $-1.29\pm0.02$ & 10.20 & 0.92 & $1054.4\pm14.4$ & $0.56^{+0.01}_{-0.01}$ & $99.25$ \\
13130714-4537438 & \object{6088027047281877120} & 7.18 & 2 & 0.8 &  & \object{6088027051573430400} & 0.7 & 98 & $1.36\pm0.35$ & $-0.22\pm0.12$ & 12.70 & - & $414.4\pm114.2$ & $0.20^{+0.05}_{-0.05}$ & $100.00$ \\
13142382-5054018 & \object{6080820298315369088} & 7.19 & 8 & 1.5 &  & \object{6080820302618880896} & 9.8 & 1369 & $1.42\pm0.08$ & $1.97\pm0.05$ & 16.93 & 3.31 & $70.1\pm11.3$ & $1.13^{+0.03}_{-0.03}$ & $0.00$ \\
13175314-5058481 & \object{6080854078239398016} & 6.36 & 3 & 0.8 &  & \object{6080854073939694592} & 33.3 & 5230 & $0.54\pm0.23$ & $-0.22\pm0.13$ & 18.59 & 3.17 & $27.1\pm4.3$ & $0.84^{+0.31}_{-0.29}$ & $69.87$ \\
13233587-4718467 & \object{6083750638577673088} & 7.74 & 5 & 1.2 &  & \object{6083750638540951552} & 4.4 & 570 & $1.20\pm0.16$ & $-0.42\pm0.13$ & 16.74 & 2.91 & $48.7\pm11.1$ & $0.40^{+0.05}_{-0.05}$ & $100.00$ \\
13270594-4856180 & \object{6082617587508723200} & 10.32 & 5 & 1.2 &  & \object{6082617591804976512} & 5.8 & 561 & $-0.22\pm0.03$ & $-3.41\pm0.03$ & 13.50 & 2.58 & $277.5\pm41.8$ & $0.73^{+0.01}_{-0.01}$ & $100.00$ \\
13335329-6536473 & \object{5863747467220735744} & 9.94 & 3 & 0.6 &  & \object{5863747462890662144} & 12.0 & 1204 & $2.64\pm0.39$ & $2.69\pm0.10$ & 14.60 & - & $106.9\pm37.7$ & $1.77^{+0.14}_{-0.13}$ & $0.00$ \\
13335329-6536473 & \object{5863747467220735744} & 9.94 & 3 & 0.6 &  & \object{5863747467220738432} & 11.1 & 1114 & $1.19\pm0.06$ & $3.59\pm0.07$ & 10.74 & 1.51 & $814.2\pm204.7$ & $1.22^{+0.02}_{-0.02}$ & $0.00$ \\
13335481-6536414 & \object{5863747467220738432} & 9.75 & 2 & 0.8 &  & \object{5863747462890662144} & 1.3 & 137 & $1.45\pm0.39$ & $-0.90\pm0.08$ & 14.60 & - & $91.9\pm31.0$ & $0.25^{+0.05}_{-0.05}$ & $99.75$ \\
13335481-6536414 & \object{5863747467220738432} & 9.75 & 2 & 0.8 &  & \object{5863747467220735744} & 11.1 & 1135 & $-1.19\pm0.06$ & $-3.59\pm0.07$ & 11.79 & 2.01 & $463.5\pm118.3$ & $1.32^{+0.02}_{-0.02}$ & $0.00$ \\
13364090-4043359 & \object{6113000942074857216} & 8.49 & 8 & 0.8 &  & \object{6113000946373209216} & 4.6 & 538 & $-1.18\pm0.03$ & $-0.79\pm0.02$ & 14.19 & 2.67 & $289.2\pm26.3$ & $0.42^{+0.01}_{-0.01}$ & $100.00$ \\
13442441-4706343 & \object{6106992866945163904} & 7.14 & 4 & 1.3 &  & \object{6106992866948449792} & 7.8 & 1097 & $-2.38\pm0.09$ & $0.10\pm0.07$ & 13.82 & 2.77 & $324.1\pm67.2$ & $0.98^{+0.04}_{-0.04}$ & $69.80$ \\
13454424-4904500 & \object{6094572753539758976} & 7.42 & 5 & 1.3 &  & \object{6094572719180018432} & 24.1 & 3245 & $0.41\pm0.15$ & $-0.73\pm0.13$ & 9.10 & 0.59 & $1737.8\pm90.8$ & $0.43^{+0.07}_{-0.07}$ & $99.75$ \\
13475054-4902056 & \object{6094529696485204864} & 7.17 & 23 & 1.0 &  & \object{6094529696485204736} & 19.0 & 2651 & $0.36\pm0.07$ & $0.89\pm0.06$ & 11.80 & 1.37 & $>1250$ & $0.56^{+0.04}_{-0.03}$ & $100.00$ \\
13540743-6733449 & \object{5850443307764629376} & 7.44 & 18 & 1.1 &  & \object{5850443303440130816} & 28.0 & 3763 & $0.95\pm0.01$ & $-0.40\pm0.02$ & 12.40 & 1.55 & $842.3\pm10.4$ & $0.69^{+0.01}_{-0.01}$ & $99.75$ \\
13540743-6733449 & \object{5850443307764629376} & 7.44 & 18 & 1.1 &  & \object{5850443303446183040} & 28.7 & 3861 & $1.27\pm0.02$ & $0.36\pm0.02$ & 13.66 & 2.08 & $555.1\pm26.7$ & $0.97^{+0.01}_{-0.01}$ & $98.96$ \\
13552552-4706563 & \object{6095161370216796928} & 8.30 & 4 & 1.0 &  & \object{6095162843384927744} & 23.4 & 2824 & $0.46\pm0.07$ & $0.18\pm0.05$ & 15.10 & 3.22 & $137.2\pm40.0$ & $0.34^{+0.04}_{-0.05}$ & $100.00$ \\
13562964-3839129 & \object{6114536929757358592} & 8.68 & 7 & 1.1 &  & \object{6114536929755100544} & 1.0 & 117 & $-0.57\pm0.04$ & $4.52\pm0.05$ & 11.97 & - & $735.9\pm70.3$ & $0.48^{+0.00}_{-0.01}$ & $100.00$ \\
14004970-4236569 & \object{6109646679996671360} & 7.46 & 20 & 1.0 &  & \object{6109646679996673024} & 11.0 & 1468 & $0.85\pm0.03$ & $0.63\pm0.04$ & 13.23 & 2.15 & $680.8\pm15.6$ & $0.47^{+0.01}_{-0.01}$ & $100.00$ \\
14213051-3845252 & \object{6116680904413813504} & 7.07 & 3 & 0.7 &  & \object{6116680908709624448} & 1.1 & 152 & $-0.82\pm0.31$ & $2.34\pm0.30$ & 12.41 & 1.63 & $556.6\pm154.3$ & $0.44^{+0.05}_{-0.05}$ & $100.00$ \\
14270556-4714217 & \object{6092201450556437632} & 8.37 & 19 & 1.1 &  & \object{6092201450556434944} & 19.1 & 2285 & $0.34\pm0.07$ & $-0.40\pm0.07$ & 15.99 & 3.69 & $152.8\pm7.4$ & $0.30^{+0.04}_{-0.04}$ & $100.00$ \\
14375022-5457411 & \object{5894194318558985984} & 8.62 & 17 & 1.1 &  & \object{5894194348576228992} & 59.7 & 6923 & $0.38\pm0.02$ & $0.37\pm0.02$ & 12.40 & 1.91 & $769.2\pm17.4$ & $0.43^{+0.01}_{-0.01}$ & $99.75$ \\
14380350-4932023 & \object{5899359721077048064} & 7.80 & 6 & 1.0 &  & \object{5899359514918611456} & 32.0 & 4103 & $-2.15\pm0.16$ & $-1.20\pm0.16$ & 15.79 & 3.57 & $112.4\pm28.0$ & $2.17^{+0.14}_{-0.13}$ & $0.00$ \\
14385440-4310223 & \object{6099803508307589632} & 7.64 & 3 & 1.2 &  & \object{6099826976008902016} & 30.2 & 3955 & $-0.16\pm0.17$ & $3.26\pm0.16$ & 15.05 & 3.14 & $120.9\pm42.5$ & $2.64^{+0.13}_{-0.13}$ & $0.00$ \\
14413499-4700288 & \object{5905886387736675840} & 8.80 & 16 & 1.2 &  & \object{5905886215936891008} & 46.5 & 5286 & $0.33\pm0.05$ & $-0.57\pm0.05$ & 15.10 & 3.02 & $205.0\pm8.8$ & $0.52^{+0.03}_{-0.04}$ & $99.75$ \\
14421590-4100183 & \object{6101758336902846208} & 8.47 & 15 & 1.0 &  & \object{6101758332601667712} & 4.4 & 521 & $1.53\pm0.03$ & $0.22\pm0.03$ & 11.84 & 1.56 & $883.5\pm7.2$ & $0.35^{+0.01}_{-0.01}$ & $100.00$ \\
14584573-3315102 & \object{6204234706098895104} & 5.77 & 52 & 0.8 &  & \object{6204234706096290688} & 2.8 & 489 & $-1.61\pm0.02$ & $-1.49\pm0.02$ & 11.87 & 1.09 & $917.1\pm5.1$ & $0.73^{+0.01}_{-0.01}$ & $100.00$ \\
14594473-3425465 & \object{6203845650778424960} & 8.66 & 7 & 1.1 &  & \object{6203845655074564864} & 0.9 & 104 & $4.57\pm0.07$ & $2.52\pm0.08$ & 11.53 & 1.33 & $871.5\pm68.6$ & $0.50^{+0.01}_{-0.01}$ & $100.00$ \\
15022600-3405131 & \object{6203959484592343936} & 6.63 & 8 & 1.0 &  & \object{6203959488887357824} & 1.8 & 264 & $-0.46\pm0.13$ & $0.13\pm0.11$ & 15.00 & - & $248.7\pm26.0$ & $0.12^{+0.03}_{-0.03}$ & $100.00$ \\
15085472-4303136 & \object{6003698907782693248} & 7.44 & 9 & 1.1 &  & \object{6003698873422956928} & 37.3 & 5011 & $-0.48\pm0.03$ & $-0.06\pm0.02$ & 9.40 & 0.65 & $1477.9\pm71.4$ & $0.33^{+0.02}_{-0.02}$ & $99.00$ \\
15110450-3251304 & \object{6201456107071825536} & 7.24 & 3 & 0.7 &  & \object{6201456107071825792} & 4.5 & 626 & $1.73\pm0.04$ & $-1.24\pm0.03$ & 13.64 & 2.30 & $313.2\pm92.5$ & $0.83^{+0.01}_{-0.01}$ & $100.00$ \\
15113968-3248560 & \object{6207460471351260160} & 7.05 & 1 & 0.4 &  & \object{6207460436991521280} & 48.6 & 6898 & $5.76\pm0.08$ & $-0.23\pm0.06$ & 12.94 & 2.35 & $294.4\pm75.3$ & $9.25^{+0.12}_{-0.12}$ & $0.00$ \\
15113968-3248560 & \object{6207460471351260160} & 7.05 & 1 & 0.4 &  & \object{6207460471351260928} & 20.6 & 2920 & $-0.70\pm0.11$ & $-0.96\pm0.08$ & 16.92 & 3.35 & $39.8\pm2.8$ & $1.55^{+0.12}_{-0.12}$ & $0.00$ \\
15124447-3116482 & \object{6210702308369949952} & 7.07 & 9 & 1.2 &  & \object{6210702415745896192} & 24.9 & 3513 & $-0.73\pm0.03$ & $-1.02\pm0.03$ & 13.98 & 2.45 & $404.6\pm29.1$ & $0.94^{+0.02}_{-0.02}$ & $99.17$ \\
15135817-4629145 & \object{5903894175757504128} & 7.25 & 3 & 0.7 &  & \object{5903894175757506176} & 12.8 & 1762 & $0.16\pm0.03$ & $0.66\pm0.02$ & 13.32 & 2.15 & $365.2\pm106.8$ & $0.43^{+0.01}_{-0.01}$ & $100.00$ \\
15152295-5441088 & \object{5886747291936159104} & 7.27 & 8 & 1.0 &  & \object{5886747291923626240} & 1.5 & 203 & $-3.22\pm0.03$ & $2.14\pm0.04$ & 13.32 & - & $509.4\pm45.0$ & $0.70^{+0.01}_{-0.01}$ & $100.00$ \\
15171083-3434194 & \object{6200310519037175040} & 7.70 & 8 & 1.1 &  & \object{6200310484677437184} & 48.9 & 6351 & $-0.16\pm0.04$ & $-0.58\pm0.03$ & 12.92 & 2.02 & $576.3\pm52.1$ & $0.55^{+0.03}_{-0.03}$ & $98.50$ \\
15171083-3434194 & \object{6200310519037175040} & 7.70 & 8 & 1.1 &  & \object{6200310514738629504} & 17.9 & 2330 & $-1.14\pm0.04$ & $-1.86\pm0.03$ & 7.97 & 0.23 & $1729.7\pm58.5$ & $0.93^{+0.01}_{-0.01}$ & $98.75$ \\
15180123-4444269 & \object{6000349589207590016} & 8.30 & 44 & 0.8 &  & \object{6000349657912636928} & 2.4 & 288 & $0.59\pm0.10$ & $2.31\pm0.09$ & 15.68 & 1.91 & $238.3\pm5.5$ & $0.54^{+0.02}_{-0.02}$ & $100.00$ \\
15182692-3738021 & \object{6007167351565715840} & 7.63 & 11 & 1.3 &  & \object{6007167179767020160} & 73.2 & 9597 & $1.17\pm0.05$ & $-1.92\pm0.04$ & 13.45 & 2.27 & $508.4\pm28.6$ & $2.43^{+0.05}_{-0.05}$ & $0.00$ \\
\hline
\end{tabular}
\end{sidewaystable*}

\begin{sidewaystable*}
\setcounter{table}{\the\numexpr\value{table}-1\relax}
\caption
{
(continued).
}
\tiny
\def\arraystretch{1.2}
\centering
\begin{tabular}{@{}llllllllllllllll@{}}
\hline \hline
\multicolumn{5}{c}{Target} & & \multicolumn{10}{c}{Companion} \\
\cline{1-5} 
\cline{7-16}
2MASS ID & \textit{Gaia} EDR3 ID & $\varpi$ & Age & Mass & & \textit{Gaia} EDR3 ID & \multicolumn{2}{c}{$\rho$} & $\Delta\mu_{\alpha *}$ & $\Delta\mu_{\delta}$ & $G$ & $G_\mathrm{BP}-G_\mathrm{RP}$ & Mass & $v_\mathrm{proj}/v_\mathrm{max}$ & $p^\mathrm{C}$\\
\cline{8-9}
 & & (mas) & (Myr) & ($M_\sun$) & & & (\arcsec) & (au) & (mas\,yr$^{-1}$) & (mas\,yr$^{-1}$) & (mag) & (mag) & ($M_\mathrm{Jup}$) & & (\%)\\
\hline
15185282-4050528 & \object{6004336074769756800} & 6.72 & 10 & 1.3 &  & \object{6004347825798261632} & 10.7 & 1592 & $-0.56\pm0.06$ & $-2.55\pm0.05$ & 14.39 & 2.64 & $365.0\pm22.8$ & $1.36^{+0.03}_{-0.03}$ & $0.00$ \\
15191600-4056075 & \object{6004331775495542656} & 7.68 & 25 & 1.0 &  & \object{6004331779800414464} & 3.9 & 503 & $-0.82\pm0.10$ & $0.00\pm0.11$ & 16.42 & 2.70 & $148.0\pm5.0$ & $0.25^{+0.03}_{-0.03}$ & $100.00$ \\
15215241-2842383 & \object{6213042069112745856} & 7.97 & 10 & 1.1 &  & \object{6213042073410590464} & 1.6 & 195 & $-2.08\pm0.15$ & $-2.12\pm0.10$ & 14.84 & - & $234.3\pm18.0$ & $0.51^{+0.02}_{-0.02}$ & $100.00$ \\
15241147-3030582 & \object{6208136391829015936} & 7.71 & 18 & 1.1 &  & \object{6208136396122878976} & 17.4 & 2261 & $-0.10\pm0.06$ & $0.71\pm0.04$ & 15.21 & 2.96 & $239.7\pm12.9$ & $0.43^{+0.02}_{-0.02}$ & $99.75$ \\
15241147-3030582 & \object{6208136391829015936} & 7.71 & 18 & 1.1 &  & \object{6208136396126956928} & 20.3 & 2625 & $0.02\pm0.03$ & $-0.50\pm0.02$ & 12.88 & 2.01 & $723.8\pm21.0$ & $0.28^{+0.01}_{-0.01}$ & $100.00$ \\
15241303-3030572 & \object{6208136396126956928} & 7.65 & 1 & 0.4 &  & \object{6208136391829015936} & 20.3 & 2648 & $-0.02\pm0.03$ & $0.50\pm0.02$ & 10.59 & 1.11 & $745.8\pm175.0$ & $0.36^{+0.01}_{-0.01}$ & $100.00$ \\
15241303-3030572 & \object{6208136396126956928} & 7.65 & 1 & 0.4 &  & \object{6208136396122878976} & 3.5 & 453 & $-0.12\pm0.06$ & $1.21\pm0.04$ & 15.21 & 2.96 & $85.8\pm8.5$ & $0.55^{+0.02}_{-0.02}$ & $99.75$ \\
15293858-3546513 & \object{6013649964744749184} & 8.57 & 7 & 1.4 &  & \object{6013649969039127552} & 2.6 & 307 & $-2.00\pm0.13$ & $-0.45\pm0.12$ & 15.64 & 2.54 & $118.4\pm22.0$ & $0.38^{+0.02}_{-0.02}$ & $100.00$ \\
15294727-3628374 & \object{6013480055841212416} & 6.12 & 5 & 1.3 &  & \object{6013480060137095296} & 1.7 & 273 & $-2.09\pm0.13$ & $0.51\pm0.10$ & 16.27 & - & $103.6\pm29.2$ & $0.55^{+0.03}_{-0.03}$ & $100.00$ \\
15295661-3135446 & \object{6208566369594256768} & 8.33 & 9 & 0.9 &  & \object{6208566373892571136} & 1.7 & 209 & $5.02\pm0.25$ & $0.41\pm0.22$ & 14.81 & - & $213.6\pm17.1$ & $0.94^{+0.05}_{-0.05}$ & $91.30$ \\
15312961-3021537 & \object{6209492506280547200} & 6.48 & 2 & 0.5 &  & \object{6209492506275602816} & 1.2 & 189 & $-1.21\pm0.18$ & $2.96\pm0.13$ & 14.34 & - & $209.8\pm60.9$ & $0.91^{+0.04}_{-0.04}$ & $98.52$ \\
15342313-3300087 & \object{6014539542668216832} & 7.84 & 2 & 0.5 &  & \object{6014539645747431680} & 43.9 & 5596 & $0.08\pm0.06$ & $-0.35\pm0.04$ & 15.16 & 3.36 & $87.5\pm29.1$ & $0.51^{+0.06}_{-0.06}$ & $99.75$ \\
15370214-3136398 & \object{6208381582919629568} & 7.95 & 8 & 1.5 &  & \object{6208381587215253504} & 4.7 & 587 & $0.90\pm0.06$ & $0.49\pm0.04$ & 11.79 & - & $878.5\pm53.4$ & $0.23^{+0.01}_{-0.01}$ & $99.75$ \\
15370214-3136398 & \object{6208381582919629568} & 7.95 & 8 & 1.5 &  & \object{6208381587220236160} & 5.9 & 742 & $-1.12\pm0.16$ & $-0.62\pm0.12$ & 11.87 & 2.60 & $856.1\pm54.7$ & $0.33^{+0.04}_{-0.04}$ & $99.75$ \\
15392776-3446171 & \object{6013399894569703040} & 6.54 & 2 & 0.7 &  & \object{6013399830146943104} & 6.4 & 973 & $-0.09\pm0.04$ & $-0.31\pm0.03$ & 13.89 & 3.09 & $265.5\pm77.7$ & $0.18^{+0.02}_{-0.02}$ & $100.00$ \\
15394637-3451027 & \object{6013398451460692992} & 6.53 & 2 & 0.9 &  & \object{6013351520350805632} & 1.1 & 170 & $-1.53\pm0.12$ & $-1.54\pm0.07$ & 13.94 & - & $260.0\pm75.2$ & $0.46^{+0.02}_{-0.02}$ & $100.00$ \\
15410679-2656263 & \object{6234377340635038848} & 8.58 & 29 & 1.0 &  & \object{6234377718592160384} & 6.3 & 733 & $-0.20\pm0.10$ & $-1.77\pm0.06$ & 16.12 & 3.07 & $161.7\pm4.5$ & $0.59^{+0.02}_{-0.02}$ & $100.00$ \\
15440376-3311110 & \object{6015107302986648064} & 6.41 & 6 & 1.3 &  & \object{6015107302990748032} & 1.3 & 210 & $-6.25\pm0.31$ & $3.07\pm0.21$ & 15.11 & - & $214.5\pm34.2$ & $1.45^{+0.06}_{-0.06}$ & $0.00$ \\
15444712-3811406 & \object{6009518657190466560} & 7.72 & 2 & 0.6 &  & \object{6009518657176115840} & 3.3 & 427 & $-1.76\pm0.09$ & $-3.22\pm0.06$ & 13.63 & 2.32 & $253.2\pm73.3$ & $1.21^{+0.02}_{-0.02}$ & $0.00$ \\
15451286-3417305 & \object{6014696841553696768} & 6.55 & 1 & 1.6 &  & \object{6014696841553696896} & 2.8 & 430 & $1.18\pm0.23$ & $-1.51\pm0.16$ & 14.74 & - & $135.1\pm34.1$ & $0.52^{+0.05}_{-0.05}$ & $99.75$ \\
15451286-3417305 & \object{6014696841553696768} & 6.55 & 1 & 1.6 &  & \object{6014696875913435520} & 49.6 & 7570 & $0.96\pm0.18$ & $0.42\pm0.11$ & 17.55 & 4.02 & $30.8\pm3.6$ & $1.24^{+0.20}_{-0.20}$ & $11.50$ \\
15462958-5217239 & \object{5885915442658792576} & 7.31 & 22 & 1.0 &  & \object{5885915442658792448} & 1.6 & 218 & $2.09\pm0.10$ & $-0.64\pm0.06$ & 14.74 & - & $355.0\pm11.4$ & $0.43^{+0.02}_{-0.02}$ & $100.00$ \\
15481299-2349523 & \object{6239880671209818880} & 9.23 & 6 & 0.5 &  & \object{6239880671208969856} & 29.4 & 3184 & $1.09\pm0.62$ & $0.55\pm0.36$ & 14.53 & 2.81 & $193.5\pm33.9$ & $1.08^{+0.46}_{-0.42}$ & $42.80$ \\
15514535-2456513 & \object{6236477064249469312} & 6.42 & 13 & 1.0 &  & \object{6236477064249562624} & 2.4 & 369 & $0.84\pm0.26$ & $1.64\pm0.15$ & 16.77 & 1.93 & $111.3\pm9.1$ & $0.60^{+0.06}_{-0.05}$ & $100.00$ \\
15545141-3154463 & \object{6039427503765943936} & 7.10 & 2 & 0.7 &  & \object{6039427503765943040} & 6.3 & 883 & $0.95\pm0.22$ & $0.04\pm0.17$ & 17.57 & 3.08 & $33.8\pm3.6$ & $0.53^{+0.11}_{-0.12}$ & $100.00$ \\
15582054-1837252 & \object{6248340348034690688} & 7.20 & 9 & 1.5 &  & \object{6248340343736161664} & 5.6 & 783 & $0.25\pm0.17$ & $0.77\pm0.09$ & 15.54 & 2.65 & $176.8\pm14.9$ & $0.28^{+0.04}_{-0.04}$ & $100.00$ \\
15591101-1850442 & \object{6247571617610119808} & 6.49 & 3 & 0.7 &  & \object{6247571583249729280} & 14.6 & 2256 & $-0.07\pm0.05$ & $0.42\pm0.03$ & 13.32 & 2.71 & $403.3\pm113.8$ & $0.34^{+0.03}_{-0.03}$ & $100.00$ \\
16011070-4804438 & \object{5984404849554176896} & 8.02 & 18 & 1.0 &  & \object{5984404849554176256} & 15.3 & 1904 & $-0.20\pm0.10$ & $-0.21\pm0.08$ & 17.29 & 3.64 & $73.8\pm4.4$ & $0.18^{+0.05}_{-0.05}$ & $100.00$ \\
16023814-2541389 & \object{6235806323497438592} & 8.94 & 5 & 0.7 &  & \object{6235806259081172736} & 31.7 & 3546 & $-0.05\pm0.03$ & $-0.11\pm0.02$ & 11.62 & 1.49 & $738.5\pm99.0$ & $0.08^{+0.01}_{-0.01}$ & $99.25$ \\
16023910-2542078 & \object{6235806259081172736} & 8.98 & 13 & 0.8 &  & \object{6235806323497438592} & 31.7 & 3529 & $0.05\pm0.03$ & $0.11\pm0.02$ & 12.36 & 1.81 & $716.8\pm23.9$ & $0.08^{+0.01}_{-0.01}$ & $99.00$ \\
16025123-2401574 & \object{6236273895118889472} & 6.92 & 6 & 1.0 &  & \object{6236273895118890112} & 7.2 & 1043 & $0.58\pm0.09$ & $0.20\pm0.05$ & 16.35 & 3.80 & $87.5\pm21.4$ & $0.31^{+0.04}_{-0.04}$ & $100.00$ \\
16032367-1751422 & \object{6249905743355599232} & 6.00 & 1 & 0.3 &  & \object{6249905743349542784} & 2.5 & 424 & $2.01\pm0.34$ & $-1.64\pm0.25$ & 15.11 & 2.90 & $123.1\pm34.2$ & $1.56^{+0.18}_{-0.18}$ & $0.12$ \\
16042165-2130284 & \object{6243393817024157184} & 6.91 & 2 & 1.4 &  & \object{6243393817024156288} & 16.2 & 2346 & $0.19\pm0.05$ & $0.95\pm0.03$ & 13.59 & 2.80 & $288.6\pm86.5$ & $0.59^{+0.02}_{-0.02}$ & $99.50$ \\
16062354-1814187 & \object{6249074718717359744} & 6.23 & 1 & 0.7 &  & \object{6249074718725644288} & 0.6 & 100 & $0.15\pm0.34$ & $0.59\pm0.20$ & 12.21 & 1.75 & $459.1\pm132.7$ & $0.12^{+0.04}_{-0.03}$ & $100.00$ \\
16065436-2416107 & \object{6242013410230303488} & 6.73 & 1 & 0.3 &  & \object{6242013414531686272} & 1.7 & 246 & $-2.67\pm0.08$ & $-1.74\pm0.05$ & 13.59 & - & $208.2\pm70.3$ & $1.19^{+0.03}_{-0.03}$ & $0.00$ \\
16065795-2743094 & \object{6042124910722287744} & 6.49 & 6 & 2.0 &  & \object{6042124807643059968} & 31.2 & 4812 & $-1.97\pm0.17$ & $-1.73\pm0.10$ & 17.93 & 4.04 & $34.9\pm5.3$ & $2.21^{+0.12}_{-0.12}$ & $0.00$ \\
16065795-2743094 & \object{6042124910722287744} & 6.49 & 6 & 2.0 &  & \object{6042124915024285440} & 1.5 & 228 & $-3.02\pm0.37$ & $-0.10\pm0.21$ & 11.78 & 0.85 & $957.4\pm83.6$ & $0.46^{+0.06}_{-0.06}$ & $100.00$ \\
16073366-3759242 & \object{5998266426996208384} & 7.32 & 10 & 0.8 &  & \object{5998266426996209408} & 3.6 & 486 & $-1.18\pm0.04$ & $1.00\pm0.03$ & 14.68 & 2.66 & $286.1\pm17.5$ & $0.51^{+0.01}_{-0.01}$ & $100.00$ \\
16080772-5041556 & \object{5983063479721274880} & 10.14 & 10 & 0.8 &  & \object{5983063479721275520} & 6.2 & 608 & $-1.03\pm0.03$ & $1.45\pm0.03$ & 13.57 & 2.98 & $352.5\pm21.7$ & $0.46^{+0.01}_{-0.01}$ & $100.00$ \\
16081824-3844052 & \object{5997477080726175744} & 7.81 & 6 & 0.7 &  & \object{5997477080724660992} & 8.3 & 1066 & $0.23\pm0.15$ & $2.83\pm0.10$ & 16.59 & 3.25 & $63.6\pm14.8$ & $1.53^{+0.05}_{-0.05}$ & $0.00$ \\
16083436-1911563 & \object{6248755585472315392} & 8.38 & 1 & 0.7 &  & \object{6248755684252562432} & 27.6 & 3300 & $1.68\pm0.51$ & $-0.46\pm0.34$ & 16.99 & 3.83 & $31.1\pm4.8$ & $1.60^{+0.44}_{-0.44}$ & $8.29$ \\
16084340-2602168 & \object{6043486660173351552} & 7.28 & 5 & 1.7 &  & \object{6043486655875354112} & 13.5 & 1855 & $0.13\pm0.33$ & $-0.23\pm0.21$ & 18.71 & 4.01 & $24.8\pm1.0$ & $0.21^{+0.12}_{-0.10}$ & $100.00$ \\
16085427-3906057 & \object{5997035351934438784} & 6.41 & 3 & 1.4 &  & \object{5997035317574700544} & 62.6 & 9761 & $1.50\pm0.10$ & $0.40\pm0.07$ & 16.15 & 3.52 & $70.8\pm21.4$ & $2.23^{+0.14}_{-0.14}$ & $0.00$ \\
16085427-3906057 & \object{5997035351934438784} & 6.41 & 3 & 1.4 &  & \object{5997035416337166976} & 48.4 & 7542 & $-0.94\pm0.43$ & $0.99\pm0.32$ & 19.30 & 2.53 & $21.3\pm4.1$ & $1.81^{+0.47}_{-0.45}$ & $3.17$ \\
\hline
\end{tabular}
\end{sidewaystable*}

\begin{sidewaystable*}
\setcounter{table}{\the\numexpr\value{table}-1\relax}
\caption
{
(continued).
}
\tiny
\def\arraystretch{1.2}
\centering
\begin{tabular}{@{}llllllllllllllll@{}}
\hline \hline
\multicolumn{5}{c}{Target} & & \multicolumn{10}{c}{Companion} \\
\cline{1-5} 
\cline{7-16}
2MASS ID & \textit{Gaia} EDR3 ID & $\varpi$ & Age & Mass & & \textit{Gaia} EDR3 ID & \multicolumn{2}{c}{$\rho$} & $\Delta\mu_{\alpha *}$ & $\Delta\mu_{\delta}$ & $G$ & $G_\mathrm{BP}-G_\mathrm{RP}$ & Mass & $v_\mathrm{proj}/v_\mathrm{max}$ & $p^\mathrm{C}$\\
\cline{8-9}
 & & (mas) & (Myr) & ($M_\sun$) & & & (\arcsec) & (au) & (mas\,yr$^{-1}$) & (mas\,yr$^{-1}$) & (mag) & (mag) & ($M_\mathrm{Jup}$) & & (\%)\\
\hline
16085673-2033460 & \object{6243922063636859008} & 7.08 & 2 & 0.8 &  & \object{6243922029277120128} & 49.2 & 6955 & $-0.09\pm0.08$ & $-2.47\pm0.06$ & 14.49 & 3.29 & $174.4\pm47.6$ & $3.34^{+0.08}_{-0.07}$ & $0.00$ \\
16090075-1908526 & \object{6245777283349430912} & 7.30 & 4 & 0.6 &  & \object{6245777283349431552} & 19.0 & 2597 & $0.06\pm0.07$ & $-0.23\pm0.04$ & 15.42 & 3.39 & $133.0\pm38.5$ & $0.23^{+0.04}_{-0.04}$ & $100.00$ \\
16101171-3226360 & \object{6035794030167079808} & 6.28 & 1 & 0.7 &  & \object{6035794030168507264} & 1.3 & 211 & $-0.51\pm0.08$ & $-0.01\pm0.05$ & 12.10 & 1.66 & $478.1\pm117.9$ & $0.12^{+0.02}_{-0.02}$ & $100.00$ \\
16101729-1910263 & \object{6245762538722565888} & 6.43 & 1 & 0.7 &  & \object{6245762543021965184} & 5.6 & 870 & $-1.35\pm0.20$ & $-1.27\pm0.13$ & 14.10 & 2.92 & $180.6\pm53.7$ & $1.03^{+0.09}_{-0.09}$ & $38.69$ \\
16101918-2502301 & \object{6049748791208799488} & 6.30 & 1 & 0.4 &  & \object{6049748786908497408} & 4.9 & 777 & $1.29\pm0.11$ & $-0.67\pm0.08$ & 15.66 & 3.15 & $84.3\pm8.5$ & $1.05^{+0.08}_{-0.07}$ & $26.17$ \\
16102174-1904067 & \object{6245781097280740864} & 7.48 & 4 & 0.5 &  & \object{6245781131640479360} & 4.6 & 615 & $-0.45\pm0.12$ & $0.72\pm0.10$ & 16.94 & 4.08 & $43.1\pm5.8$ & $0.44^{+0.05}_{-0.05}$ & $100.00$ \\
16102653-2756293 & \object{6042235793900264448} & 18.45 & 2 & 0.3 &  & \object{6042235725180784768} & 24.6 & 1333 & $-1.00\pm0.03$ & $1.65\pm0.02$ & 13.15 & 2.47 & $95.4\pm32.3$ & $0.69^{+0.01}_{-0.01}$ & $100.00$ \\
16104202-2101319 & \object{6243833724749589632} & 7.14 & 1 & 0.7 &  & \object{6243833724749589760} & 3.2 & 448 & $0.21\pm0.11$ & $1.26\pm0.08$ & 15.51 & 3.25 & $78.9\pm12.5$ & $0.49^{+0.03}_{-0.03}$ & $100.00$ \\
16111534-1757214 & \object{6249000566108106112} & 7.39 & 2 & 0.4 &  & \object{6249001841715440512} & 40.5 & 5477 & $-1.21\pm0.12$ & $-0.93\pm0.08$ & 13.67 & 3.09 & $259.9\pm74.7$ & $2.14^{+0.15}_{-0.15}$ & $0.00$ \\
16114387-2526350 & \object{6049537165279381632} & 7.11 & 7 & 1.1 &  & \object{6049532737170518016} & 32.9 & 4618 & $2.26\pm0.11$ & $-0.25\pm0.07$ & 15.49 & 3.47 & $167.6\pm20.0$ & $2.18^{+0.10}_{-0.10}$ & $0.00$ \\
16114387-2526350 & \object{6049537165279381632} & 7.11 & 7 & 1.1 &  & \object{6049537169580902272} & 46.9 & 6597 & $-3.88\pm0.16$ & $3.52\pm0.10$ & 16.40 & 4.32 & $88.5\pm16.5$ & $6.18^{+0.16}_{-0.16}$ & $0.00$ \\
16123916-1859284 & \object{6245821092014031616} & 7.42 & 10 & 1.1 &  & \object{6245821126373768832} & 19.1 & 2578 & $-0.09\pm0.03$ & $-3.57\pm0.02$ & 10.40 & 1.57 & $1261.8\pm35.1$ & $1.81^{+0.01}_{-0.01}$ & $0.00$ \\
16124051-1859282 & \object{6245821126373768832} & 7.47 & 1 & 2.0 &  & \object{6245821092014031616} & 19.1 & 2559 & $0.09\pm0.03$ & $3.57\pm0.02$ & 12.99 & 2.33 & $265.4\pm72.1$ & $1.81^{+0.01}_{-0.01}$ & $0.00$ \\
16125265-2319560 & \object{6242176829446854656} & 6.60 & 9 & 1.1 &  & \object{6242176653347275136} & 38.3 & 5799 & $1.07\pm0.04$ & $-0.23\pm0.03$ & 9.06 & 0.99 & $1567.9\pm87.1$ & $0.88^{+0.03}_{-0.03}$ & $98.24$ \\
16135801-3618133 & \object{6022499010422921088} & 8.02 & 28 & 1.0 &  & \object{6022499006128860928} & 5.7 & 716 & $1.22\pm0.03$ & $0.80\pm0.02$ & 13.34 & 2.23 & $652.7\pm7.5$ & $0.43^{+0.01}_{-0.01}$ & $99.75$ \\
16135801-3618133 & \object{6022499010422921088} & 8.02 & 28 & 1.0 &  & \object{6022499010440068864} & 7.5 & 931 & $2.39\pm0.09$ & $1.25\pm0.05$ & 15.51 & - & $225.7\pm9.2$ & $1.05^{+0.03}_{-0.03}$ & $6.44$ \\
16140035-2108439 & \object{6243089247998315264} & 6.96 & 7 & 0.7 &  & \object{6243089217937775360} & 45.8 & 6582 & $-0.30\pm0.05$ & $-0.39\pm0.04$ & 15.53 & 3.14 & $168.2\pm19.0$ & $0.70^{+0.06}_{-0.06}$ & $99.25$ \\
16145207-5026187 & \object{5935099415271699072} & 8.36 & 4 & 1.5 &  & \object{5935099415267651328} & 2.2 & 261 & $-3.57\pm0.05$ & $0.55\pm0.04$ & 11.42 & - & $792.0\pm137.5$ & $0.52^{+0.01}_{-0.01}$ & $100.00$ \\
16153456-2242421 & \object{6242598526515737728} & 7.25 & 2 & 0.8 &  & \object{6242598526515738112} & 1.9 & 264 & $1.64\pm0.10$ & $0.36\pm0.05$ & 13.25 & - & $320.5\pm91.2$ & $0.40^{+0.02}_{-0.02}$ & $100.00$ \\
16161423-2643148 & \object{6042418858284146688} & 8.26 & 5 & 0.9 &  & \object{6042418828224074752} & 71.2 & 8627 & $1.05\pm0.12$ & $1.39\pm0.07$ & 14.24 & 2.94 & $242.2\pm39.7$ & $2.08^{+0.11}_{-0.11}$ & $0.00$ \\
16161423-2643148 & \object{6042418858284146688} & 8.26 & 5 & 0.9 &  & \object{6042418862581174016} & 2.5 & 304 & $-0.54\pm0.05$ & $1.54\pm0.03$ & 12.64 & 1.94 & $519.0\pm75.7$ & $0.33^{+0.01}_{-0.01}$ & $100.00$ \\
16175569-3828132 & \object{5997311359393377280} & 9.39 & 8 & 0.6 &  & \object{5997311363710561280} & 6.8 & 721 & $-2.11\pm0.10$ & $1.44\pm0.06$ & 16.88 & 3.19 & $50.7\pm10.1$ & $1.02^{+0.03}_{-0.03}$ & $25.66$ \\
16183723-2405226 & \object{6050056478369804288} & 6.45 & 33 & 1.0 &  & \object{6050056482664014976} & 4.5 & 692 & $-1.59\pm0.09$ & $1.49\pm0.07$ & 11.47 & 1.28 & $919.0\pm8.9$ & $0.73^{+0.03}_{-0.03}$ & $100.00$ \\
16191217-2550383 & \object{6048571072519663872} & 6.47 & 5 & 1.2 &  & \object{6048571076818643968} & 1.3 & 207 & $-0.57\pm0.13$ & $2.45\pm0.09$ & 16.12 & - & $107.0\pm29.8$ & $0.55^{+0.02}_{-0.02}$ & $100.00$ \\
16204468-2431384 & \object{6049266036882633856} & 6.02 & 10 & 1.6 &  & \object{6049266002522895232} & 33.4 & 5547 & $-1.57\pm0.10$ & $3.24\pm0.07$ & 16.24 & 3.80 & $155.2\pm11.6$ & $3.79^{+0.08}_{-0.07}$ & $0.00$ \\
16204468-2431384 & \object{6049266036882633856} & 6.02 & 10 & 1.6 &  & \object{6049266414839756416} & 42.9 & 7126 & $-3.20\pm0.09$ & $3.44\pm0.07$ & 16.28 & 3.64 & $152.3\pm11.7$ & $5.61^{+0.09}_{-0.09}$ & $0.00$ \\
16235484-3312370 & \object{6024816025017339392} & 7.78 & 4 & 0.7 &  & \object{6024816020718366464} & 42.6 & 5473 & $0.81\pm0.49$ & $-1.72\pm0.38$ & 7.46 & 0.06 & $2185.5\pm311.4$ & $1.25^{+0.26}_{-0.25}$ & $15.45$ \\
16235484-3312370 & \object{6024816025017339392} & 7.78 & 4 & 0.7 &  & \object{6024816059387932416} & 45.5 & 5849 & $1.44\pm0.49$ & $-1.62\pm0.38$ & 7.00 & -0.02 & $2273.4\pm318.0$ & $1.45^{+0.28}_{-0.28}$ & $4.85$ \\
16240289-2524539 & \object{6048929792484158208} & 5.78 & 5 & 2.6 &  & \object{6048929792479889920} & 2.1 & 356 & $-5.93\pm0.14$ & $4.84\pm0.11$ & 13.50 & - & $500.4\pm79.8$ & $1.60^{+0.03}_{-0.03}$ & $0.00$ \\
16250991-3047572 & \object{6037691855943122176} & 7.76 & 14 & 1.3 &  & \object{6037691855943131520} & 40.9 & 5267 & $9.89\pm0.20$ & $-0.41\pm0.17$ & 14.61 & 2.84 & $301.2\pm19.3$ & $8.27^{+0.17}_{-0.17}$ & $0.00$ \\
16263591-3314481 & \object{6024849594490361856} & 6.73 & 5 & 1.3 &  & \object{6024849594490360320} & 5.6 & 839 & $3.39\pm0.03$ & $2.88\pm0.02$ & 13.50 & 2.05 & $432.2\pm63.2$ & $1.65^{+0.01}_{-0.01}$ & $0.00$ \\
16265280-2343127 & \object{6050875614529786752} & 7.25 & 3 & 1.6 &  & \object{6050875618822791936} & 30.0 & 4137 & $-4.47\pm0.49$ & $-1.13\pm0.41$ & 19.05 & - & $21.1\pm4.1$ & $3.63^{+0.38}_{-0.37}$ & $0.00$ \\
16265700-3032232 & \object{6037784004457590528} & 9.16 & 9 & 0.8 &  & \object{6037784004457597568} & 10.6 & 1158 & $0.04\pm0.03$ & $-1.25\pm0.02$ & 12.58 & 2.12 & $592.3\pm44.2$ & $0.45^{+0.01}_{-0.01}$ & $100.00$ \\
16265700-3032232 & \object{6037784004457590528} & 9.16 & 9 & 0.8 &  & \object{6037784008760031872} & 9.3 & 1013 & $1.65\pm0.03$ & $-1.29\pm0.03$ & 11.13 & 1.45 & $1001.3\pm23.0$ & $0.62^{+0.01}_{-0.01}$ & $99.75$ \\
16265763-3032279 & \object{6037784008760031872} & 9.20 & 1 & 0.5 &  & \object{6037784004457590528} & 9.3 & 1009 & $-1.65\pm0.03$ & $1.29\pm0.03$ & 11.64 & 1.66 & $406.6\pm101.5$ & $0.87^{+0.01}_{-0.01}$ & $100.00$ \\
16265763-3032279 & \object{6037784008760031872} & 9.20 & 1 & 0.5 &  & \object{6037784004457597568} & 3.6 & 391 & $-1.61\pm0.03$ & $0.04\pm0.02$ & 12.58 & 2.12 & $260.3\pm71.7$ & $0.45^{+0.01}_{-0.01}$ & $99.75$ \\
16272794-4542403 & \object{5942410686571895424} & 5.46 & 7 & 1.5 &  & \object{5942410682235163648} & 24.5 & 4495 & $-0.92\pm0.03$ & $-3.75\pm0.03$ & 13.48 & 1.87 & $596.2\pm61.7$ & $3.71^{+0.03}_{-0.02}$ & $0.00$ \\
16303796-2954222 & \object{6031823178275372800} & 8.71 & 11 & 1.1 &  & \object{6031823173966757120} & 5.0 & 569 & $3.18\pm0.05$ & $-0.68\pm0.04$ & 14.86 & 1.69 & $208.6\pm12.0$ & $0.88^{+0.01}_{-0.01}$ & $100.00$ \\
16310436-2404330 & \object{6050627747672189312} & 7.37 & 1 & 0.9 &  & \object{6050627782031927296} & 13.1 & 1771 & $-2.34\pm0.16$ & $0.13\pm0.12$ & 13.02 & 2.33 & $266.6\pm76.8$ & $1.40^{+0.10}_{-0.09}$ & $0.01$ \\
16320058-2530287 & \object{6045791575844270208} & 7.13 & 5 & 0.7 &  & \object{6045791953801392128} & 14.2 & 1998 & $-0.87\pm0.02$ & $-0.25\pm0.02$ & 12.66 & 1.81 & $591.3\pm83.7$ & $0.57^{+0.01}_{-0.01}$ & $100.00$ \\
16320160-2530253 & \object{6045791953801392128} & 7.15 & 3 & 0.8 &  & \object{6045791575844270208} & 14.2 & 1994 & $0.87\pm0.02$ & $0.25\pm0.02$ & 12.47 & 1.72 & $537.2\pm139.6$ & $0.55^{+0.01}_{-0.01}$ & $100.00$ \\
16345314-2518167 & \object{6047289699098449920} & 6.94 & 7 & 1.4 &  & \object{6047289699098450944} & 10.2 & 1465 & $-0.25\pm0.13$ & $0.86\pm0.10$ & 14.83 & 2.81 & $245.6\pm33.7$ & $0.44^{+0.05}_{-0.05}$ & $99.75$ \\
16345314-2518167 & \object{6047289699098449920} & 6.94 & 7 & 1.4 &  & \object{6047289733458188032} & 44.4 & 6402 & $0.36\pm0.14$ & $1.57\pm0.10$ & 15.12 & 2.96 & $205.1\pm25.0$ & $1.66^{+0.11}_{-0.11}$ & $0.00$ \\
16430140-4405275 & \object{5967552634824947456} & 5.19 & 16 & 1.2 &  & \object{5967552634824945152} & 10.8 & 2075 & $-0.29\pm0.12$ & $-0.43\pm0.10$ & 17.09 & 2.71 & $135.4\pm8.9$ & $0.45^{+0.09}_{-0.09}$ & $100.00$ \\
16452615-2503169 & \object{6046749289143173504} & 6.93 & 1 & 1.1 &  & \object{6046749319188591104} & 48.1 & 6932 & $-0.04\pm0.04$ & $0.02\pm0.03$ & 14.39 & 2.90 & $150.5\pm39.3$ & $0.07^{+0.04}_{-0.03}$ & $99.25$ \\
16473710-2014268 & \object{4130416623473262720} & 9.48 & 11 & 1.0 &  & \object{4130416726552477440} & 22.1 & 2335 & $0.40\pm0.10$ & $-0.68\pm0.07$ & 14.57 & 2.58 & $223.9\pm14.7$ & $0.41^{+0.04}_{-0.04}$ & $100.00$ \\
\hline
\end{tabular}
\end{sidewaystable*}

\section{Posterior distribution of the MCMC analysis}
\label{sec:mcmc_posterior}

\begin{figure*}
\resizebox{\hsize}{!}{\includegraphics{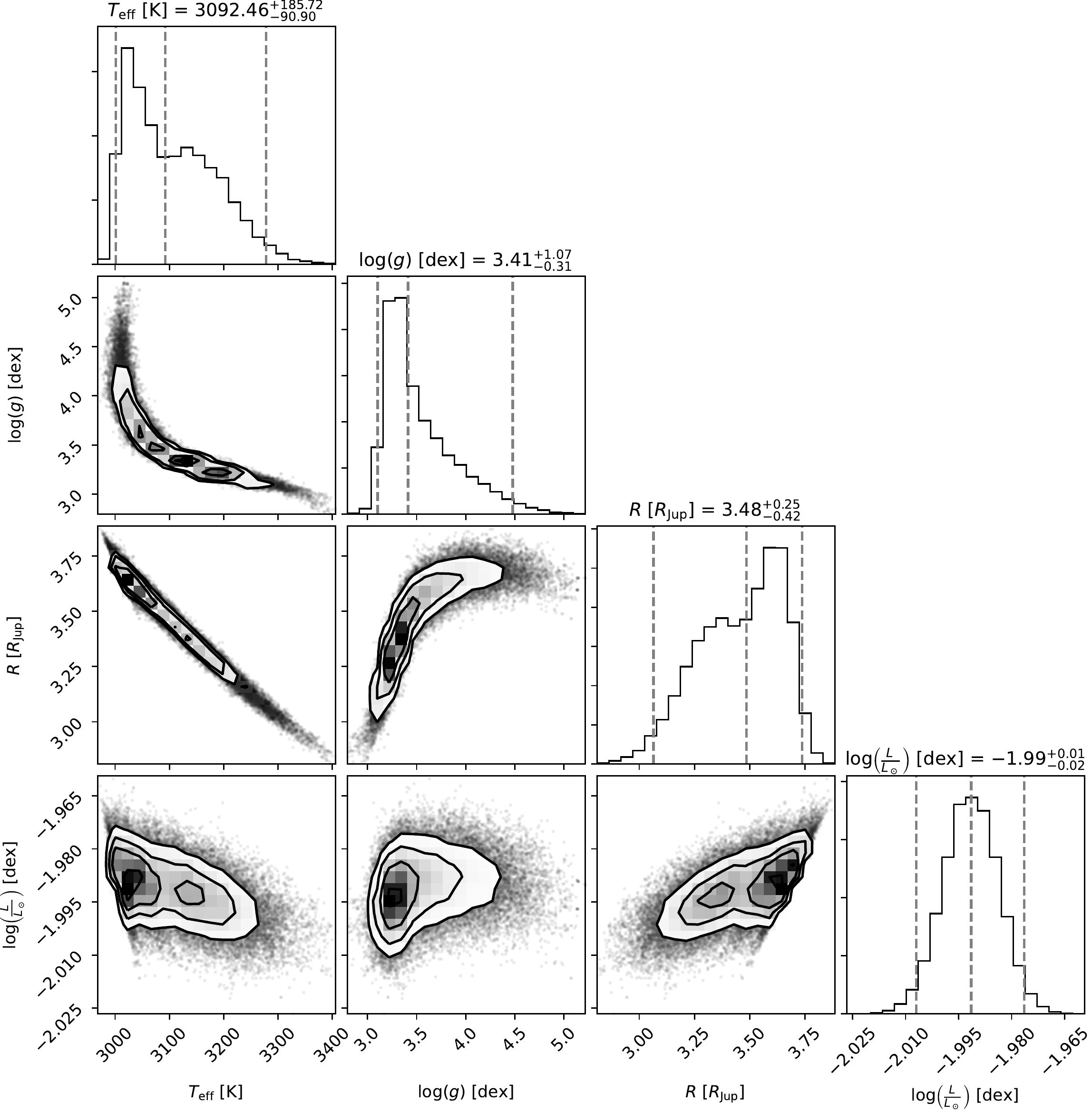}}
\caption{
Posterior distribution of the MCMC-based fit to the SED of TYC~8252-533-1~B.
The blue, dashed lines represent the 2.5th, 50th, and 97.5th percentiles of the marginalized distributions.
}
\label{fig:sed_mcmc_post_yses_companion}
\end{figure*}

In Fig.~\ref{fig:sed_mcmc_post_yses_companion} we present the posterior distribution of our MCMC-based SED fitting of TYC~8252-533-1~B.
After discarding the burn-in phase and thinning the chains, the visualized 45,000 posterior samples remained.
We additionally present the luminosities that are derived from the input parameters ($T_\mathrm{eff}$, $\log\left(g\right)$, $R$) as these quantity is used for the determination of the object's mass.
We derived $T_\mathrm{eff}=3092^{+186}_{-91}\,K$, $\log\left(g\right)=3.41^{+1.07}_{-0.31}$\,dex, $R=3.5^{+0.3}_{-0.4}\,R_\mathrm{Jup}$, and $\log\left(L_*/L_\sun\right)=-1.99^{+0.01}_{-0.02}$ as the 95\,\% quantiles around the medians of the distributions.

\end{appendix}

\end{document}